%% file: arxiv_Dec_16_2022.tex
\documentclass[onefignum,onetabnum]{siamart190516}

\input{ex_shared_arxiv}

\begin{document}

\maketitle

\begin{abstract}
We provide a numerical platform for the analysis of particle shape and topology effect on the macroscopic behavior of granular media. We work within a Discrete Element Method (DEM) framework and apply a peridynamic model for deformable particles accounting for deformation and damage of individual particles. To accommodate arbitrary particle shapes including nonconvex ones as well as particle topology, an efficient method is developed to keep intra-particle peridynamic interaction within particle boundaries. Particle contact with the rigid boundary wall is computed analytically to improve accuracy. To speed up simulations with particles of different shapes and sizes the initial configuration is chosen using security disks containing different particle shapes that are placed in a jammed state using an optimization-based method. The effect of particle shape and topology on settling and compaction of the aggregate for deformable particles is analyzed. \end{abstract}

\begin{keywords}
 peridynamics, damage, granular media, grain shape, DEM
\end{keywords}

\begin{AMS}
    70-08, 70-10, 74A70, 74R10
\end{AMS}

\section{Introduction}%
\label{sec:introduction}

Granular media simulations are important for many industrial and geophysical applications. The discrete element method (DEM) introduced by Cundall and Strack \cite{CundallStrackDiscrete79}, provides one framework for granular simulation assuming that particle shapes are rigid. 
{\rev DEM is a molecular dynamics (MD) model for particles on a representative volume and is used to capture macroscopic transport properties \cite{LudingIntroduction08} of particle aggregates.
The method discussed here is a direct generalization of the DEM method introduced by Cundal and Strack \cite{CundallStrackDiscrete79} for modeling granular assemblies.
The scope of the DEM method addresses applications ranging from rock crushing \cite{HARMON2020112961} and powder rheology \cite{{DesaiEtAlRheometry19}}, to the modeling of vehicles traveling over gravel pavements \cite{RECUERO201739}. Here each rigid particle in the aggregate satisfies Newton's second law of motion applied to particle centers and interacts with other particles through contact forces. 
}

While DEM-based approaches capture the rigid motion of the particle boundaries they do not account for the deformation of individual grains. In addition DEM does not capture complex damage propagation based on each particle’s geometry, nor the effects of {\rev  notched} or pre-cracked particles.  Hence capturing the elastic and inelastic deformation of individual grains will lead to improved models for granular media. With this in mind the ``peridynamic'' models introduced by Silling \cite{SillingReformulation00} are a nonlocal reformulation of continuum mechanics that model elastic deformation but also model fracture growth as an emergent behavior. Recently, peridyamic models have been used for mesoscale modeling of granular media, especially for capturing elastic and inelastic deformation and intra-granular force within individual grains.  Behzadinasab et. al. \cite{BehzadinasabEtAlPeridynamics18} studied shockwave perturbation decay in particle beds of circular grains. Zhu and Zhao \cite{ZhuZhaoModeling19,ZhuZhaoPeridynamic19} used a Weibull statistics-peridynamics approach to investigate crushing piles of sand.  In recent joint work the authors combined the direct element method (DEM) with peridynamics (PeriDEM) to the study of granular flows see, Jha et. al. \cite{JhaEtAlPeridynamicsbased21}. 

The effects of particle shapes on large aggregates have been of recent interest.
In \cite{athanassiadis2014particle,MurphyEtAlTransforming19} authors performed triaxial compression tests on 3D printed shapes to study the stress response of granular packing of various shapes.
Hafez et al \cite{HafezEtAlEffect21} also experimentally studied the effect of particle shapes in particle discharge or clogging.
Using LS-DEM simulations  \cite{kawamoto2018all} captured the shear banding of sand by modeling the shapes of individual sand grains. Motivated by  these considerations we enhance and flesh out the PeriDEM method as a computational platform to assess the effect of particle shape and topology on the aggregate motion of particle beds as influenced by intra-granular elastic deformation and damage. This provides the opportunity to investigate the  motion of the aggregate as a function of the physical properties of the particles and their geometry.

{\rev  In the context of this paper, particles are no longer rigid and all points inside each particle interact with each other peridynamically. This allows for both elastic and inelastic particle deformation. As in DEM the particles interact with other particles through frictional contact forces but now rephrased for deformable particles in an appropriate way. We systematically address all contact forces, friction, and damping to account for interaction  between deformable particles in \Cref{sub:repulsive_contact_force} through \Cref{sub:normal_damping_force}. The transition between static and dynamic friction, i.e. sliding-sticking friction is captured in  \Cref{sub:modification_of_kinetic_friction_force_for_small_velocity}. Force interaction between newly formed  components of a shattered particle are given in \Cref{sub:self_contact}.  The force interaction between particles and container walls is given in  \Cref{sec:intersection_with_the_wall}. Our treatment of peridynamic forces for non convex particles is given in \Cref{sec:peridynamic_bonds_in_non_convex_material}. {\rev  
We note that in order to compare the macroscopic effects associated with assemblages of particles of different shape, it is essential to start the dynamics from an initial particle configuration that is agnostic to particle shape. To realize such a initial configuration we require the maximum cross-sectional diameter of every particle to be a fixed constant across all shapes. Additionally the location of the center point of this diameter is prescribed so each particle experiences no interaction force from any other particle or the container walls. Lastly the particles are randomly oriented. To accomplish this we
employ  the notion of non overlapping security spheres containing one particle each. The methodology behind this construction is given in  \Cref{sec:generating_particle_distribution}.}
In summary the combination of intra-particle and inter-particle interaction  allows for particle comminution and crushing as part of the dynamic rheology. As an example one can think of dense but loosely packed particle aggregates as seen in pebble or sandy roadbeds subject to vehicular traffic. }

As with MD and DEM our simulations  illustrate how  macroscopic properties can be obtained from microscopically dynamic simulations.
{\rev  We begin by discussing time integration and the choice of time step to insure stability in simulations in \Cref{sec:time_integration}.} 
{\revv 
It is pointed out here that the current numerical simulations focus on two dimensional problems.
}
A series of numerical experiments on particle beds/aggregates involving different particle shapes are carried out. The goal is to provide a new opportunity to examine effect of shape and topology on  macroscopic properties from microstructural dynamics. Here we illustrate how the use of particles with convex shapes such as spheres and squares differ from non convex shapes such as crosses and more generally  rough shapes with re-entrant corners.
We start computing the solution to initial value problems for two and three particles in Section \Cref{sec:simulations}. Here we execute two particle collisions with fracture corresponding to the Kalthoff Winkler experiment. In this experiment a rectangular particle {\em the impactor} collides head on with a stationary notched (hence non-convex) particle, see \Cref{sub:experiment_collision_with_fracture}. The experimentally observed fracture pattern is recovered by our numerical simulation of two particles using the peridynamic intra-particle model and the nonlocal contact model.  This provides a corroboration between experiment and the intra-particle deformation and particle to particle contact model for convex and non-convex particles.  Next we illustrate the effect of horizon size on the damage zone for particle fracture for colliding cross shaped particles in \Cref{sub:convergence}.  The generation of non-interpenetrating child particles and non-interpenetration of fissures under compression fracture is illustrated over a range of fracture toughness's in \Cref{sub:experiment_fracture_toughness_and_damage_propagation}.

{\rev  As examples involving particle aggregates  we provide numerical simulations showing the relative effect of particle geometry and topology on macroscopic quantities. The dynamic settling simulations of \cite{Liu,Sahu} using DEM are carried out and are found to produce realistic structural information as obtained through experiment \cite{Finney}. Motivated by this we perform numerical simulations for the settling of particle columns  made from deformable particles with different shapes and topology under gravitational forces. The reaction force on the wall of the particle containers is computed and the effect of particle shape is illustrated, see \Cref{sub:settle}.}  
Next we apply our approach to particles of variable shape and topology exhibiting both inelastic and elastic behavior inside each grain. This allows us to incorporate effects of grain shape and topology through both particle deformation and particle damage. We simulate aggregates subject to dynamic compaction undergoing deformation and damage that highlight these features are provided in \Cref{sub:compaction-damage}. The new methods introduced here provide the tools to study and to design particle shape and topology for desired macroscopic effects not just in the elastic regime, but also when individual grains suffer damage. 

%

\section{Overview of capturing intra-particle deformation and inter-particle interaction and particle interaction with container}%
\label{sec:overview-inter-intra}

    We introduce a particle model of hybrid type for modeling particle aggregates. This method was initiated by the authors together with coauthors in the joint work \cite{JhaEtAlPeridynamicsbased21}. The method includes elastic and inelastic effects inside each particle as well as well as inter particle interaction and boundary effects. This paper extends the methodology and provides new modeling capability for particle crushing and domains containing the aggregate that change shape with time, as well as  eliminating numerical instabilities inherent in sliding friction. 
Let $\Omega\in \mathbb{R}^d$ denote the domain containing the particle assemblage where $d=2$ or $3$ is the dimension. 
{\rev
The assemblage consists of $N$ particles $D_i \subset \Omega$, $i=1,2, ...,N$.
The time-evolution of particle $D_i$ in the media is given by $D_i(t)$, $t\in[0,T]$ with $D_i(0) = D_i$. Let $\xx\in D_i(0)$ denote the coordinates of a material point in the particle in the reference configuration which is taken to be the initial configuration and let $\uu : D_i(0) \times [0,T] \to \mathbb{R}^d$ and $\vv : D_i(0) \times [0,T] \to \mathbb{R}^d$ denote the displacement and velocity fields. 
{\revv At any time $t\in [0,T]$ the coordinates of the material point $\pp=\pp(\xx,t)$ inside $D_i(t)$} is given by $\pp(\xx,t) = \xx + \uu(\xx,t)$ and $\vv(\xx,t)=\dot{\uu}(\xx,t)$. 
}
{\revv The particles are subjected to external forces such as gravitational acceleration and moving container walls that dynamically alter the position of particles.} There are two different interactions in the particle media: intra-particle interaction in which each particle reacts to forces on its boundary, these drive the evolution of internal forces inside each particle; and the inter-particle interaction in which particles come into contact and exchange forces at their interface as well as boundary forces imparted on the particles by moving rigid domain walls. For the former we consider the peridynamic description of solid deformation. For the latter, we propose a Peridynamics-DEM like model to account for exchange of force between particles and domain walls. Since contact in the model is defined at the level of material points in the neighborhood of the contact region, the model can be used to describe contact for arbitrarily shaped particles and particles of different topology.

The dynamics of the particle assemblage is given by an interacting particle system and in this way described by the dynamics of each particle.
The motion of a particle aggregate inside $\Omega$ is given by Newton's second law of motion:
\begin{align}\label{eq:pdmotion}
  {\revv \rho} \ddot{\uu}(\xx,t) = \boldsymbol{F}_{i}^{int}(\xx, t; \uu) + \boldsymbol{F}^{ext}_i(\xx,t; \uu), \ 
  \forall (\xx,t) \in {\revv D_i }\times [0,T], \hbox{for $i=1.\ldots,N$}.
\end{align}
where {\revv $\rho$ is the mass density of the particle,} $\boldsymbol{F}^{int}_i(\xx,t,\uu(\xx,t))$ are the forces inside {\revv$D_i(t)$} and $\boldsymbol{F}^{ext}_i(\xx,t)$ are the external forces on the $i^{th}$ particle  such as inter-particle contact force or a moving container wall acting on the particles. 
We close the above system by specifying an initial condition on the displacement 
$\uu(\xx, 0) = \uu_0(\xx)$ and velocity $\dot{\uu}(\xx, 0) = \vv_0(\xx)$ for all $\xx \in {\revv D_i }$, $i=1,\ldots,N$.
In this way we have framed the particle dynamics for the assemblage as an initial value problem for a displacement field $\uu(\xx,t)$ for $\xx \in 
{\revv D_i }$, $i=1,\ldots,N$ and $t\in[0,T]$ where the details of the intra and inter-particle forces acting on {\revv $D_i(t)$ depend on $\boldsymbol{F}^{int}_i(\xx, t; \uu)$ and $\boldsymbol{F}^{ext}_i(\xx, t; \uu)$.}
In the next sections we show the specific form of $\boldsymbol{F}^{int}_i$ and $\boldsymbol{F}^{ext}_i$ used for capturing intra-particle deformation and inter-particle interaction and interaction with container.

\section{Peridynamic intra-particle force model}%
\label{sec:peridynamic_model}
To model both elastic and inelastic effects inside a particle viewed as a continuum we opt for a nonlocal modelling approach.
The forces acting on a point $\pp(\xx,t)$ inside the $i^{th}$ particle domain $D_i(t) \subset \R^2$ is given by the integro-differential equation
\begin{align}
    \label{eq:peri}
    \boldsymbol{F}^{int}_i(\xx, t; \uu) = \int\limits_{H_\epsilon(\xx) \cap {\revv D_i }}^{} \ff(\uu(\xx', t), \uu(\xx, t), \xx', \xx, t) dV_{\xx'},
\end{align}	
where $\ff$ is the force density function between pairs of points. 
The peridynamic \textit{horizon} is defined as the set $H_\epsilon(\xx) = \{ \xx' \in \R^2 : |\xx' - \xx| \le \epsilon \}$. For a material point $\xx' \in H_\epsilon(\xx) \cap {\revv D_i }$, $V_{\xx'}$ denotes the volume element associated with $\xx'$.
Here the force density has the units  force per unit volume$^2$ or $area^2$ depending on the dimension and the first term on the right hand side  of \Cref{eq:peri} is the total force exerted on $\xx$ by its surrounding neighborhood. 
Here the force density is given for two-point interactions. This is called bond-based peridynamics \cite{SillingReformulation00}. 
Given $\xx \in D_i$ and $\xx' \in H_\epsilon(\xx)\cap D_i$, the vector $\xiB = \xx' - \xx$ is referred to as a \textit{bond}. Defining $\etaB = \uu(\xx', t) - \uu(\xx, t)$, the \textit{stretch} $s$ associated with a bond $\xiB$ is defined as
$
s=s(\uu(\xx', t), \uu(\xx,t), \xx', \xx):= \frac{\abs{\xiB + \etaB } - \abs{\xiB} }{\abs{\xiB} }{\revv=\frac{\abs{\pp(\xx',t)-\pp(\xx,t) } - \abs{\xx'-\xx} }{\abs{\xx'-\xx} } }.
$

{\revvv Next we describe the constitutive law relating force density to stretch between two points.}
{\rev A microelastic material considered in
\cite{BobaruEtAl09}
is given by the pairwise force density function $\ff$ of the form
}
\begin{align}
    \label{pmb}
    \ff = 
    \begin{cases}
	c_w\ w(\abs{\xiB} )\ s\  \frac{\xiB + \etaB}{\abs{\xiB + \etaB} }  & \text{ if } \abs{\xiB} < \epsilon \\
	0 & \text{ otherwise},
    \end{cases}
\end{align}	
where the \textit{micromodulus function} $w(r)$ is a non-negative scalar function that is non-increasing in $r$. The peridynamic spring constant $c_w$ is chosen such that the integral operator  agrees with the Cauchy-Navier operator up to the second order. 
We list the value of the peridynamic spring constant and two micromodulus functions in \Cref{tab:2d-peri}.
{\rev 
The model with constant micromodulus is called the Prototype Microelastic Brittle (PMB) material and was introduced in \cite{SillingAskariMeshfree05}.
}

\subsection{Irreversible damage and memory}%
\label{sec:damage_model}
The bond $\xiB$ between $\xx$ and $\xx'$ is broken at time $t$ if the stretch $s(\uu(\xx', t), \uu(\xx,t), \xx', \xx)$ exceeds the \textit{critical stretch} $s_0$ in the absolute value, i.e. when $\abs{s} > \abs{s_0}$.  The value of $s_0$ is determined by equating the critical energy release rate $G_c$ with the total energy required to sever all bonds across a crack surface of unit area. 
In \Cref{tab:2d-peri}, we  list the value of the critical stretch for both ``constant'' and ``conic''  micromodulus functions \cite{HaBobaruStudies10}.
Once a bond is broken at time $t = t_0$, it remains broken for all time $t > t_0$.
The \textit{damage} {\revv of a material point $\xx$ is defined as the ratio of the number of broken bonds connected to $\xx$ at time $t$ to the number of bonds connected to $\xx$ in the reference configuration (i.e. at $t=0$)}.
\begin{table}[ht]
    \centering
    \caption{Peridynamic spring constant for various choices the micromodulus function in \Cref{pmb}.}
    \label{tab:2d-peri}
    \begin{tabular}{| c |  c  | c | c | c |}
	\hline
	Type & 
 $w(\abs{\xiB} )$ & $c_w$  & $c_w w(\abs{\xiB})$ & $s_0$ \\
	\hline
	\hline
	Constant &
	$1$ & $\frac{6 E}{\pi \epsilon^3(1 - \nu)}$  & $\frac{6 E}{\pi \epsilon^3(1 - \nu)}$ & $\sqrt{\frac{4 \pi G_c}{9 E \epsilon} }$ \\
	\hline
	Conic & 
	 $ \left(1 - \frac{\abs{\xiB} }{\epsilon}\right) $ & $\frac{24 E}{\pi \epsilon^3(1 - \nu)}$  & $  \frac{24 E}{\pi \epsilon^4(1 - \nu)}\left(\epsilon - \abs{\xiB} \right)$ & $\sqrt{\frac{5 \pi G_c}{9 E \epsilon} }$  \\
	 \hline
    \end{tabular}
\end{table}

\begin{rem}
Because bond-based peridynamics is a two-point interaction model the Poisson ratio is $\frac{1}{3}$ in 2D and $\frac{1}{4}$ in 3D \cite{TrageserSelesonBondBased20}. While this limitation can be easily overcome by using a state-based model \cite{SillingEtAlPeridynamic07}, we do not pursue that here. 
\end{rem}
In the following section we introduce the different inter particle forces and wall forces then combine them with \Cref{eq:peri} to get the equation of evolution for every particle in the aggregate given in \Cref{sub:combined_model}. 

\section{Contact model}%
\label{sec:contact_model}
To capture the inter grain interactions we apply the short-range contact force model used by \cite{SillingAskariMeshfree05} and \cite{BehzadinasabEtAlPeridynamics18}.
In the short-range contact model, two material points belonging to two different peridynamic bodies are said to be in contact if they are within a certain distance $R_c$, called the \textit{contact radius}.
Let $D_i$ and $D_j$ be two particles ($i \ne j$) and $\pp(\xx,t) \in D_i(t)$ and $\pp(\yy,t) \in D_j(t)$. 
We define the \textit{normal direction} $\ee(\yy, \xx, t)$ by the unit vector
$
    \ee(\yy, \xx, t) = \frac{\pp(\yy, t) - \pp(\xx, t)}{\abs{\pp(\yy, t) - \pp(\xx, y)} }
    $
which denotes the direction from $\xx$ to $\yy$.

\subsection{Repulsive contact force}%
\label{sub:repulsive_contact_force}

 The short-range repulsive force $\FF_r(\yy, \xx, t)$ exerted on $\pp(\xx,t) \in D_i(t)$ by  $\pp(\yy,t) \in D_j(t)$ (see \cite{SillingAskariMeshfree05, BehzadinasabEtAlPeridynamics18, JhaEtAlPeridynamicsbased21}) is given by 
\begin{align}
    \label{eq:law-repulsive}
    \begin{split}
    &\FF_r(\yy, \xx, t)
    \\
    & = 
    \begin{cases}
     -K_n (R_c - |\pp(\yy, t) - \pp(\xx,t)|) V_{\xx} V_{\yy} \ee(\yy, \xx,t) & \text{if } |\pp(\yy,t) - \pp(\xx,t) | < R_c \\
    0 & \text{ otherwise } 
    \end{cases}
    \end{split}
\end{align}
where the normal contact stiffness is $K_n = \frac{18k}{\pi \epsilon^5}$, where $k$ is the bulk modulus, $V_{\xx}$ and $V_{\yy}$ are volume elements associated with $\xx$ and $\yy$, respectively \cite{BehzadinasabEtAlPeridynamics18,JhaEtAlPeridynamicsbased21}.
When the participating peridynamic bodies have bulk moduli $k_1$ and $k_2$, respectively, an effective bulk modulus is used and  given by the harmonic mean, i.e. $k = \frac{2 k_1 k_2}{k_1 + k_2} $ \cite{JhaEtAlPeridynamicsbased21}.
Here the repulsive force is chosen to be  linear in the distance between the points in contact. 
A nonlinear relation is also possible but over short distances they are comparable \cite{DesaiEtAlRheometry19}.

We conclude from \Cref{eq:law-repulsive} that the total repulsive force on a point $\pp(\xx.t)\in D_i(t)$ due to all neighboring particles is given by
\begin{align}
\label{eq:sum-repulsive}
	\sum_{j \ne i}^{} 
	\int
	\limits_{{\revv \{ \yy \in D_j : \abs{\pp(\yy,t) - \pp(\xx,t)} < R_c \}}}^{}  
	\ff_r(\yy, \xx, t)  dV_{\yy},
\text{ where } \ff_r = \frac{\FF_r}{V_{\xx}}. 
\end{align}

\subsection{Nonlocal friction}%
\label{sub:modification_of_kinetic_friction_force_for_small_velocity}

{\revv
\begin{figure}[htpb]
 \centering
     \subfloat[]{\label{fig:drawing-fric}
         \includegraphics[width=0.48\linewidth]{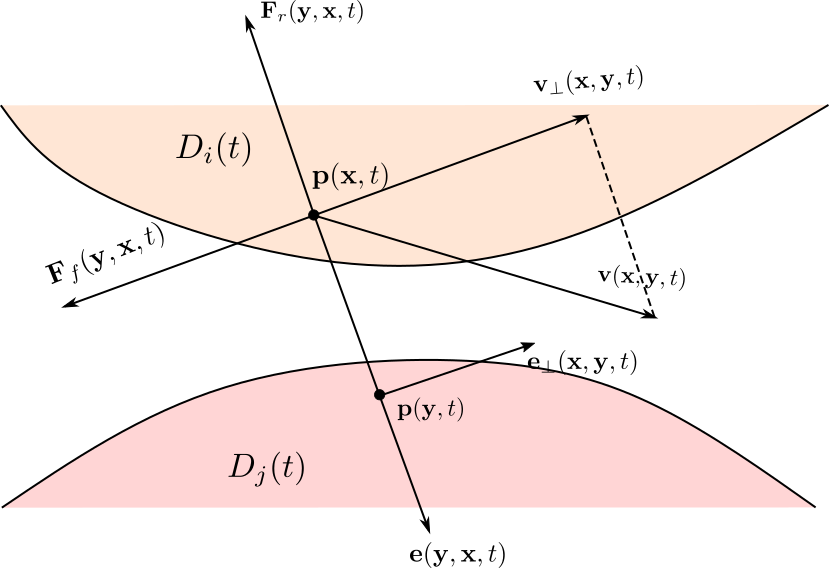}
     }
     \subfloat[]{\label{fig:drawing-regime}
         \includegraphics[width=0.48\linewidth]{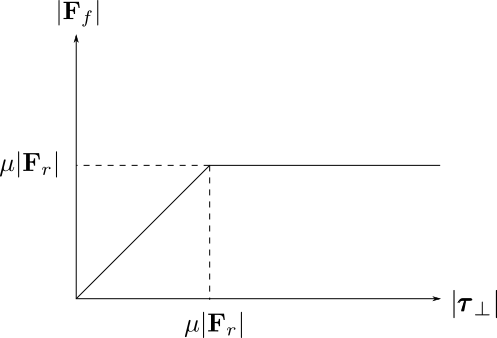}
     }
 \caption{\revv (a) Geometric picture for friction force (b) Friction force vs the prescribed tangential force and the transition between sticking and slipping regimes.}
 \label{fig:fric}
 \end{figure}
A tangential damping force is incorporated for nodes in contact to model energy dissipation due to friction using Coulomb's law.
Such forces are incorporated in DEM models in terms of virtual spring displacement (see, for example, \cite{LudingIntroduction08}).
To capture the stick/slip transition in our dynamic friction model, we follow the approach of regularized Coulomb's model \cite{martinsOden1983numerical,campos1982numerical} adapted to the nonlocal setting. 
Let the relative velocity of $\pp(\xx, t) \in D_i(t)$ with respect to $\pp(\yy, t) \in D_j(t)$ be $\vv(\xx, \yy, t) = \dot{\uu}(\xx, t) - \dot{\uu}(\yy, t)$.
The tangential component of the relative velocity of $\pp(\xx,t)$ with respect to $\pp(\yy,t)$ is therefore given by
$
    \vv_\perp(\xx, \yy, t) = \vv(\xx, \yy, t) - \left(\vv(\xx, \yy, t) \cdot \ee(\yy, \xx, t)\right) \ee(\yy, \xx, t)
$.
We define the tangential contact direction by
$
    \ee_\perp(\xx, \yy, t) = \frac{\vv_\perp(\xx, \yy, t)}{\abs{\vv_\perp(\xx, \yy, t)} },
    $
    see Figure \ref{fig:drawing-fric}.
The friction force $\FF_f(\yy, \xx, t)$ on $\pp(\xx,t)$ due to  $\pp(\yy,t)$ 
depends on the prescribed tangential force $\pmb{\tau}_\perp(\xx, \yy, t)$  on $\pp(\xx,t)$ in the direction $\ee_\perp(\xx, \yy, t)$. 
In the sticking regime i.e., when $\abs{\pmb{\tau}_\perp(\xx, \yy, t)} \le \mu \abs{\FF_r{\yy, \xx, t}}$, the friction force acts as a restoring force to prevent the motion of $\pp(\xx, t)$ with respect to $\pp(\yy, t)$. In the slipping regime, i.e., when $\abs{\pmb{\tau}_\perp(\xx, \yy, t)} > \mu \abs{\FF_r{\yy, \xx, t}}$, the friction force has a constant magnitude of $\mu \abs{\FF_r(\yy, \xx, t)}$ and acts in the opposite direction of $\vv_\perp(\xx, \yy, t)$. The transition of the friction force between the sticking and slipping regime depending on the prescribed tangential force is shown in \Cref{fig:drawing-regime}.
As in \cite{martinsOden1983numerical}, the stick/slip transition in our model is characterized by the relative speed $\abs{\vv_\perp(\xx, \yy,t)}$ crossing a speed threshold $v_\text{thr}$. Here, $v_\text{thr}$ is taken to be depending on the numerical time step $\Delta t$ and is given by $v_{\text{thr}}(\xx, \yy, t, \Delta t) = \frac{\mu}{\rho V_{\xx}} \abs{\FF_r(\yy, \xx, t)} \Delta t$,
which is obtained by approximating the maximal magnitude of $\vv_\perp(\xx, \yy, t)$ resulting from the prescribed tangential force $\pmb{\tau}_\perp(\xx, \yy, t) = \mu \abs{\FF_r(\yy, \xx, t)} \ee_\perp(\xx, \yy, t)$ in the absence of friction.
Note that $v_{\text{thr}}(\xx, \yy, t, \Delta t) \to 0$ as $\Delta t \to 0$.

The frictional force $\FF_f$ on $\pp(\xx,t) \in D_i(t)$ due to  $\pp(\yy,t) \in D_j(t)$ is given by 
\begin{align}
\label{eq:friction}
    \FF_f(\yy, \xx, t) = 
    \begin{cases}
	- \mu \abs{\FF_r(\yy, \xx, t)} \ee_\perp(\yy, \xx,t), &\text{ if } \abs{\vv_\perp(\xx, \yy, t)} > v_{\text{thr}}(\xx, \yy, t, \Delta t)
	\\
	- \frac{1}{\epsilon_f} \abs{\vv_\perp(\xx, \yy, t)} \ee_\perp(\yy, \xx,t), &\text{ if } 0 \le \abs{\vv_\perp(\xx, \yy, t)} \le v_{\text{thr}} (\xx, \yy, t, \Delta t)
,
    \end{cases}
\end{align}	
where $\epsilon_f$ is a  $\Delta t$-dependent regularization parameter given by $\epsilon_f = \frac{\Delta t}{\rho V_{\xx}}$.
$\epsilon_f$ is chosen so that once $\abs{\pmb{\tau}_\perp(\xx, \yy, t)} \le \mu |\FF_r(\yy,\xx,t)|$, the relative velocity of $\pp(\xx,t)$ reduces to zero in the next iteration. 
Related regularized nonlocal tangential dynamic friction models that employ \cite{martinsOden1983numerical} between particles are introduced in \cite{kamensky2019peridynamic}.

}

Consequently, the total friction force on the point $\pp(\xx,t) \in D_i(t)$ due to all the neighboring particles is given by
\begin{align}
\label{eq:sum-friction}
	\sum_{j \ne i}^{} \int\limits_{{\revv \{ \yy \in D_j }: \abs{\pp(\yy,t) - \pp(\xx,t)} < R_c \}}^{}  \ff_f(\yy, \xx, t)  dV_{\yy},
\text{ where } \ff_f = \frac{\FF_f}{V_{\xx}}. 
\end{align}

Note that the frictional force is dependent on relative velocity between points taken from different particles in contact and is implemented using the velocity-Verlet scheme described in \Cref{sec:time_integration}. 

\subsection{Normal damping force}%
\label{sub:normal_damping_force}

In this implementation we incorporate normal damping to allow energy dissipation upon normal contact between particles.
The damping force results in shortened relaxation times and therefore lowers computational costs for simulations approaching mechanical equilibrium.
The damping force on $\pp(\xx,t) \in D_i(t)$ due to $\pp(\yy,t) \in D_j(t)$ is given by
\begin{align}
\label{eq:normal-damp}
\FF_d(\yy,\xx,t) = 
\begin{cases}
    -\beta_d \left( \frac{\vv(\yy, \xx, t)}{\abs{\vv(\yy, \xx, t)} } \cdot \ee(\yy, \xx, t) \right) \ee(\yy,\xx,t) V_{\xx} V_{\yy} & \text{ if } \abs{\pp(\yy,t) - \pp(\xx,t) } < R_c \\ 
    0 & \text{ otherwise } 
\end{cases}
\end{align}
where $\beta_d$ is the damping coefficient given by $\beta_d = 2 r_d \sqrt{\frac{K_n \rho}{V_\yy}}$ and $r_d \in [0, 1]$ is the damping ratio.
The combined damping forces acting on the point $\pp(\xx,t) \in D_i(t)$ due to all other particles is therefore given by
\begin{align}
\label{eq:sum-damping}
	\sum_{j \ne i}^{} \int\limits_{ \{ {\revv\yy \in D_j} : \abs{\pp(\yy,t) - \pp(\xx,t)} < R_c \}}^{}  \ff_d(\yy, \xx, t)  dV_{\yy},
\text{ where } \ff_d = \frac{\FF_d}{V_{\xx}}. 
\end{align}

{\revv
The expression for the damping coefficient $\beta_d$ is derived from the Kelvin-Voigt model for a damped spring \cite{HuntCrossleyCoefficient75}.
Combining \Cref{eq:law-repulsive} and \Cref{eq:normal-damp}, the magnitude of the pairwise damped repulsion force on $\pp(\xx,t)$ due to $\pp(\yy,t)$ in the direction $\ee(\yy, \xx,t)$ is given by
$
-K_n V_{\yy}V_{\xx} (R_c - \abs{\pp(\yy, t) - \pp(\xx,t)}) - \beta_d V_{\yy}V_{\xx}  \vv_{n}(\yy, \xx, t),
$
where $\vv_n(\yy, \xx, t) = \frac{\vv_(\yy, \xx,t)}{\abs{\vv(\yy, \xx,t)}} \cdot \ee(\yy, \xx,t)$
and is equivalent to that of a viscoelastic spring connecting $\pp(\xx,t)$ and $\pp(\yy,t)$ with reference length $R_c$, spring constant $k = K_n V_{\yy}V_{\xx}$, and damping coefficient $c = \beta_d V_{\xx}V_{\yy}$. 
%
{\revv In terms of the damping ratio $r_d$, the damping coefficient is written as $ c = 2 r_d \sqrt{K_n V_{\xx}^2 V_{\yy} \rho}$,
which implies $\beta_d   = 2 r_d \sqrt{\frac{K_n \rho}{V_{\yy}}}$.
   When damping is absent, (i.e. $r_d = 0$), we have pairwise conservation of energy
    \begin{align*}
	\frac{d}{dt} \left( \frac{1}{2} \rho V_{\xx} \abs{\vv_n(\yy, \xx, t) }^2 + \frac{1}{2} k \left( R_c - \abs{\pp(\xx, t) - \pp(\yy,t)} \right)^2 \right) = 0.
    \end{align*}	
In the presence of damping, the energy dissipation rate is given by
    \begin{align*}
	\frac{d}{dt} \left( \frac{1}{2} \rho V_{\xx} \abs{\vv_n(\yy, \xx, t)}^2 + \frac{1}{2} k \left(R_c - \abs{\pp(\xx, t) - \pp(\yy,t)} \right)^2 \right) = - c \abs{\vv_n(\yy, \xx,t)}^2
.
    \end{align*}	
    }
}
The damping force acts in the opposite direction of $\ee(\yy,\xx,t)$ and the magnitude of the damping force depends on the magnitude of normal projection of relative velocity.
A nonlinear damping model such as \cite{JankowskiAnalytical06} can be considered but we do not consider that here.
We remark that introducing such damping forces between the material points within the same particle leads to a viscoelastic material model \cite{SillingAttenuation19}.

It is remarked that our contact model is an improvement of the one used in \cite{JhaEtAlPeridynamicsbased21}, where a particle  damping is implemented on the centroid of the particle and is dependent on the mean velocity of the nodes in a particle. The damping model presented here enables us to specify the physical law of damping at the nodal (local) level that manifests in particle level (global) damping. The law also applies to nodes that have undergone damage and are isolated from its parent particle.

\subsection{Self-contact}%
\label{sub:self_contact}

The presence of a peridynamic bond between two material points from the same parent particle provides the necessary repulsive force to ensure that the points do not overlap numerically.
However, such repulsive forces are absent when the peridynamic bond between the points is broken.
Therefore, we specify a self-contact law between nodes of the same parent particle which are not connected by a peridynamic bond but are close to each other due to large deformations. 
This is especially important for preventing the numerical inter-penetration of different parts of nonconvex particle shapes, and for modeling the contact between various broken segments of a parent particle where a peridynamic force is absent.

Our self-contact law depends on the distance between nodes in the reference (undeformed) configuration. 
If there is no peridynamic bond between two nodes $\xx$ and $\yy$ from the same parent particle $D_i$ at time $t$ and the current distance between them is within $R_c$ (i.e. if $ \abs{\xiB + \etaB} < R_c$), the normal repulsive force on $\xx$ due to $\yy$ is given by
\begin{align*}
    \ff_r^{\text{self}}(\xx, \yy, t) = 
    \begin{cases}
	c_w \frac{\abs{\xiB + \etaB} - \abs{\xiB}}{\abs{\xiB}} \frac{\xiB + \etaB}{\abs{\xiB + \etaB}} \chi_{ \{ \abs{\xiB + \etaB} < \abs{\xiB} \}}, 
	&\text{ if } \abs{\xiB} < R_c
	\\
	c_w \frac{\abs{\xiB + \etaB} - R_c}{R_c} \frac{\xiB + \etaB}{\abs{\xiB + \etaB}} \chi_{ \{ \abs{\xiB + \etaB} < R_c \}}, 
	&\text{ if } R_c < \abs{\xiB} ,
    \end{cases}
\end{align*}	
where $\chi_S$ is the characteristic function of the set $S$.

Note that in the small reference length scale $\abs{\xiB} < R_c < \epsilon$, the repulsive contact force is modeled using a repulsive-only peridynamic bond force. This ensures that two nodes with reference distance $\abs{\xiB} < R_c$ do not experience any repulsive force from each other unless 
they come closer than their reference distance $ \abs{\xiB} $.
If $\abs{\xiB} > R_c$, the contact force between nodes from the same parent particle is same as the contact force between nodes from different parent particles if $\frac{c_w}{R_c} = K_n$.

\subsection{Contact with wall}%
\label{sec:intersection_with_the_wall}
{\rev 
 \begin{figure}[htpb]
 \centering
     \subfloat[]{\label{fig:data/2d-segment.pdf}
         \includegraphics[width=0.45\linewidth]{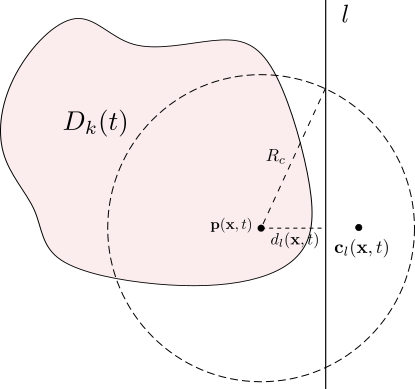}
     }
     \subfloat[]{\label{fig:data/2d-segment-intersect.pdf}
      \includegraphics[width=0.45\linewidth]{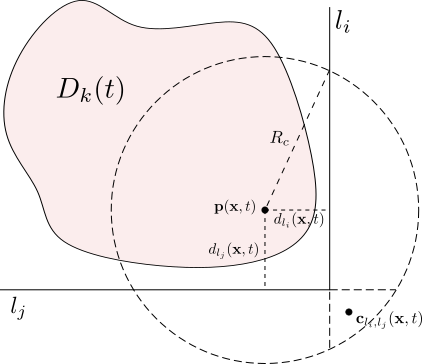}
     }\
 \caption{{\revv Computation of wall contact volume and the centroid of wall volume responsible for exerting contact forces for contact with (a) single and (b) double wall boundaries. The contact radius is exaggerated to show the quantities of interest.}}%
 \label{fig:chords}
 \end{figure}
{\revv
In earlier works \cite{JhaEtAlPeridynamicsbased21,BehzadinasabEtAlPeridynamics18}, the wall of the container containing all the particles is treated as a peridynamic domain. For our purposes, the wall is considered to be rigid (i.e., not deformable) and of thickness at least $\frac{R_c}{2}$. The inner boundary of the container wall is assumed to be rectangular, consisting of straight lines $L = \{l_i\}_{i=1}^4$.
The point $\pp(\xx, t)$ inside the particle $D_k(t)$ experiences contact force due the wall if there exists $l \in L$ such that the perpendicular distance from $\pp(\xx,t)$ to $l$ is smaller than the contact radius $R_c$. 
In this case, we denote the set of all wall points exerting contact force on $\pp(\xx,t)$ by the set $S_l(\xx,t)$, which is a circular segment of the disk $B_{R_c}(\pp(\xx,t))$.
}
In this case, the repulsive force due to the wall on $\pp(\xx,t)$ is given by
\begin{align}
    \label{eq:wall-repulsive}
    \FF_r^{l }(\xx, t) = 
    \begin{cases}
	-K_n (R_c - \abs{\cc_l(\xx,t) - \pp(\xx,t)}) V_{\xx} \abs{S_l(\xx,t)} \ee^l(\xx,t) & \text{ if } d_l(\xx,t) < R_c \\
    0 & \text{ otherwise },
    \end{cases}
\end{align}	
{\revv where $|S_l(\xx, t)|$ is the volume of $S_l(\xx,t)$}, $\cc_l(\xx,t)$ is the centroid of $S_l(\xx,t)$, and $d_l(\xx,t)$ is the distance from $\pp(\xx,t)$ to $l$ (see \Cref{fig:data/2d-segment.pdf}). 
$\ee_{l}(\xx,t)$ is the unit vector in the direction $\cc_l(\xx,t)$ from $\pp(\xx,t)$ given by
$
    \ee_{l}(\xx,t) = \frac{\cc_l(\xx,t) - \pp(\xx,t)}{\abs{\cc_l(\xx,t) - \pp(\xx,t)}}
$
and determines the direction of repulsive wall force on $\pp(\xx,t)$.

Near the corner of a rectangular container, the point $\pp(\xx,t)$ {\revv will experience contact forces from two inner wall boundary segments $l_i$ and $l_j$ if $d_{l_i,l_j}(\xx, t) := \sqrt{d_{l_i}^2(\xx, t) + d_{l_j}^2(\xx, t)} < R_c$ (see \Cref{fig:data/2d-segment-intersect.pdf}). In this case, the repulsive force on $\pp(\xx,t)$ due to the wall is given by}
$
    \FF_{r}^{l_i \cup l_j}(\xx, t) 
    = 
	\FF_r^{l_i}(\xx, t) + \FF_r^{l_j}(\xx,t) - \FF_r^{l_i,l_j}(\xx, t)
$,
where
\begin{align*}
    \FF_r^{l_i,l_j}(\xx, t) 
    =
    -K_n (R_c - \abs{\cc_{l_i,l_j}(\xx,t)
	- \pp(\xx,t)}) V_{\xx} \abs{S_{l_i,l_j}(\xx,t)} \ee_{l_i,l_j}(\xx,t),
\end{align*}	
$S_{l_i,l_j} = S_{l_i} \cap S_{l_j}$, $\cc_{l_i,l_j}(\xx,t)$ is the centroid of $S_{l_i,l_j}$, and 
$
    \ee_{l_i,l_j}(\xx,t) = \frac{\cc_{l_i,l_j}(\xx,t) - \pp(\xx,t)}{\abs{\cc_{l_i,l_j}(\xx,t) - \pp(\xx,t)}}.
$
Therefore, for a rectangular wall boundary with straight line segments $ \{ l_i \}_{i=1}^4$, the repulsive contact force due to the wall boundary is given by
\begin{align}
    \label{eq:wall-total}
    \FF^{\text{wall}}_r(\xx,t) = \sum_{i=1}^{4} \FF_r^{l_i}(\xx,t) - \sum_{i < j}^{} \FF_r^{l_i,l_j}(\xx,t).
\end{align}	
The total friction and damping forces $\FF^{\text{wall}}_f(\xx,t)$ and $ \FF^{\text{wall}}_d(\xx,t) $ due to the wall boundary can be 
{\revv computed accordingly from \Cref{eq:friction} and \Cref{eq:normal-damp}, respectively, by replacing $\ee(\yy, \xx,t)$ by $\ee^l(\xx,t)$, $V_{\yy}$ by $\abs{S_l(\xx,t)}$, and $\pp(\yy,t)$ by $\cc_l(\xx,t)$.
}

Since the wall boundaries are not deformable, the wall volumes $|S_l|$ and $|S_{l_i, l_j}|$ participating in exerting contact forces due to single and double wall boundaries can be computed analytically.
{\revv 
This reduces the computational cost of the simulations as one does not need to discretize the wall during simulations.} For the convenience of the reader, we present the analytical expressions here.

For a single-wall contact, the area of the circular segment $S_l$ is given by 
\begin{align*}
    \abs{S_l(\xx, t)} =  \frac{1}{2} \, \pi R_c^{2} - R_c^{2} \arcsin\left(\frac{d_{l}(\xx,t)}{R_c}\right) - \sqrt{R_c^{2} - d_{l}(\xx,t)^{2}} d_l(\xx,t)
\end{align*}	
and the distance from the point $\pp(\xx,t)$ to the centroid of $S_l(\xx,t)$ is
\begin{align*}
    \abs{\cc_l(\xx,t) - \pp(\xx,t)} = \frac{2}{3}  \frac{1}{\abs{S_l(\xx,t)} }
	(R_c^2 - d_l(\xx,t)^2)^{\frac{3}{2}}.
 \end{align*}	
Note that when $\pp(\xx,t)$ touches the wall boundary segment $l$ we have $d_l(\xx,t)=0$, therefore the effective wall contact volume is $\abs{S_l(\xx,t)} = \frac{\pi R_c^2}{2}$, the area of the half-circle.

For the contact with two wall boundary segments $l_i$ and $l_j$, the effective wall volume responsible for exerting contact force on $\pp(\xx,t)$ is $ \abs{S_{l_i}(\xx,t) \cup S_{l_j(\xx,t)}}$. {\revv We show how to compute the effective wall volume for this case}. }Using the inclusion-exclusion principle, we have
$
 \abs{S_{l_i}(\xx,t) \cup S_{l_j(\xx,t)}} = 
    \abs{S_{l_i}(\xx,t)} + \abs{S_{l_j}(\xx,t)} - \abs{S_{l_i}(\xx,t) \cap S_{l_j}(\xx,t)}.
$
Integrating in polar coordinates, we obtain
\begin{align*}
    &
    \abs{S_{l_i}(\xx,t) \cap S_{l_j}(\xx,t)}
    \\
    & = -\frac{1}{2} \, R_c^{2} \arcsin\left(\frac{d_{l_i}(\xx,t)}{R_c}\right)  
+ \frac{1}{2} \, R_c^{2} \arcsin\left(\frac{\sqrt{R_c^{2} - d_{l_j}(\xx,t)^{2}}}{R_c}\right) 
    \\
    & \quad \quad
    + d_{l_i}(\xx,t) d_{l_j}(\xx,t) - \frac{1}{2} \, \sqrt{R_c^{2} - d_{l_i}(\xx,t)^{2}} d_{l_i}(\xx,t) - \frac{1}{2} \, \sqrt{R_c^{2} - d_{l_j}(\xx,t)^{2}} d_{l_j}(\xx,t)
\end{align*}	
Let the centroid of $ S_{l_i}(\xx,t) \cap S_{l_j}(\xx,t) $ be denoted by $\cc_{l_i,l_j}(\xx,t)$. Then, we have
\begin{align*}
    \cc_{l_i,l_j}(\xx,t) - \pp(\xx,t) =  \frac{1}{\abs{ S_{l_i}(\xx,t) \cap S_{l_j}(\xx,t) } } \begin{bmatrix}
    I_x \\ I_y
\end{bmatrix},
\end{align*}	
where the moments of inertia $I_x$ and $I_y$ are given by
\begin{align*}
    I_x & = \frac{1}{6} \, d_{l_j}(\xx,t)^{3} - \frac{1}{2} \, {\left(R_c^{2} - d_{l_i}(\xx,t)^{2}\right)} d_{l_j}(\xx,t) + \frac{1}{3} \, {\left(R_c^{2} - d_{l_i}(\xx,t)^{2}\right)}^{\frac{3}{2}}
    \\
    I_y &= -\frac{1}{2} \, R_c^{2} d_{l_i}(\xx,t) + \frac{1}{6} \, d_{l_i}(\xx,t)^{3} + \frac{1}{2} \, d_{l_i}(\xx,t) d_{l_j}(\xx,t)^{2} + \frac{1}{3} \, {\left(R_c^{2} - d_{l_j}(\xx,t)^{2}\right)}^{\frac{3}{2}}.
\end{align*}	
Note that when $d_{l_i}(\xx,t) = d_{l_j}(\xx,t) =0$, i.e., when the point $\pp(\xx,t)$ touches the wall corner, we have
$ \abs{S_{l_i}(\xx,t) \cap S_{l_j}(\xx,t)} = \frac{\pi R_c^2}{4} $,
the area of the quarter of the contact circle $B_{R_c}(\pp(\xx,t))$.

\subsection{Combined model for the particle aggregate}%
\label{sub:combined_model}

The equation of motion for the particle aggregate can now be given explicitly. Combining all the forces, the equation of motion of the points $\mathbf{p}(\xx,t) \in D_i(t)$, $i=1,\dots,N$ is given by
\begin{align}
\label{eq:full}
\begin{split}
    & \rho \ddot{\uu}(\xx, t) 
     = 
    \int\limits_{H_\epsilon(\xx) \cap D_i}^{} \ff(\xx', \xx, t) dV_{\xx'}
    \\
    + & \int\limits_{\{\yy  \in D_i: \abs{\pp(\yy,t) - \pp(\xx,t)} < R_c \}}^{} \left(  \ff^{\text{self}}_r(\xx', \xx, t)
    +\ff^{\text{self}}_d(\xx', \xx, t) 
    +\ff^{\text{self}}_f(\xx', \xx, t) 
    \right) dV_{\xx'}
    + \Bb(\xx, t) \\
    + & 	\sum_{j \ne i}^{} \int\limits_{ \{ {\revv \yy \in D_j} : \abs{\pp(\yy,t) - \pp(\xx,t)} < R_c \}}^{} \left( \ff_r(\yy, \xx, t) + \ff_d(\yy, \xx, t) + \ff_f(\yy, \xx, t) \right) dV_{\yy}\\
    + & \ff_r^{\text{wall}}(\xx, t)+\ff_d^{\text{wall}}(\xx, t)+\ff_f^{\text{wall}}(\xx, t),\hbox{  for $\xx\in D_i(0),\,\,i=1,\ldots,N$}
\end{split}
\end{align}	
where $\ff_r = \frac{\FF_r}{V_{\xx}}$, $\ff_d = \frac{\FF_d}{V_{\xx}}$, and $ \ff_f = \frac{\FF_f}{V_{\xx}}$ are the repulsive, damping, and friction force density functions, respectively, and they all have the unit of  force/volume$^2$. Here the wall forces are expressed as the body force densities $\ff_r^{\text{wall}} = \frac{\FF_r^{\text{wall}}}{V_{\xx}}$, $\ff_r^{\text{wall}} = \frac{\FF_d^{\text{wall}}}{V_{\xx}}$, and $\ff_r^{\text{wall}} = \frac{\FF_f^{\text{wall}}}{V_{\xx}}$ .


\section{Peridynamic bonds in nonconvex domains}
\label{sec:peridynamic_bonds_in_non_convex_material}
Nonconvex particle contacts were handled in DEM-based approaches recently using level sets \cite{kawamoto2018all} and using convex-gluing methods 
\cite{GovenderEtAlHopper18}\cite{RakotonirinaEtAlGrains3D19}.
However, in the peridynamic treatment of nonconvex particles, a technical challenge arises where one needs to identify bonds that extend outside the domain.
If the peridynamic domain $D$ is nonconvex, there exists a material point $\xx \in D$ such that for some $\yy \in H_\epsilon(\xx)$ the straight line segment joining $\xx$ and $\yy$ extends beyond the domain $D$.
In other words, there exists  $t \in [0,1]$ such that the convex combination  
$
    l_{\xx,\yy}(t) := \xx + t(\yy - \xx) \notin D.
    $
Since peridynamic force cannot extend beyond the domain boundary, the bond between $\xx$ and $\yy$ is considered broken in the reference configuration.
For general peridynamic domain $D$, the definition of peridynamic horizon can thus be modified to
$
    H_\epsilon(\xx) = \{ \yy \in D: \abs{\yy - \xx} \le \epsilon \} \setminus C_{\xx},
$
where
$
    C_{\xx} = \{ \yy \in D : \exists\ t \in [0, 1] \text{ such that } l_{\xx, \yy}(t) \notin D \}.
    $
Note that when $D$ is convex, the set $C_x$ is empty for all $\xx \in D$.

Here, we outline a method to characterize the set $C = \cup_{\xx \in D} C_{\xx}$ by proving a method to determine whether a line segment with length less than $\epsilon$ between two points in the domain extends outside the domain. We refer to such line segments as \textit{non-bonds}.
As a first attempt, note that if a line segment extends outside the domain $D$, it intersects the domain boundary $\partial D$. However, the converse is not true, in particular, for line segments starting and ending on the boundary $\partial D$.
Moreover, numerically detecting line segments that are part of the domain boundary $\partial D$ by solving a linear system is sensitive to round-off error.
Therefore, we take a different approach to characterize $C$.
Our method involves checking the intersection with an extended domain boundary and the angle bisectors of nonconvex cusps of the boundary.

\begin{figure}[htpb]
\centering
    \subfloat[]{\label{fig:unit-bisector}
     \includegraphics[width=0.25\linewidth]{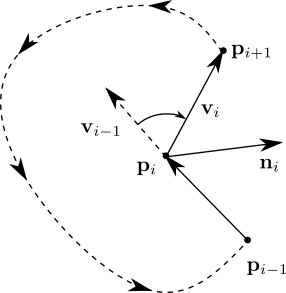}
    }
    \subfloat[]{\label{fig:data-nonconvex-demo}
        \includegraphics[width=0.25\linewidth]{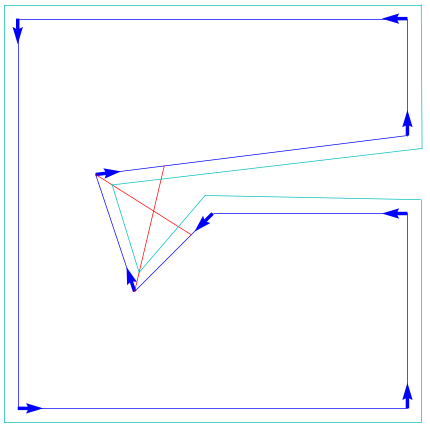}
    }
    \subfloat[]{\label{fig:data/nonconvex-demo-ring.png}
     \includegraphics[width=0.25\linewidth]{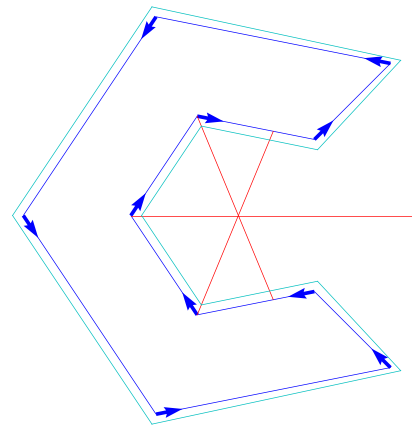}
    }
\caption{(a) Unit outward normal $\nn_{i}$ at a domain boundary vertex (b),(c) Extended boundary and the angle bisector at nonconvex cusps for example particle boundary shapes.}%
\label{fig:nonconvex-demo}
\end{figure}

We work with a piecewise straight line approximation of the domain boundary $\partial D$.
Let $\partial D \in \R^2$ be a closed polygon $\pp^1 \pp^2 \dots \pp^n \pp^1$ oriented in the counterclockwise direction. In other words, the boundary is given by the set 
$
    \partial D = \cup_{i \in \Z_n} \{ l_{\pp^{i}, \pp^{i+1}}(t) : t \in [0,1] \}.
    $
Define the unit tangent to the line segment $\pp^i \pp^{i+1}$ as
$
\vv_i = \frac{\pp^{i+1} - \pp^{i}}{\abs{\pp^{i+1} - \pp^{i}}}.
$
We say the vertex $\pp^i$ is a \textit{nonconvex cusp} if the angle between the vectors $\vv_{i-1}$ and $\vv_{i}$ is obtuse, i.e., if
$
\vv_{i-1} \times \vv_{i} < 0.
$
At vertex $\pp^i$ we define the outward `normal' $\nn_i$ given by
(see \Cref{fig:unit-bisector})
\begin{align*} 
    \nn_i =
    \begin{cases}
    \frac{-\vv_{i-1} + \vv_{i} }{\abs{\vv_{i-1} + \vv_{i}}}, &\text{ if } \vv_{i-1} \times 
\vv_{i} < 0\\
    \frac{\vv_{i-1} + \vv_{i} }{\abs{\vv_{i-1} + \vv_{i}}}, &\text{ if } \vv_{i-1} \times \vv_{i} > 0\\
	\begin{bmatrix}
	    0 & -1 \\
	    1 & 0
	\end{bmatrix} \vv_{i}, &\text{ if } \vv_{i-1} \times \vv_{i} = 0.
    \end{cases}
\end{align*}	
We remark here that the notion of unit normal at the point $\pp^i$ is not well-defined due to non-uniqueness. Therefore, we choose $\nn_i$ to be simply the unit vector in the direction of the bisector of the outer angle at $\pp^i$.
Extending $\partial D$ in the outward normal direction, we obtain the extended boundary $ \overline{\partial D} $ given by the polygon $ \overline{\pp}^1 \overline{\pp}^2 \dots \overline{\pp}^n \overline{\pp}^1$, where  
$\bar{\pp}^i$ is a $\delta$-perturbation of the vertex $\pp^i$ in the outward normal direction $\nn_i$ given by
$
\bar{\pp}^i = \pp^i + \delta \nn_i
$
for some $\delta > 0$.

For a nonconvex cusp $\pp^i$, we also define $\qq^i$ to be the point of intersection with the domain boundary $\partial D$, if it exists. i.e.,
$
    \qq^i = \pp^i + t_0 \nn^i
    $
such that there exists $j \in \Z_n$ and $t_0, s \in [0,1]$ such that 
$
    \pp^i + t_0 \nn^i =  \pp^j + s ( \pp^{j+1} - \pp^j).
    $
If no such intersection exists, we define
$
    t_{0} = \abs{\pp^i} + \sup \{ \abs{\xx} : \xx \in D \}.
$
In that case, using the reverse triangle inequality we have
$
\abs{\qq^i} \ge \sup \{ \abs{\xx} : \xx \in D \},
$
and therefore the line segment $\pp^i \qq^i$ does not extend beyond the minimal bounding circle of the domain $D$.

Now, if $\yy \in C_{\xx}$, the line segment $\xx \yy$ intersects either
the outer angle bisectors ${\pp^i \qq^i}$ at a nonconvex cusp $\pp^i$ or the extended boundary segments ${\overline{\pp}^i \overline{\pp}^{i+1}}$ for some $i \in \Z_n$.
This method leads to an effective detection of non-bonds to construct the peridynamic horizon, which is especially useful for a nonconvex domain $D$.
\Cref{fig:nonconvex-demo} demonstrates the construction of lines segments to intercept non-bonds in various nonconvex peridynamic domains.
{\revvv In the following sections, the finite element mesh generating software \texttt{Gmsh} \cite{geuzaine2009gmsh} is used to obtain the nodes on and within the particle boundaries. The mesh size is defined as the minimum distance between any two nodes in all the particles in the simulation.}

\section{Simulations: two and three particle collisions}%
\label{sec:simulations}
{\rev 
In this section, we first illustrate our method for handling intra-particle deformation and particle to particle interaction by solving the initial value problem \Cref{eq:full} for simple problems with two and three particle interactions, $N=2,3$. \Cref{sec:time_integration} provides a discussion of time integration and choice of  time step used in all simulations. In  \Cref{sub:experiment_collision_with_fracture} we validate our dynamic fracture and contact model for nonconvex domains against the well-known Kalthoff-Winkler experiment \cite{kalthoff}.
In \Cref{sub:convergence} we illustrate the choice of peridynamcic horizon and mesh size on elastic and inelastic deformation. In the last subsection \Cref{sub:experiment_fracture_toughness_and_damage_propagation} we investigate crack initiation under compression for a pre-notched particle. We illustrate non-interpenetration of cracks and the calving of child particles by simulating a three particle initial value problem. Such behavior is essential to model comminution and crushing seen in aggregrates.
}

We conclude the introduction listing particle dimensions and material properties used in all simulations used both in this section and the following section.
The material properties of particles used in the simulations in \Cref{sec:simulations} and \Cref{sec:simulations_bulk_test} are listed as in \Cref{tab:material} as $M_1$  and $M_2$, respectively.
\begin{table}[htpb]
    \centering
    \caption{Common material properties used across simulations.}
    \label{tab:material}
    \begin{tabular}
	{ 
	|c | c | c | c |
	}
	\hline
	Material & Young's modulus ($E$)       & Bulk modulus ($k$)   & Density ($\rho$) 
    \\
    \hline
    $M_1$ & $ 191 \times 10^{9} $  Pa  & $159.2 \times 10^{9}$ Pa & 8000 kg/m$^3$
    \\
    \hline
    $M_2$ & $ 1.23 \times 10^{9} $  Pa  & $2 \times 10^{9}$ Pa & 1200 kg/m$^3$
    \\
    \hline
    \end{tabular}
\end{table}

\subsection{Time integration steps for $N$ particles}%
\label{sec:time_integration}

To simulate the dynamics of the particle nodes, we use the Velocity-Verlet scheme \cite{HairerEtAlGeometric03}.
In the reference configuration, we discretize all particles $D_k$ $k=1, 2, \dots, N$ in space. $\xx$ and $\yy$ denote generic nodes inside two distinct particles.  A complete algorithm depicting the time integration steps  with time step $\Delta t >0$ involving neighborhood search and force computations is given in \Cref{alg:buildtree}.
 
\begin{algorithm}
\caption{Time integration}
\label{alg:buildtree}
\begin{algorithmic}[1]
\STATE{Define $\uu(\xx), \dot{\uu}(\xx), \ddot{\uu}(\xx), \Bb(\xx)$ for $\xx \in D_k$ at time $t=0$ for $k=1,2,\dots,N$}
\STATE{Compute intra-particle neighbors $S^{\text{ref}}(\xx) = \{\xx' \in D_k : \abs{\xx - \xx'} < \epsilon \}$ for all $\xx \in D_k$, $k=1,2, \dots, N$
    }
\WHILE{$t < T$}
  \FOR{$k = 1, 2, \dots, N$ and $\xx \in D_k$}
    \STATE{
    $\uu(\xx) \leftarrow \uu(\xx) + \Delta t \dot{\uu}(\xx) + \frac{\Delta t^2}{2} \ddot{\uu}(\xx)$ for $\xx \in D_k$
    }
    \STATE{Compute current position $\pp(\xx) = \xx + \uu(\xx)$
    }
  \ENDFOR
  \FOR{$k = 1, 2, \dots, N$ and $\xx \in D_k$}
    \STATE{Compute inter-particle neighbors $S^{\text{nbr}}_{j}(\xx) = \{\yy \in D_j : \abs{\pp(\xx) - \pp(\yy)} < R_c \}$, $j=1, 2, \dots, N$
    }
    \STATE{Compute peridynamic force density
    $
    \ff^{\text{peri}}(\xx)  = \sum\limits_{\xx' \in S^{\text{ref}}(\xx)}  \ff(\xx', \xx)    V_{\xx'}
    $
    }
    \STATE{Compute self-contact force density \\
    $
    \ff^{\text{self}}(\xx)  = \sum\limits_{\xx' \in S^{\text{nbr}}_{k}}^{}  \left( \ff^{\text{self}}_r(\xx', \xx) + \ff^{\text{self}}_d(\xx', \xx) + \ff^{\text{self}}_f(\xx', \xx) \right)    V_{\xx'}
    $
    }
    \STATE{Compute neighboring-particle force density \\
    $
    \ff^{\text{nbr}}(\xx)  = \sum_{j \ne k} \sum\limits_{\yy \in S^{\text{nbr}}_{j}}^{}  \left( \ff_r(\yy, \xx) + \ff_d(\yy, \xx) + \ff_f(\yy, \xx) \right)    V_{\yy}
    $
    }
    \STATE{Compute wall-contact force density \\
    $
    \ff^{\text{wall}}(\xx)  =   \ff^{\text{wall}}_r(\yy, \xx) + \ff^{\text{wall}}_d(\yy, \xx) + \ff^{\text{wall}}_f(\yy, \xx)  
    $
    }
    \STATE{Back up acceleration from previous time step: $\ddot{\uu}^{\text{old}} := \ddot{\uu}(\xx)$}
    \STATE{
    $
    \ddot{\uu}(\xx) \leftarrow \frac{1}{\rho} 
    \left(
    \ff^{\text{peri}} (\xx)
    +\ff^{\text{self}} (\xx)
    +\ff^{\text{nbr}} (\xx)
    +\ff^{\text{wall}} (\xx)
    + \Bb(\xx, t)
    \right)
    $
    }
    \STATE{
    $
    \dot{\uu}(\xx)  \leftarrow \dot{\uu}(\xx) + \frac{\Delta t}{2}  \left(\ddot{\uu}^{\text{old}} + \ddot{\uu}(\xx)  \right)
    $
    }
  \ENDFOR
    \STATE{$t \leftarrow t + \Delta t$}
\ENDWHILE
\end{algorithmic}
\end{algorithm}

{\rev 
We take our time step $\Delta t$ and mesh size $h$ small enough to resolve all particle interactions for numerical stability as described in  \cite{burns2019critical} and to simultaneously satisfy the nonlocal CFL condition that depends explicitly on the peridynamic horizon $\delta$ \cite{jha2018numerical}. 
 Applying this criterion we have chosen $\Delta t = 2.5$ ns for the two-particle tests in \Cref{sub:experiment_collision_with_fracture,sub:convergence}, 
$\Delta t = 20$ ns for the three-particle test in \Cref{sub:experiment_fracture_toughness_and_damage_propagation},
and $\Delta t = 70$ ns for the bulk settling and compression tests in \Cref{sub:settle,sub:compaction-damage}.} 
To fix ideas in this section we have chosen the friction and damping forces to be absent so $\mu = 0$ and $\beta_d = 0$. All friction and damping forces are active in the following \Cref{sec:simulations_bulk_test}.

\subsection{Collision with fracture and validation}%
\label{sub:experiment_collision_with_fracture}

\begin{figure}[htpb]
\centering
    \subfloat[]{\label{fig:kw-sc}
        \includegraphics[width=0.35\linewidth]{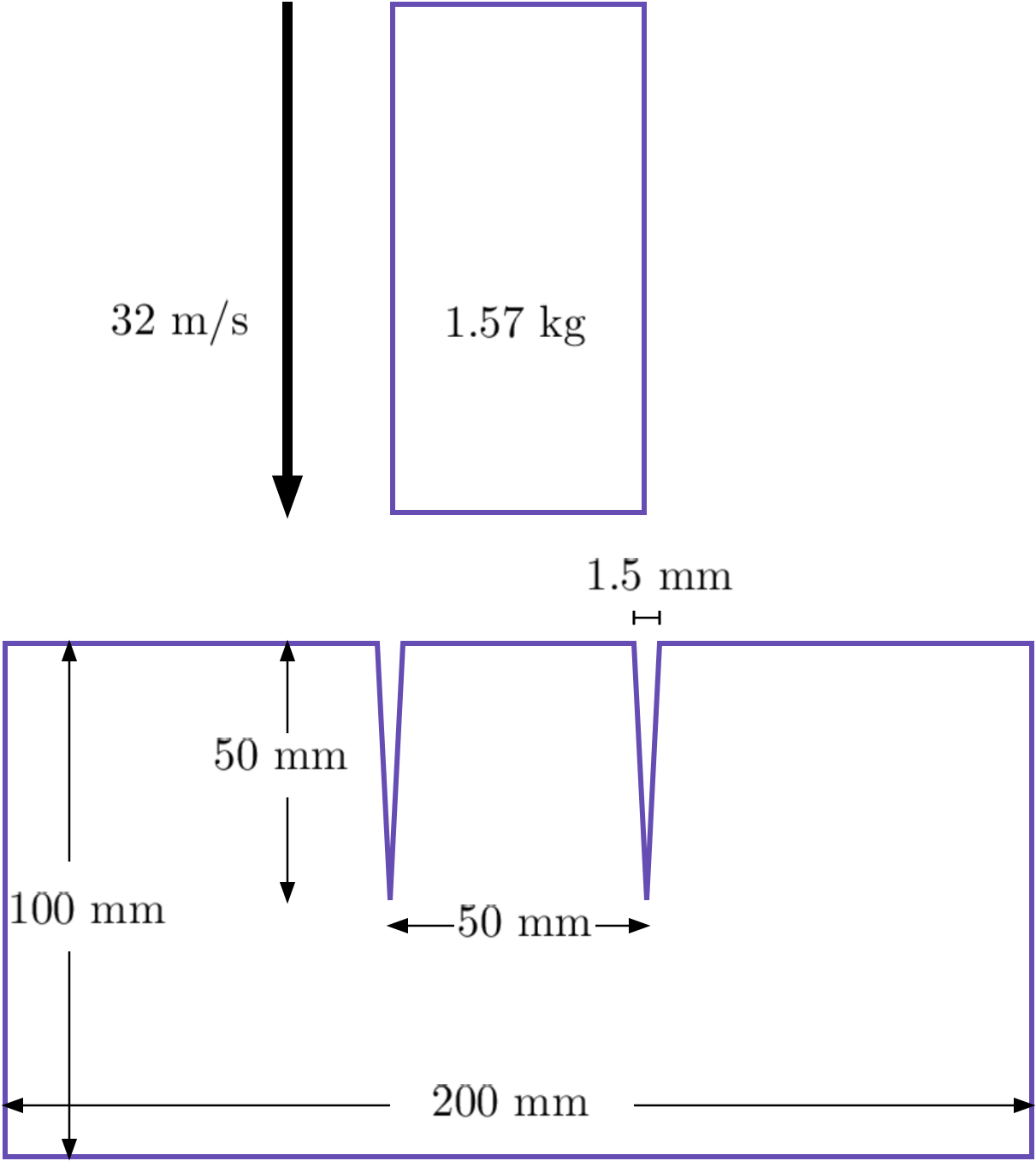}
    }
    \subfloat[]{\label{fig:kw-sim}
        \includegraphics[width=0.45\linewidth]{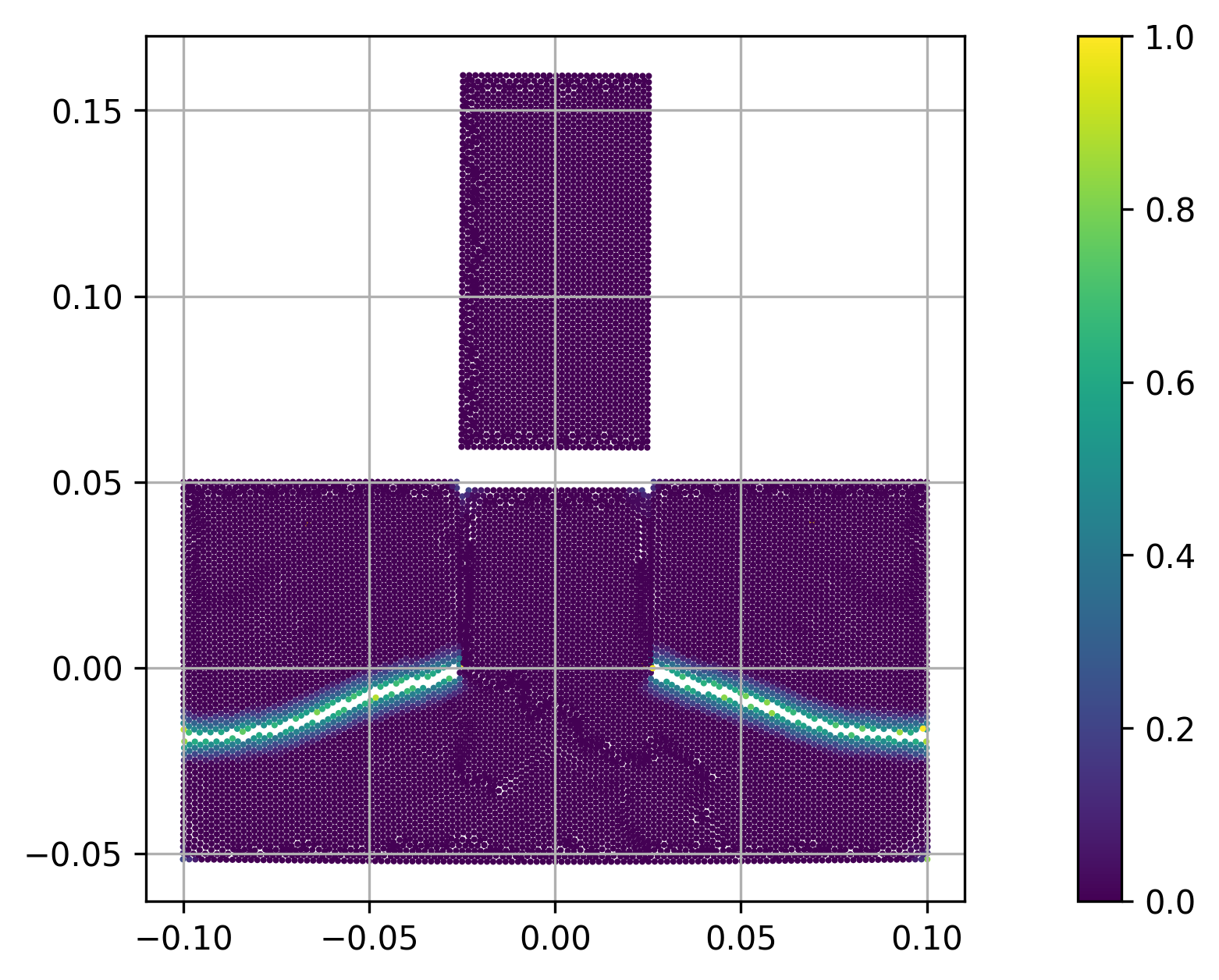}
    }
        \caption{
        (a) A schematic diagram of the Kalthoff-Winkler experiment (b) Simulation producing a crack angle of 68 degrees matching experimental results.}%
\label{fig:kw-schematics}
\end{figure}

Here, we validate our dynamic fracture and contact model for nonconvex domains against the well-known Kalthoff-Winkler experiment \cite{kalthoff}, where a cylindrical impactor strikes a plate with two existing notches, leading to a crack pattern that is experimentally reproducible.
The schematic diagram is shown in \Cref{fig:kw-sc}.
{\revv The experiment has been simulated in \cite{SILLING2003641,TRASK2019151} 
with a mode II dynamic displacement condition which replaces the effect of the impactor.}
Here, we simulate the experiment as a two-particle collision problem, i.e., we solve \Cref{eq:full} for $N=2$. Initially the impactor is traveling toward the plate at $32$ meters per second. The notched particle is stationary with the corner left and right ligaments held fixed at the top. The impactor hits the central ligament.  Using  \Cref{sec:peridynamic_bonds_in_non_convex_material} we remove the peridynamic bonds across the notches.
Our simulation shown in \Cref{fig:kw-sim} is in good agreement with the experiment where the crack angle is observed to be roughly 68 degrees.

\subsection{Study of peridynamic horizon size and mesh size on the damage of microscopic particles}%
\label{sub:convergence}

We study the effect of the peridynamic horizon size $\epsilon$ and the mesh size $h$ on {\bf elastic and inelastic deformation} inside a particle.  This effect is clearly exhibited for extreme particle deformation to the point of fracture and affects the location and extent of damage.
 
\begin{figure}
    \centering
    \subfloat[]{\label{fig:sc-2part}
    \includegraphics[width=0.32\linewidth]{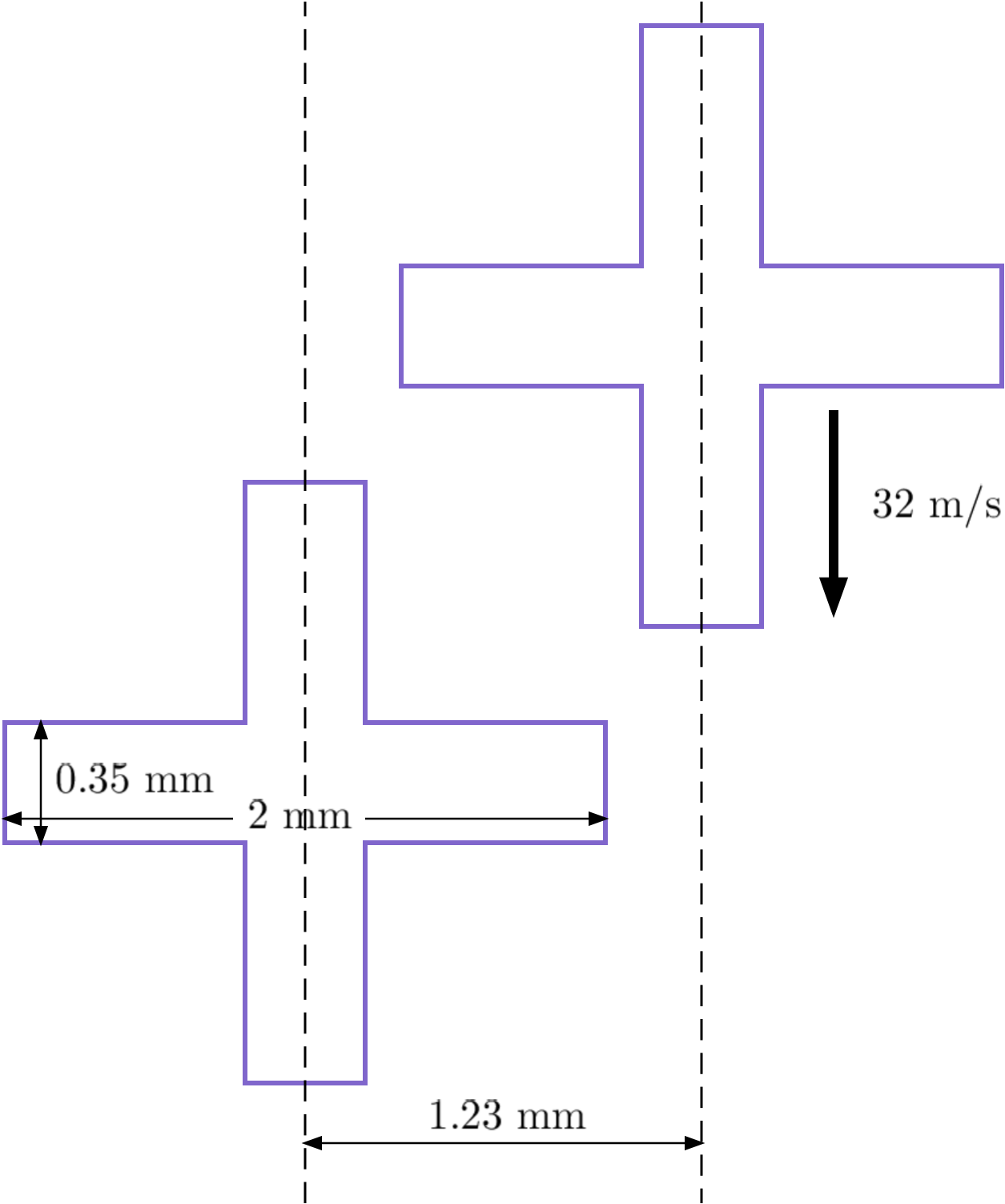}
    }
    \subfloat[]{\label{fig:sc-3part}
    \includegraphics[width=0.22\linewidth]{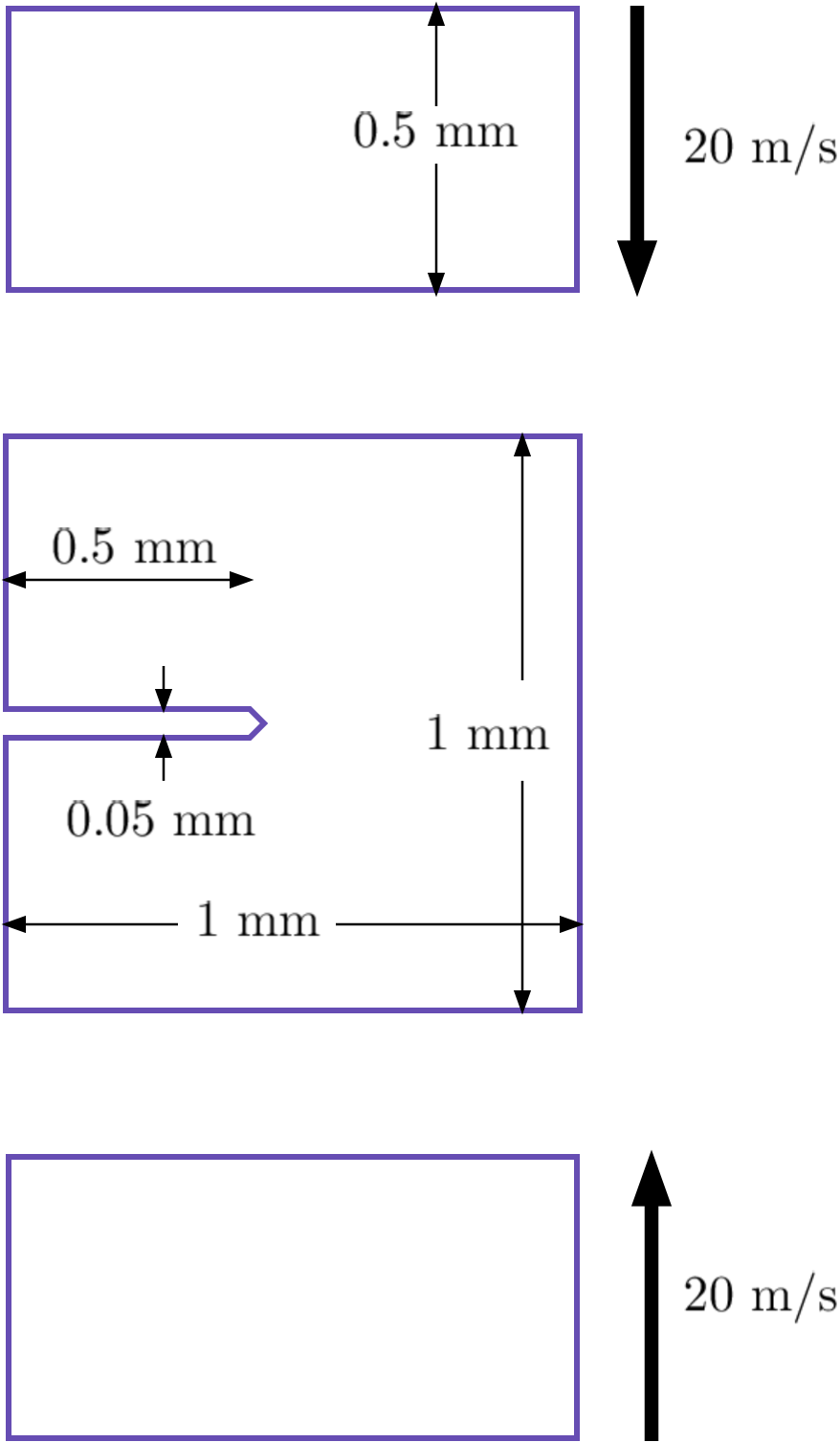}
    }
    \caption{Schematics of the two and three-particle collision tests.}
    \label{fig:sc-2-3}
\end{figure}

\begin{figure}
    \centering
    \subfloat[$\epsilon=4h$]{\label{fig:del4}
    \includegraphics[width=0.32\linewidth]{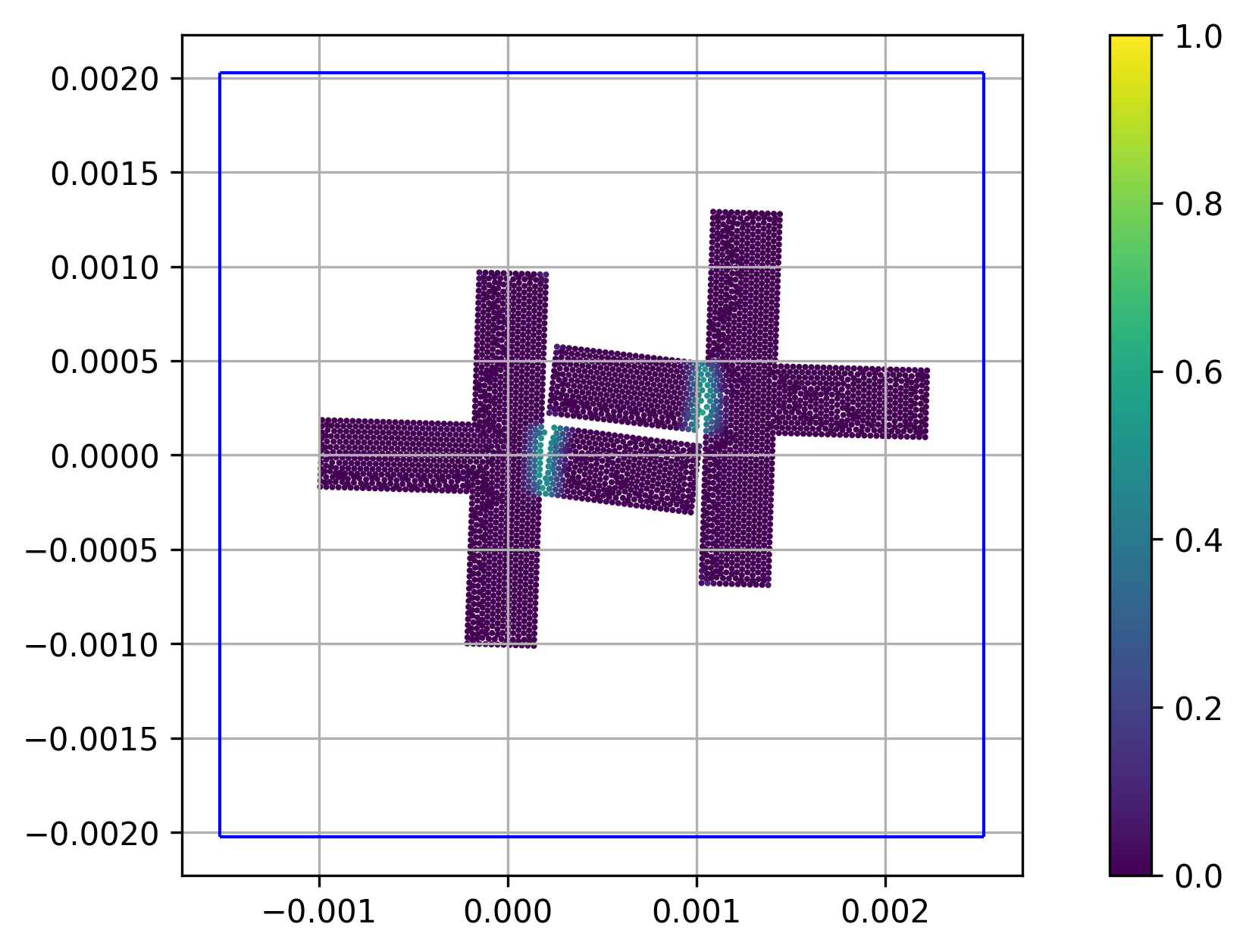}
    }
    \subfloat[$\epsilon=15h$]{\label{fig:del15}
    \includegraphics[width=0.32\linewidth]{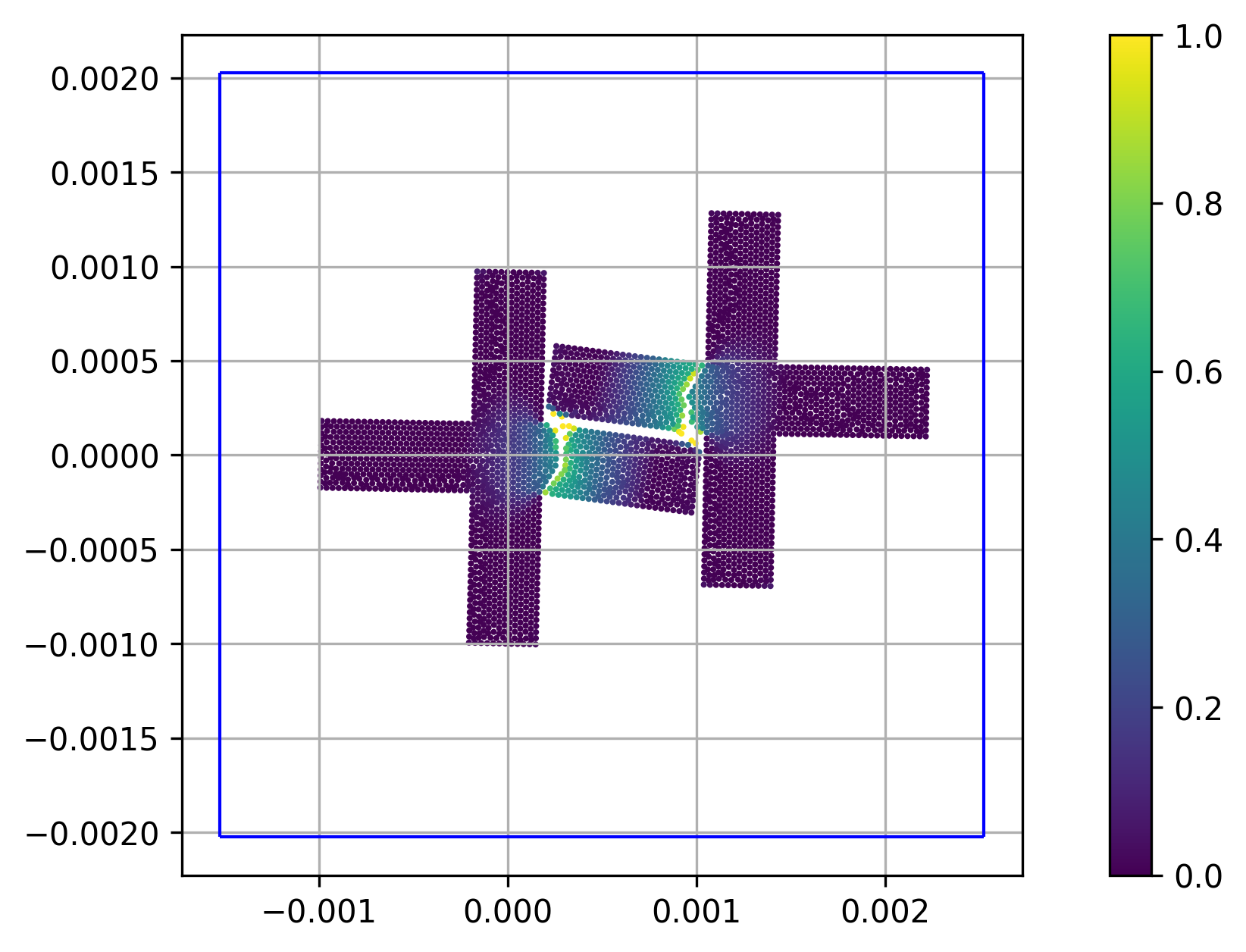}
    }
    \subfloat[$\epsilon=25h$]{\label{fig:del25}
    \includegraphics[width=0.32\linewidth]{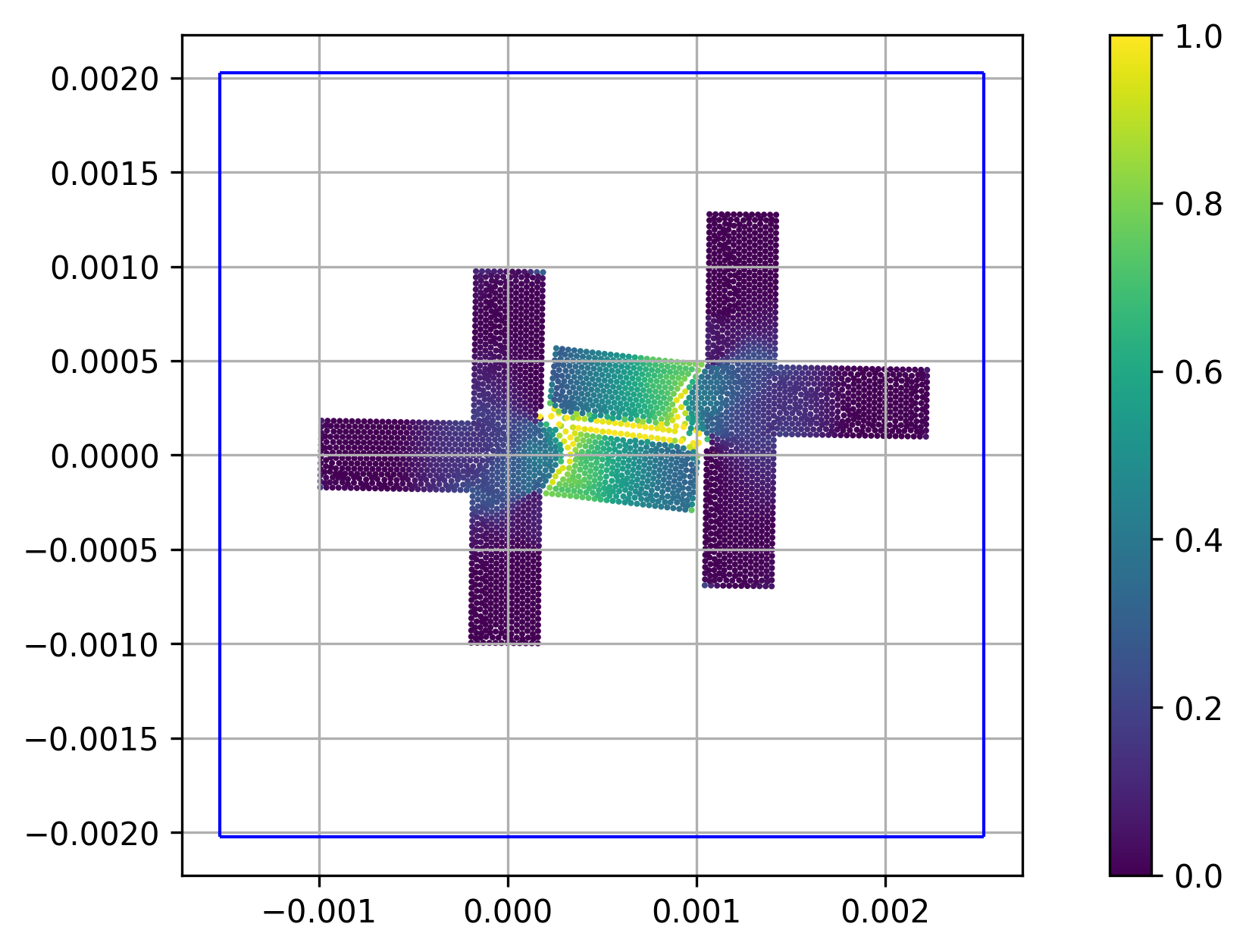}
    }
    \caption{Fracture patterns for varying peridynamic horizon size $\epsilon$ while the mesh size is fixed. Damage zone grows with horizon.}
    \label{fig:kw-mini-fixed-ms}
\end{figure}

\begin{figure}
    \centering
    \subfloat[$h=\frac{R}{30}$]{\label{fig:ms30}
    \includegraphics[width=0.32\linewidth]{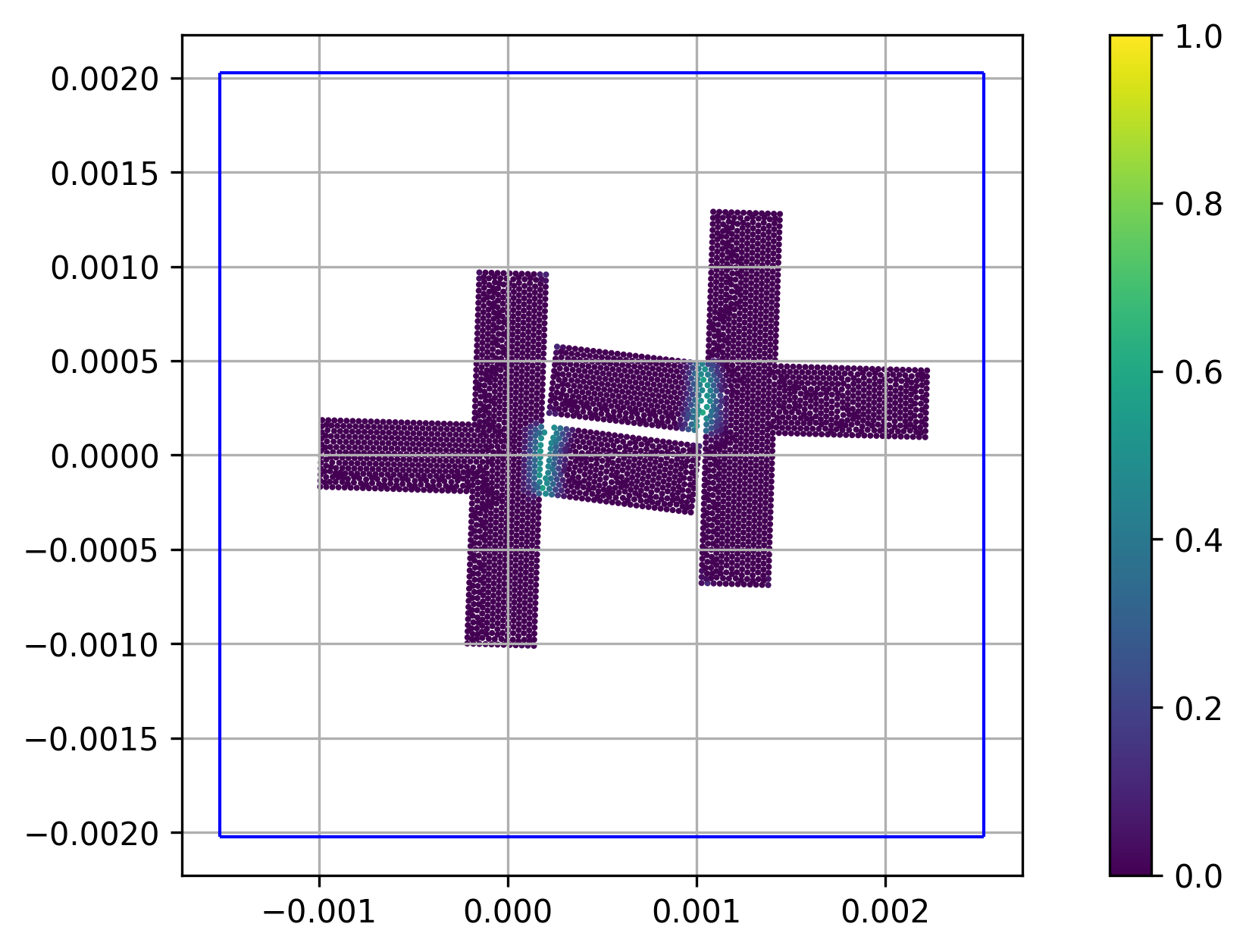}
    }
    \subfloat[$h=\frac{R}{40}$]{\label{fig:ms40}
    \includegraphics[width=0.32\linewidth]{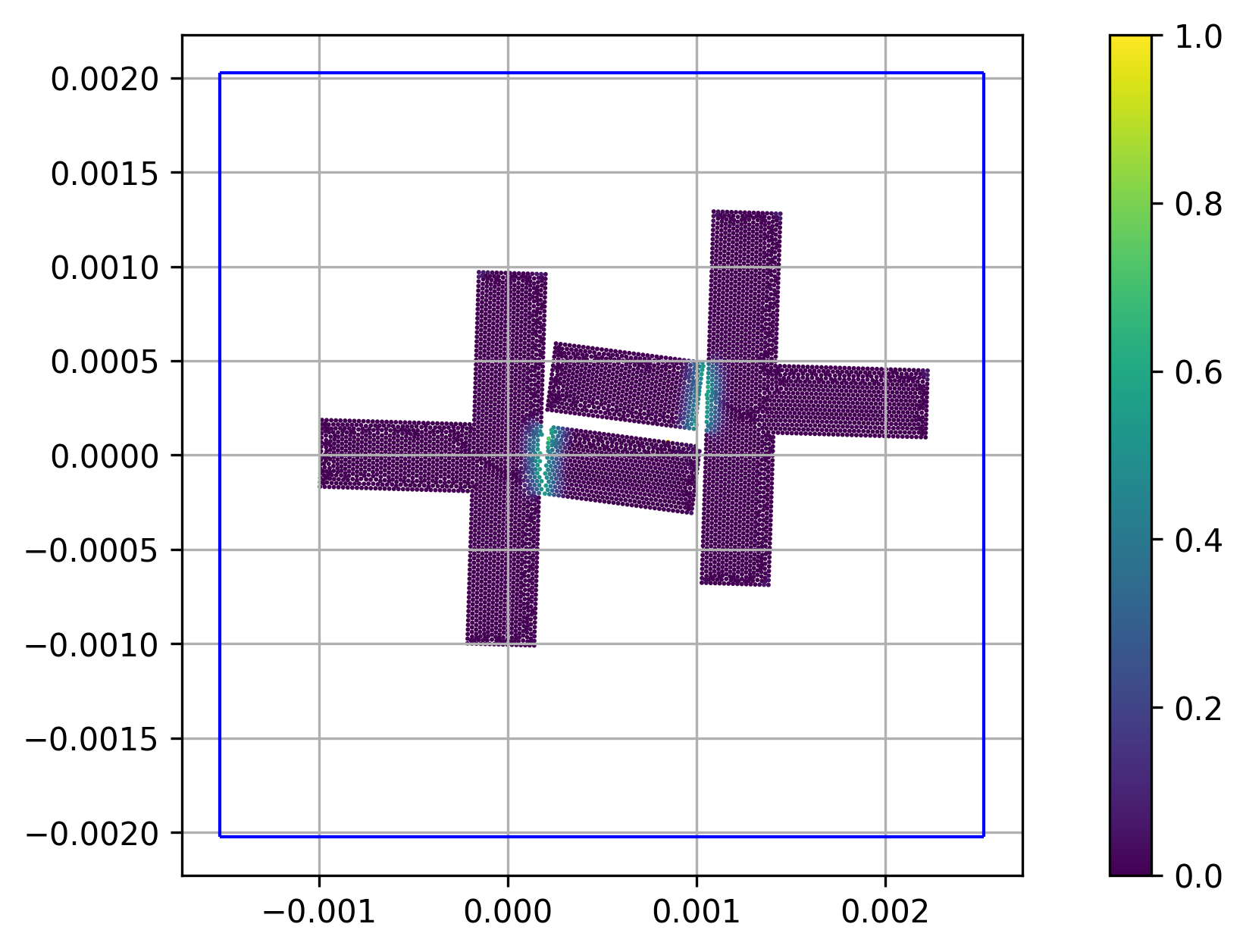}
    }
    \subfloat[$h=\frac{R}{50}$]{\label{fig:ms50}
    \includegraphics[width=0.32\linewidth]{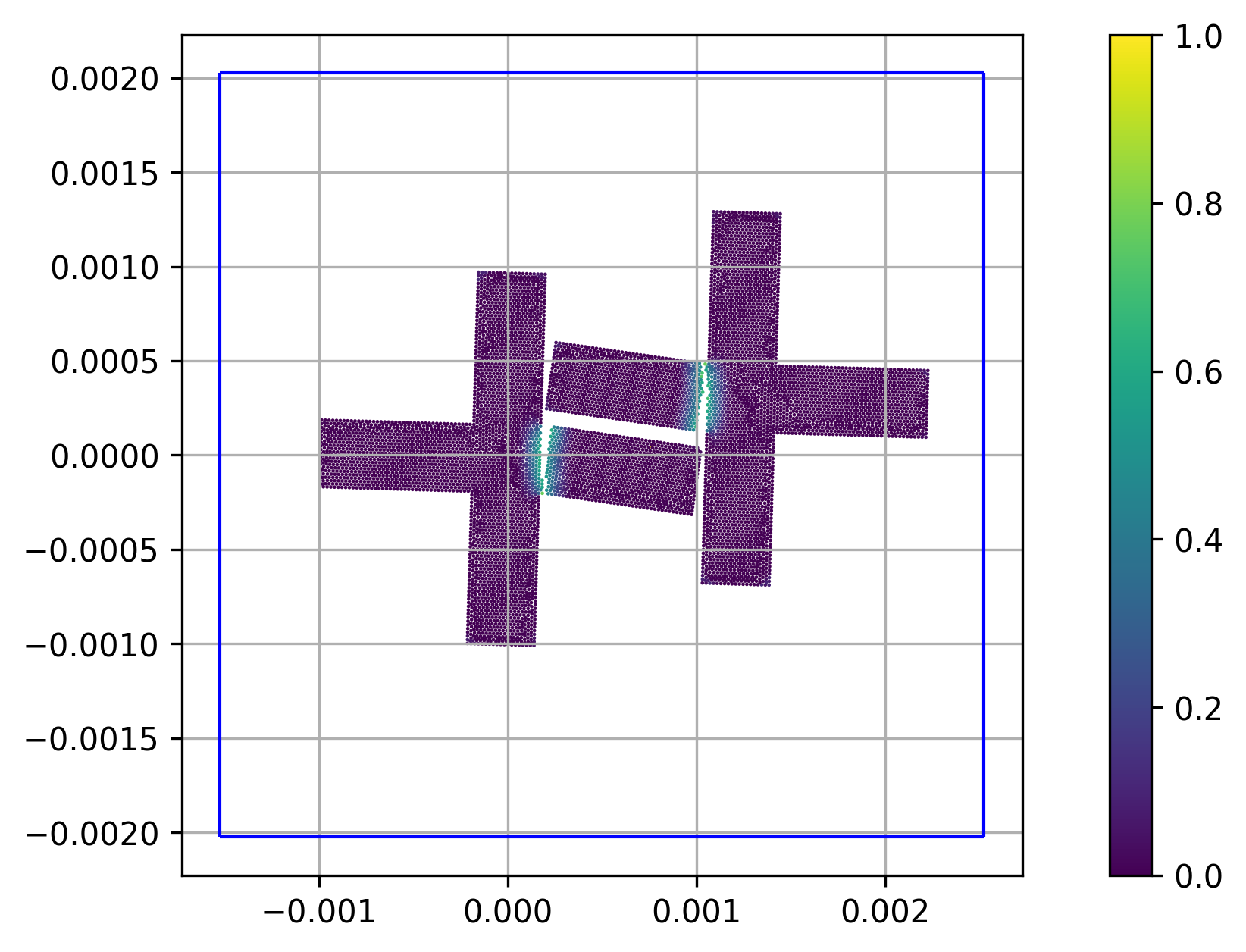}
    }
    \caption{Fracture patterns for varying mesh size $h$ while the peridyanamic horizon size is fixed. Damage zone remains localized in nearly the same location for this choice of mesh sizes.}
    \label{fig:kw-mini-fixed-delta}
\end{figure}

We consider a two-particle collision experiment in the millimeter length scale. In this experiment  (see \Cref{fig:sc-2part}), a plus-shaped particle of half length $R = 1$ mm with arm width $0.35$ mm traveling at the speed of 32 m/s collides with another cross at rest. The horizontal distance between the center of the particles is taken to be 1.23 mm. The fracture toughness of each particle is taken to be 424 Jm$^{-2}$.
The mesh size is kept fixed at $h = \frac{R}{30}$, and we take the peridynamic horizon to be variable with  $\epsilon = 4h, 15h$, and $25h$. The damage of the particle for each $\epsilon$ is shown in  \Cref{fig:kw-mini-fixed-ms}.
For the next simulation, we fix the peridynamic horizon at $\epsilon = 0.133 R$ mm, and decrease the mesh size. \Cref{fig:kw-mini-fixed-delta} shows the damage of the particles for mesh size $h=\frac{R}{30}, \frac{R}{40}$, and $\frac{R}{50}$. 
We observe that for a fixed mesh size and increased peridynamic horizon size, the fracture patterns are less localized and spread over larger regions in the particle.
However, particle-to-particle contact is dramatically influenced by the particle geometry, in particular, the presence of re-entrant corners influence the location of the fracture zone, as the contact forces are determined by a contact radius that is is significantly smaller than the peridynamic horizon size.

{\revv
\begin{figure}[htbp]
\centering
\subfloat[]{\label{fig:conv-delta}
    \includegraphics[width=0.37\linewidth]{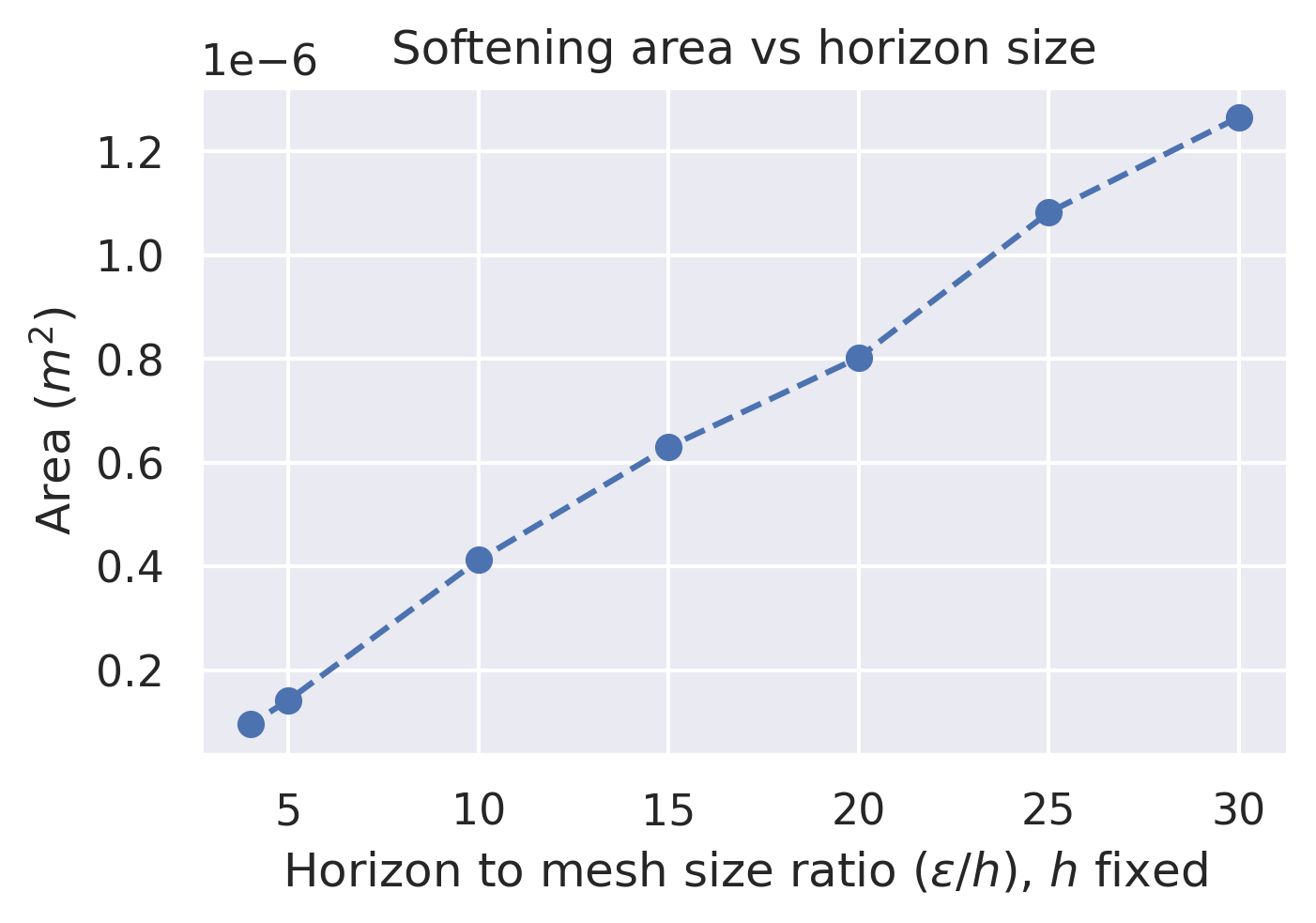}
    }
\subfloat[]{\label{fig:conv-meshsize}
    \includegraphics[width=0.37\linewidth]{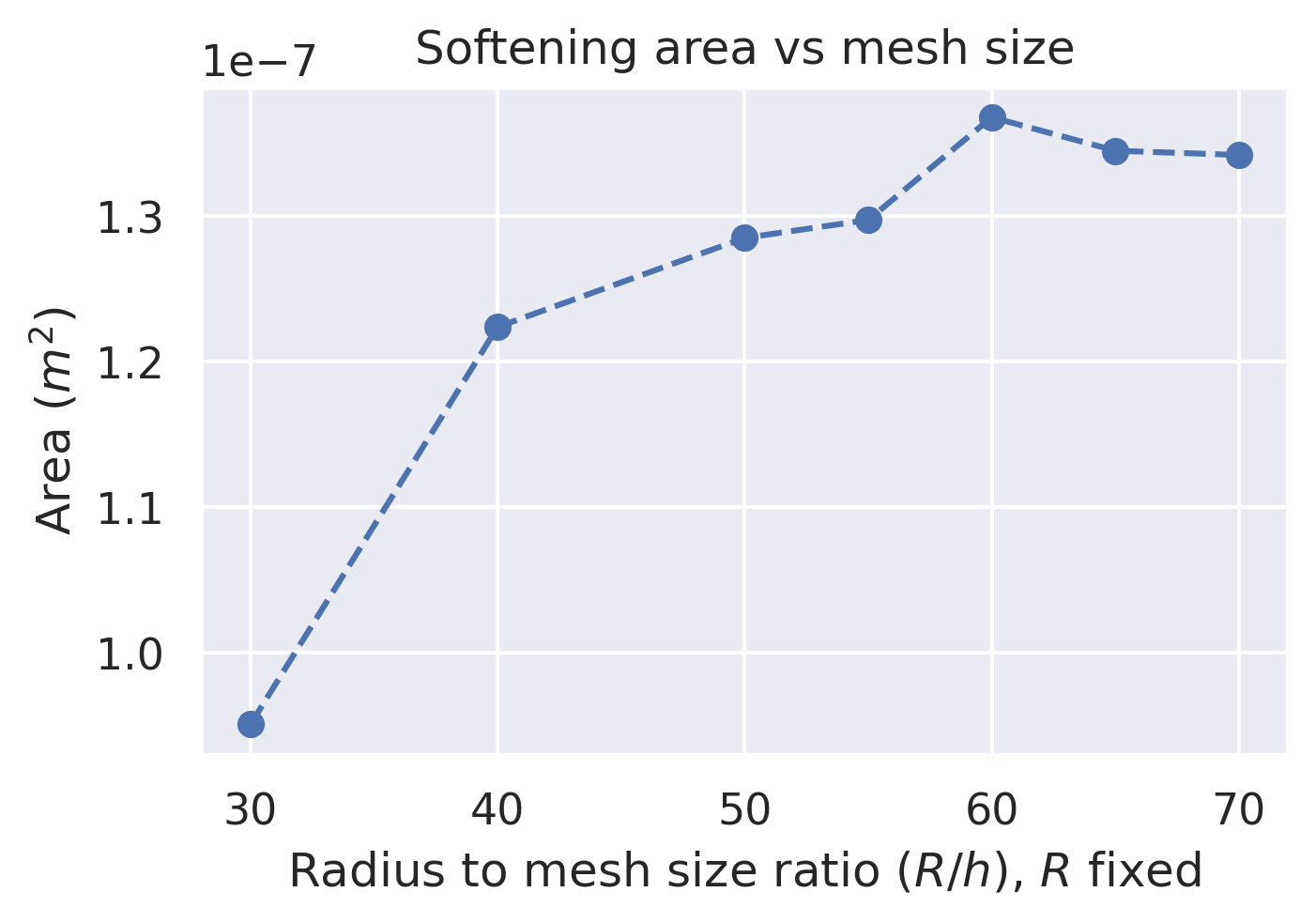}
    }
\caption{
Convergence of softening zone for (a) vanishing horizon $\epsilon$ and (b) vanishing mesh size $h$.
}
\label{fig:convergence}
\end{figure}
After 20 {\revvv time steps (i.e. after 400 ns)}, the area of the softening zone in the first particle is computed for each simulation. The softening zone is defined by all points in the particle with nonzero damage. Consistent with theory \cite{lipton2016cohesive}, in the vanishing horizon limit $\epsilon \to 0$, the area of the softening zone converges to zero (see \Cref{fig:conv-delta}) when the mesh size remains fixed at $h = \frac{R}{30}$, leading to a fracture path of codimension 1. For the fixed horizon size ($\epsilon = 0.133 R$) simulations (\Cref{fig:conv-meshsize}), the area of the softening zone
converges to a fixed value as the mesh size $h$ approaches zero.
}

\subsection{Simulation: fracture toughness and damage propagation}%
\label{sub:experiment_fracture_toughness_and_damage_propagation}
{\rev 
Here, we study the effect of fracture toughness on crack patterns on particles with an existing pre-notch observed from a symmetric impact.
}
In this experiment, (see \Cref{fig:sc-3part}) we consider a square-shaped particle (particle A) with length $1$ mm with a pre-existing notch of length $0.5$ mm that extends to the center of the particle. All peridynamic bonds that run across the pre-notch are removed in the reference configuration.
Two rectangular particles (particles B and C) with dimension $1$ mm $\times$ $0.5$ mm with velocities $20$ m/s and $-20$ m/s, respectively, collide vertically with particle A simultaneously. Here damping and friction forces are turned off.
We have taken 2 different values of fracture toughness for particle A.
\begin{figure}[htpb]
\centering
    \subfloat[]{\label{fig:crack_1_29.png}
    \includegraphics[width=0.2\linewidth, trim={{0.2\linewidth} {0.4\linewidth} {0.2\linewidth} {0.4\linewidth}},clip]{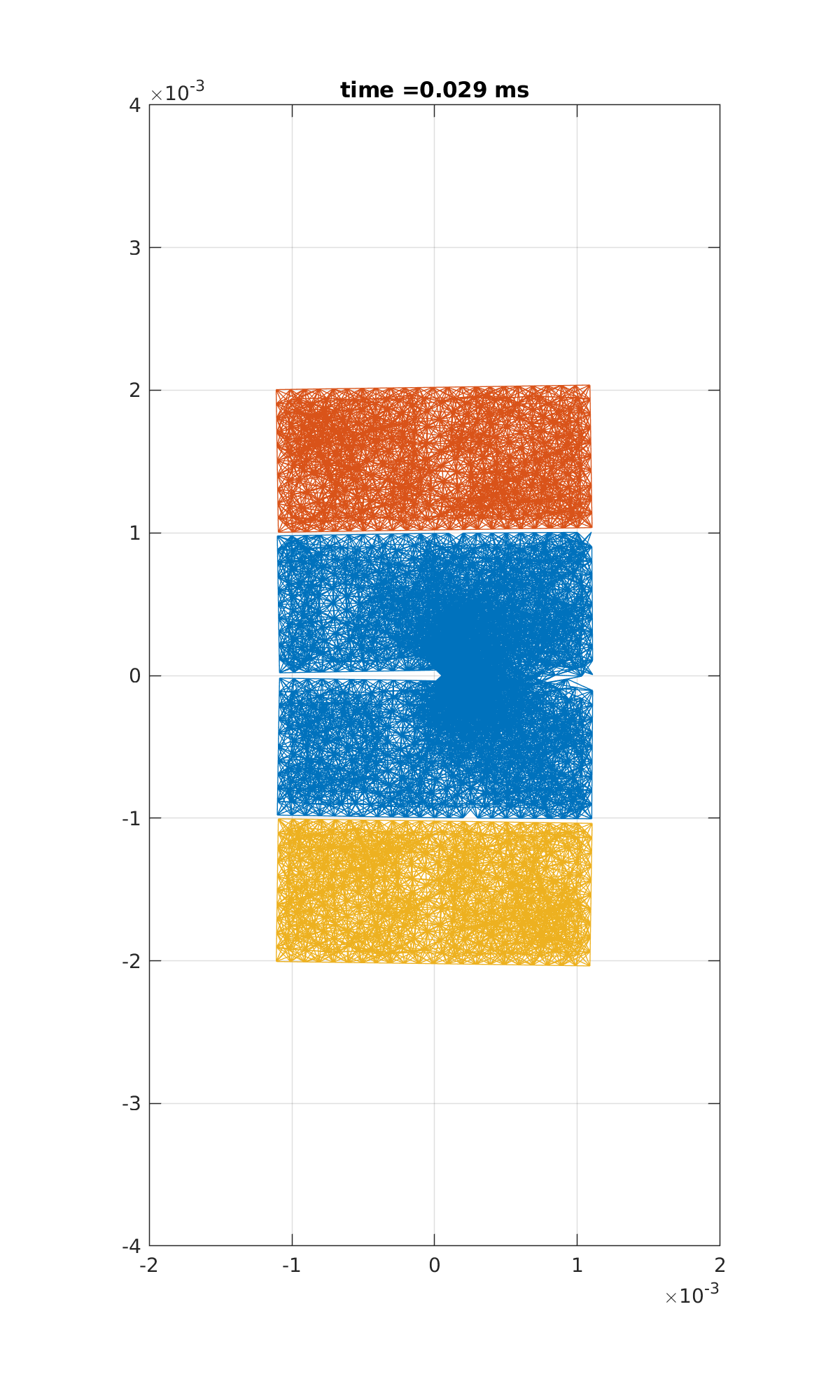}
    }
    \subfloat[]{\label{fig:crack_1_30.png}
    \includegraphics[width=0.2\linewidth, trim={{0.2\linewidth} {0.4\linewidth} {0.2\linewidth} {0.4\linewidth}},clip]{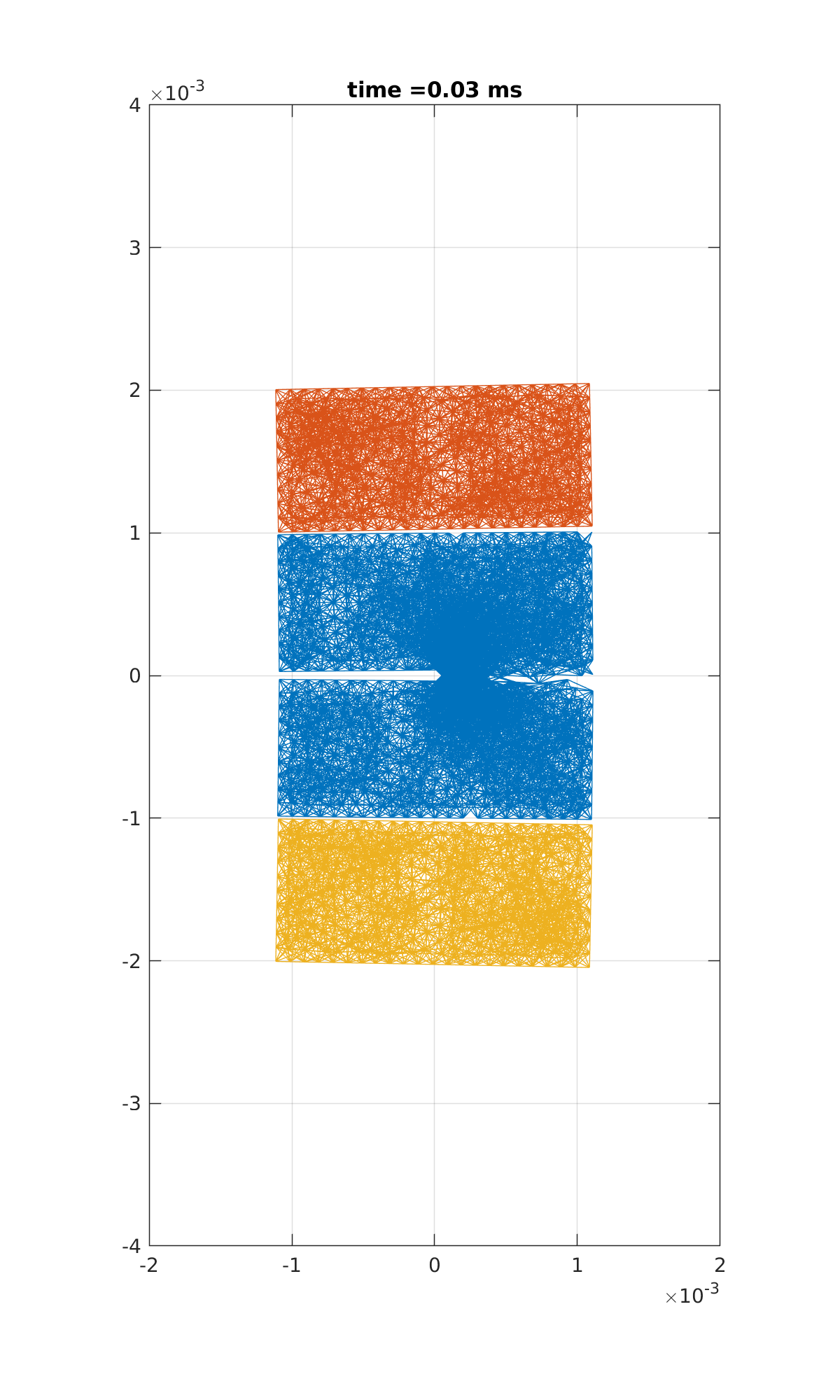}
    }
    \subfloat[]{\label{fig:crack_1_31.png}
    \includegraphics[width=0.2\linewidth, trim={{0.2\linewidth} {0.4\linewidth} {0.2\linewidth} {0.4\linewidth}},clip]{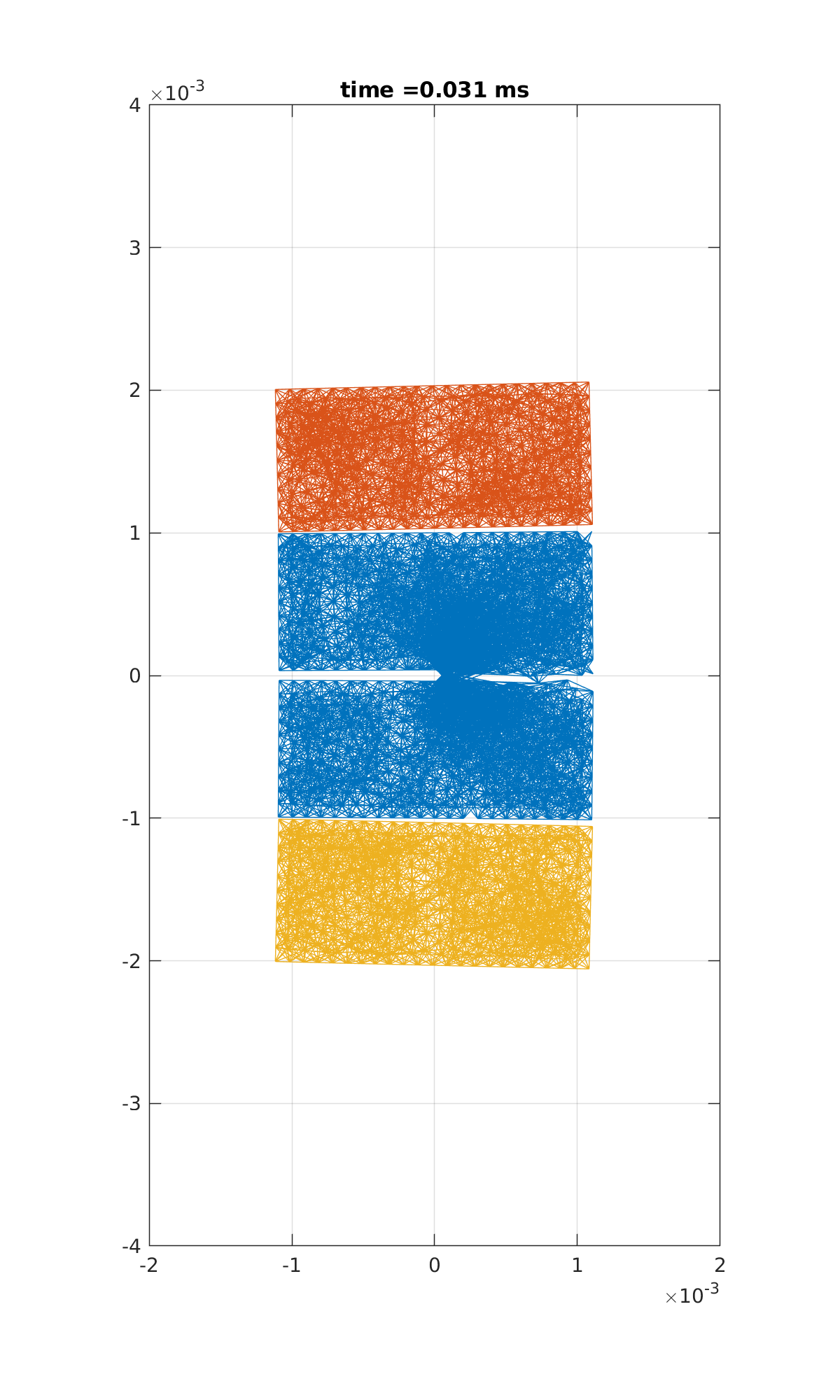}
    }
    \subfloat[]{\label{fig:crack_1.png}
    \includegraphics[width=0.2\linewidth, trim={{0.2\linewidth} {0.4\linewidth} {0.2\linewidth} {0.4\linewidth}},clip]{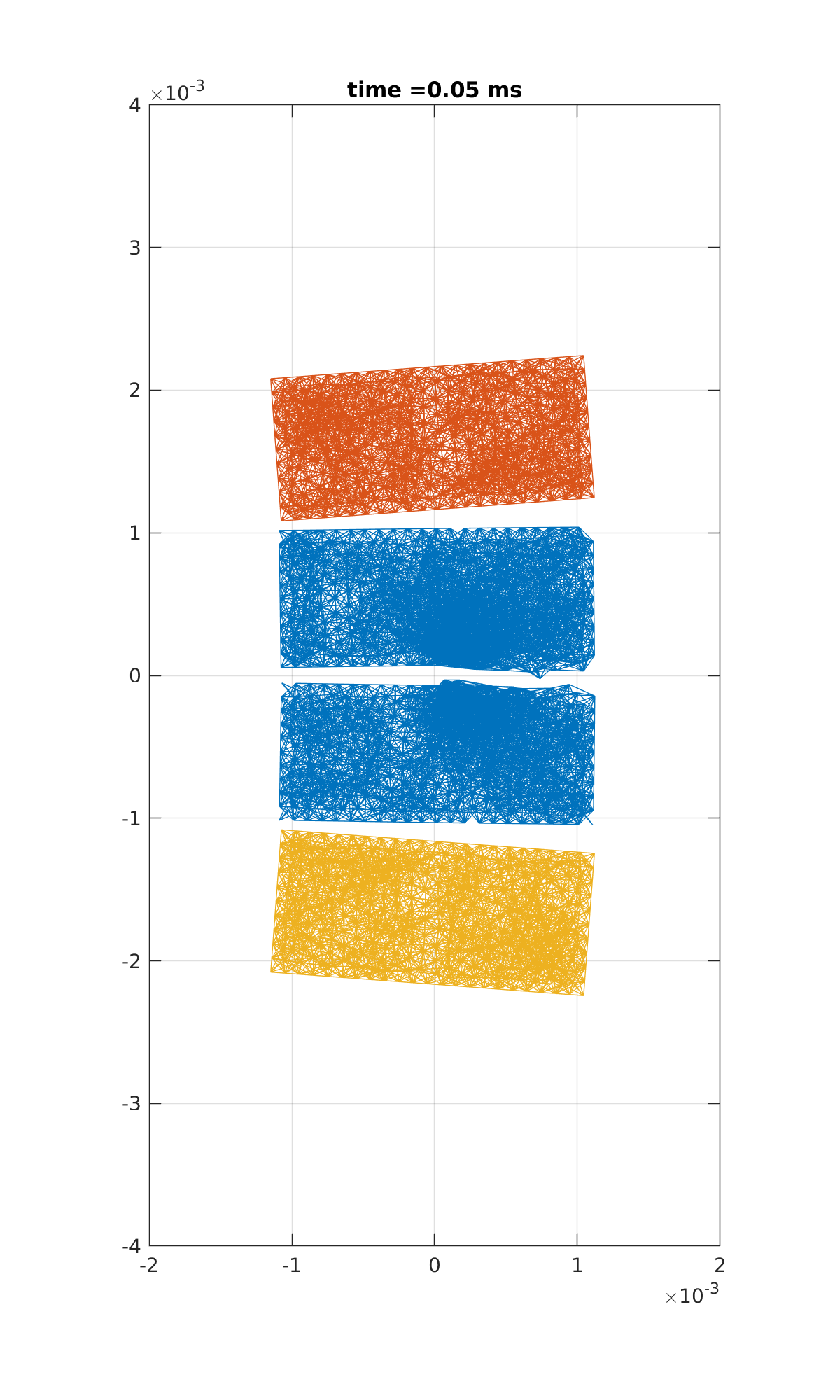}
    }
    \\
    \subfloat[]{\label{fig:crack_3_28.png}
    \includegraphics[width=0.2\linewidth, trim={{0.2\linewidth} {0.4\linewidth} {0.2\linewidth} {0.4\linewidth}},clip]{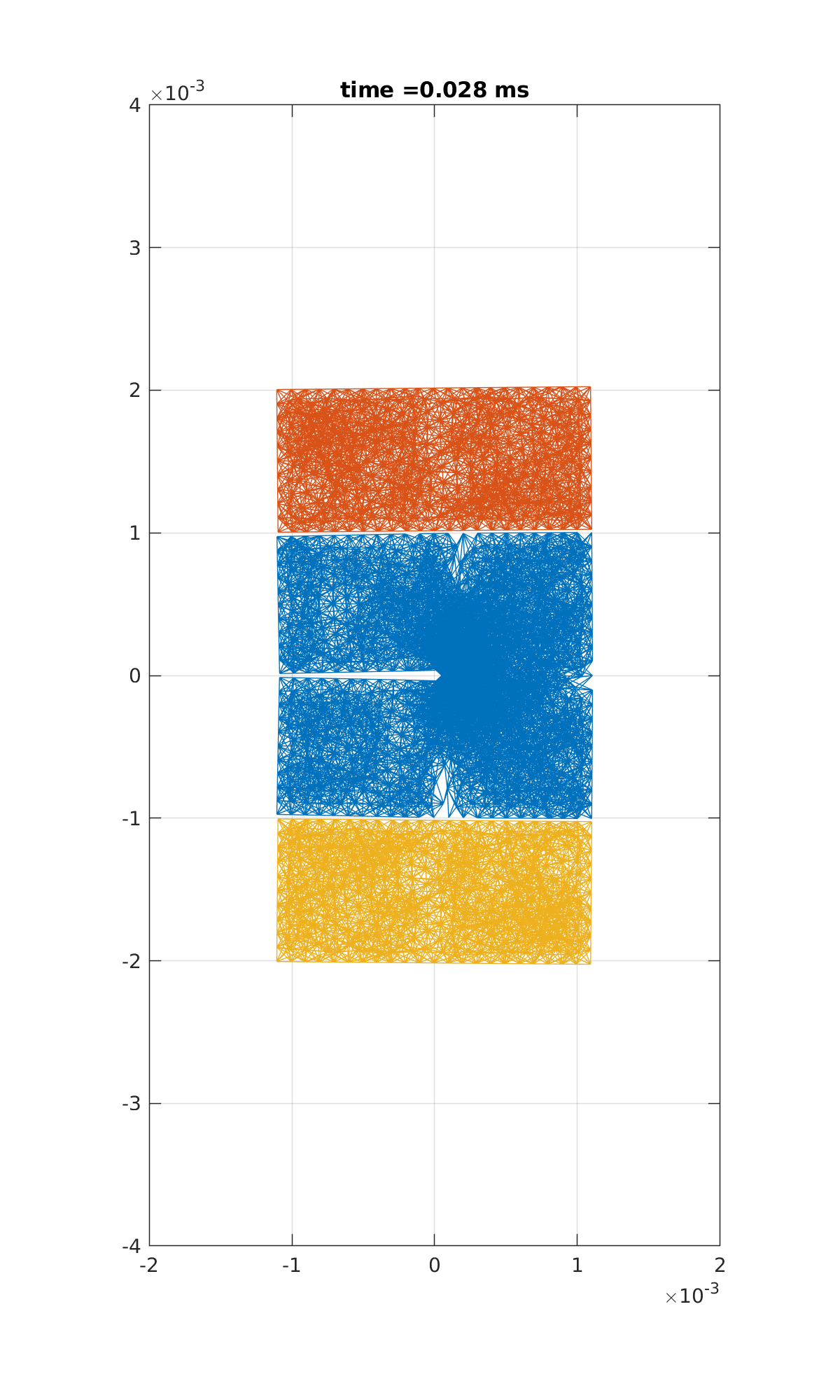}
    }
    \subfloat[]{\label{fig:crack_3_29.png}
    \includegraphics[width=0.2\linewidth, trim={{0.2\linewidth} {0.4\linewidth} {0.2\linewidth} {0.4\linewidth}},clip]{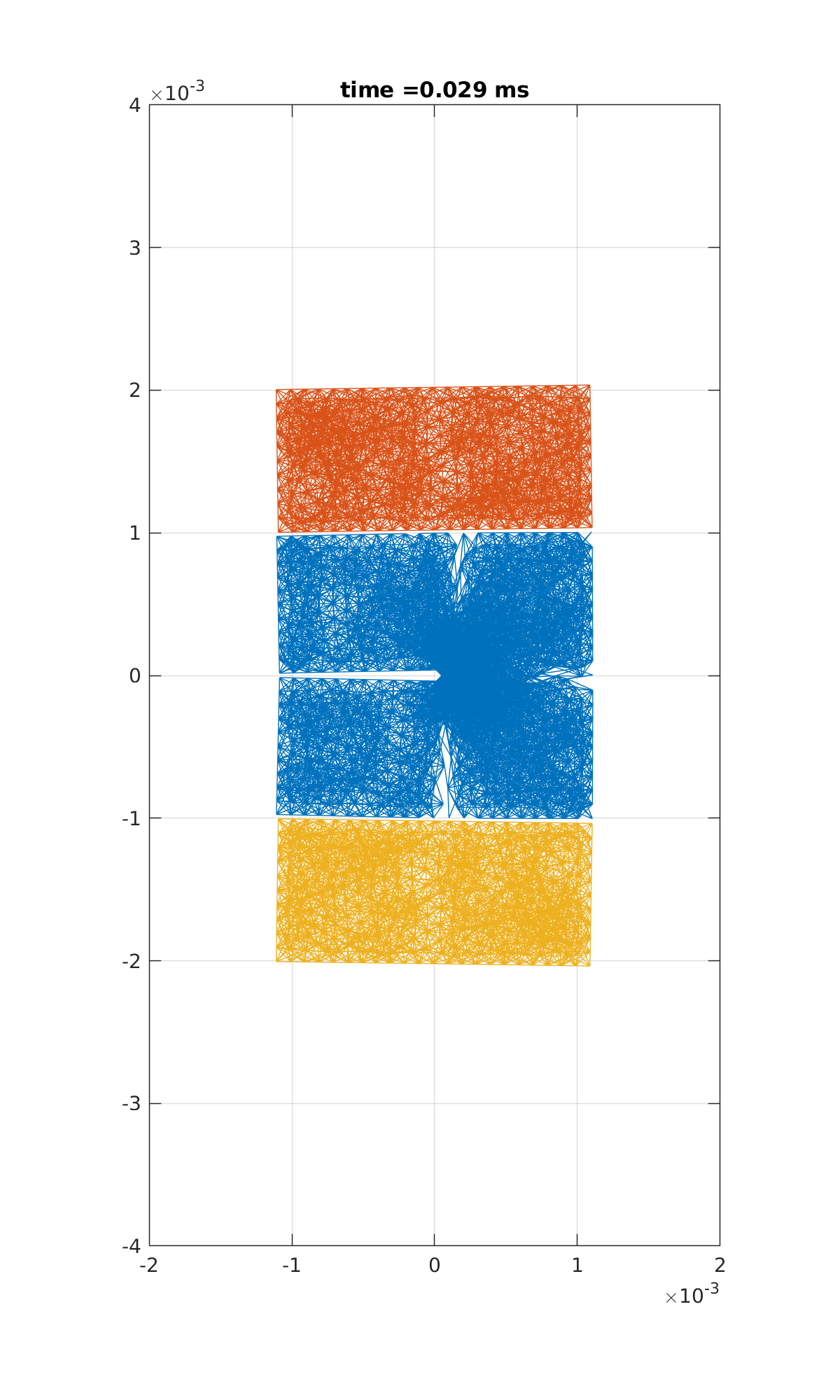}
    }
    \subfloat[]{\label{fig:crack_3_30.png}
    \includegraphics[width=0.2\linewidth, trim={{0.2\linewidth} {0.4\linewidth} {0.2\linewidth} {0.4\linewidth}},clip]{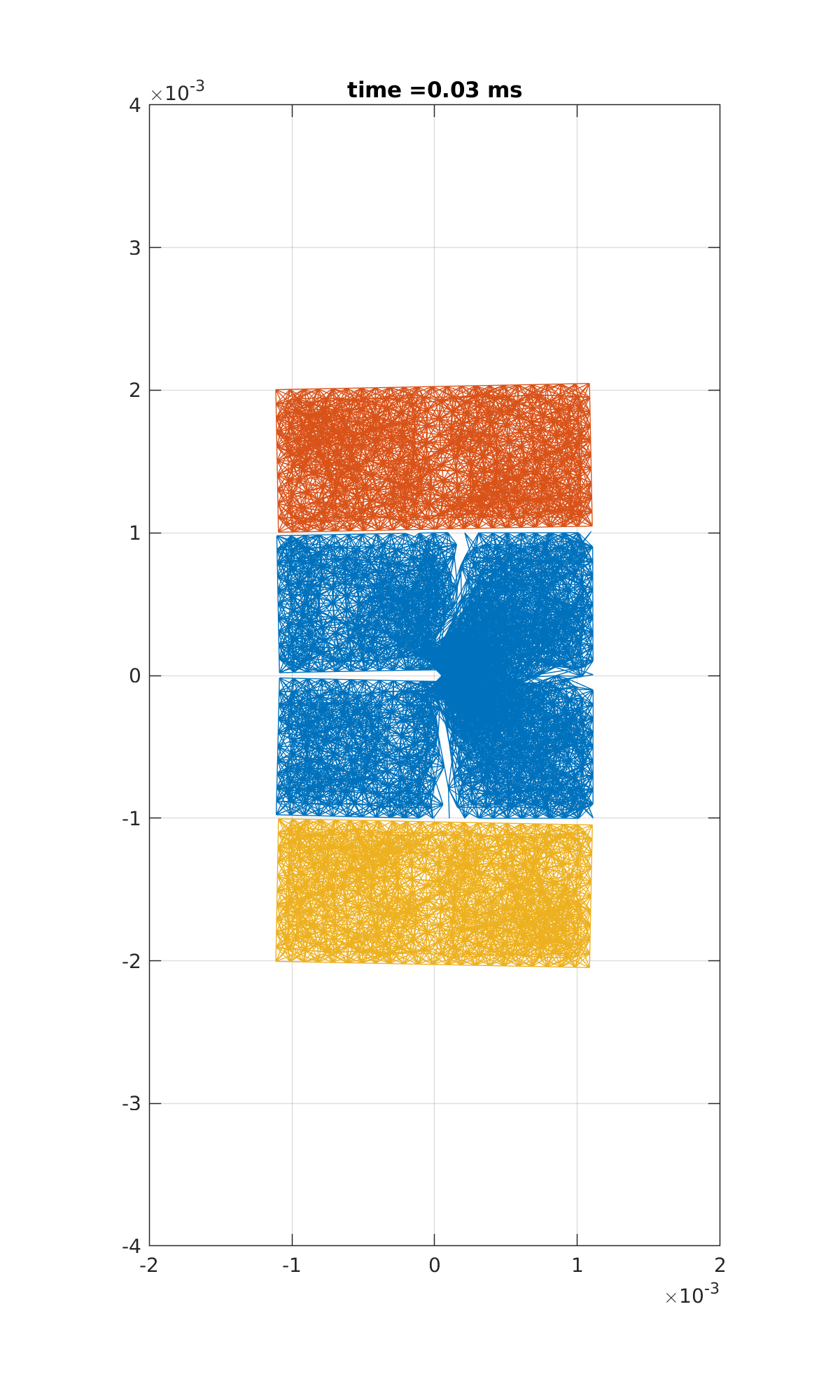}
    }
    \subfloat[]{\label{fig:crack_3.png}
    \includegraphics[width=0.2\linewidth, trim={{0.2\linewidth} {0.4\linewidth} {0.2\linewidth} {0.4\linewidth}},clip]{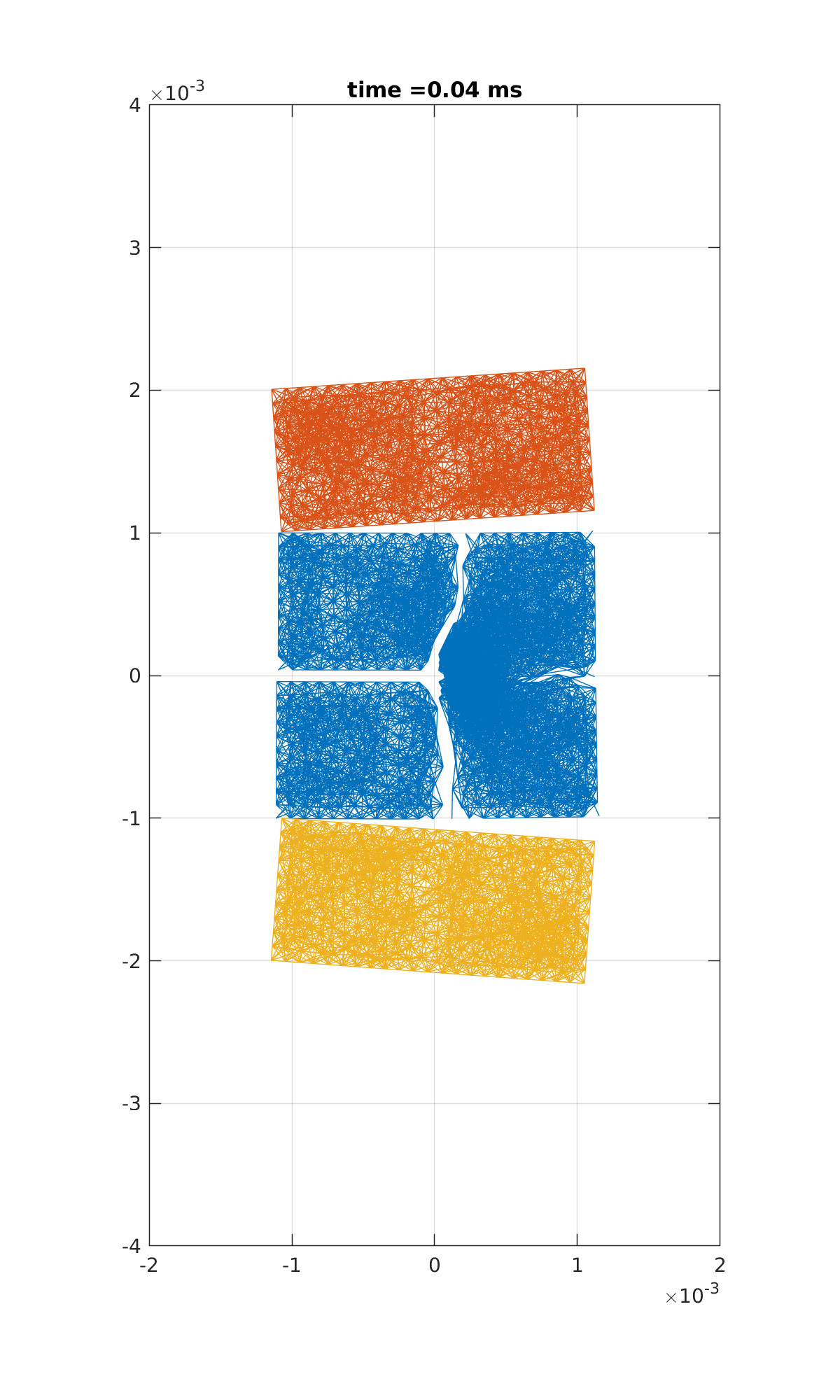}
    }
\caption{Crack formation in domains with pre-notch for different fracture toughness. Figures in the top row correspond to $t = 29, 30, 31$, and 50 $\mu$s after the simulation starts, and $t =28, 29, 30$, and 40 $\mu$s for the bottom row.}%
\label{fig:crack_test}
\end{figure}
In \Cref{fig:crack_1_29.png,fig:crack_1_30.png,fig:crack_1_31.png,fig:crack_1.png}, the fracture toughness is taken to be $G_c = 135$ J/m$^2$.
We observe that a crack formed on the right edge of the particle propagates toward the tip of the pre-notch. When the crack tip reaches the tip of the pre-notch, particle A produces 2 disjoint child particles.
In \Cref{fig:crack_3_28.png,fig:crack_3_29.png,fig:crack_3_30.png,fig:crack_3.png}, the fracture toughness is taken to be $G_c = 13.5$ J/m$^2$, implying a softer particle.
After the symmetric impact, 3 cracks are formed on the right, top, and bottom of the particle that propagate toward the crack tip. The cracks starting at the top and the bottom meet with the tip pre-notch simultaneously and before the crack from the right reaches the tip of the pre-notch.
As a result, particle A is divided into 3 child particles, one of which (the one on the right) is partially cracked.
Due to the stress concentration at the corners of the domains the corners are smoothed out upon impact.
In \Cref{fig:crack_test} we have showed all the intact bonds present in the particles.
{\revv 
The crack paths visualized here as an absence of peridynamic bonds emerge naturally from the progressive failure of bonds.} Under the same experiment setup, the two different fracture patterns and the number of subsequent child particles generated after the collision are entirely determined by the fracture toughness $G_c$, which is a material property.  {\rev  It is important to observe that there is no interpenetration at the notch of particle A nor in the newly formed free surface of the cracks and child particles. This is due to the presence of self contact forces in the peridynamic model \Cref{sub:self_contact} developed here.}

\section{Simulations: particle beds}%
\label{sec:simulations_bulk_test}
{\rev In this section, we simulate the dynamic settling and compression of particle beds and illustrate the relative effects of different particle shapes and particle topology.
We restrict ourselves to the two-dimensional case. The walls and floor of the particle bed container are modeled using straight lines and
for consistency we will refer to surface area as ``volume".
Wall-particle contact is computed in the way described in \Cref{sec:intersection_with_the_wall}. 
The domain containing the initial aggregate in a columnar configuration is denoted by $\Omega$, see \Cref{fig:data/plus_arrangement..png}. It has prescribed height and is bounded by a top horizontal edge, side walls and floor. The boundary is denoted by $\partial\Omega$. As mentioned earlier, it is essential to start the dynamics from an initial particle configuration that is agnostic to particle shape.
To realize such a initial configuration we require the maximum cross-sectional diameter of every particle to be a fixed constant across all shapes. Additionally, the location of the center point of this diameter is prescribed so each particle experiences no interaction force from any other particle or the container walls. Lastly, the particles are randomly oriented. {\revv More generally, we will consider polydispersed aggregates consisting of particles of the same shape, which are initially free from contact forces and are randomly rotated about their centroids.} Here ``diameter'' is short for the maximum diameter of a particle. The method accomplishes the following goals:
\begin{itemize}
    \item A fast way to construct particle agnostic initial conditions for aggregates.
    \item A means to generate initial conditions to better understand macroscopic properties from microscopic dynamics of differently shaped particles.
\end{itemize}
The initial configuration is realized using the notion of security disks inside which the particle is placed. 
The size of the particle is chosen such that only its boundary has points in common with the boundary of the security disk. Each particle is randomly rotated about the center of its respective security disk. Next, we define the notion of jammed state for a disk (or sphere) packing of domains with disks  of a finite number of different radii.
\begin{definition}
\label{Jammed}
A domain packed with disjoint disks is in a jammed state if a displacement of any disk from its location  causes it to overlap with another disk.
\end{definition}
With this in mind the location of the security disks are chosen such that by increasing all their radii by half the interaction radius $R_c$ recovers a sphere packing that is in a jammed state. This provides the starting configuration for the dynamics. Other methods for doing this using more involved 
mesh-based packing of disks are possible resulting in higher packing densities, \cite{LabraOnateHighdensity09} but we do not apply them here.
The specifics of the construction are given in the next section.

}

\subsection{Generating an initial particle distribution}%
\begin{figure}[htbp]
\centering
\subfloat[]{\label{fig:inc_orig}
    \includegraphics[width=0.40\linewidth]{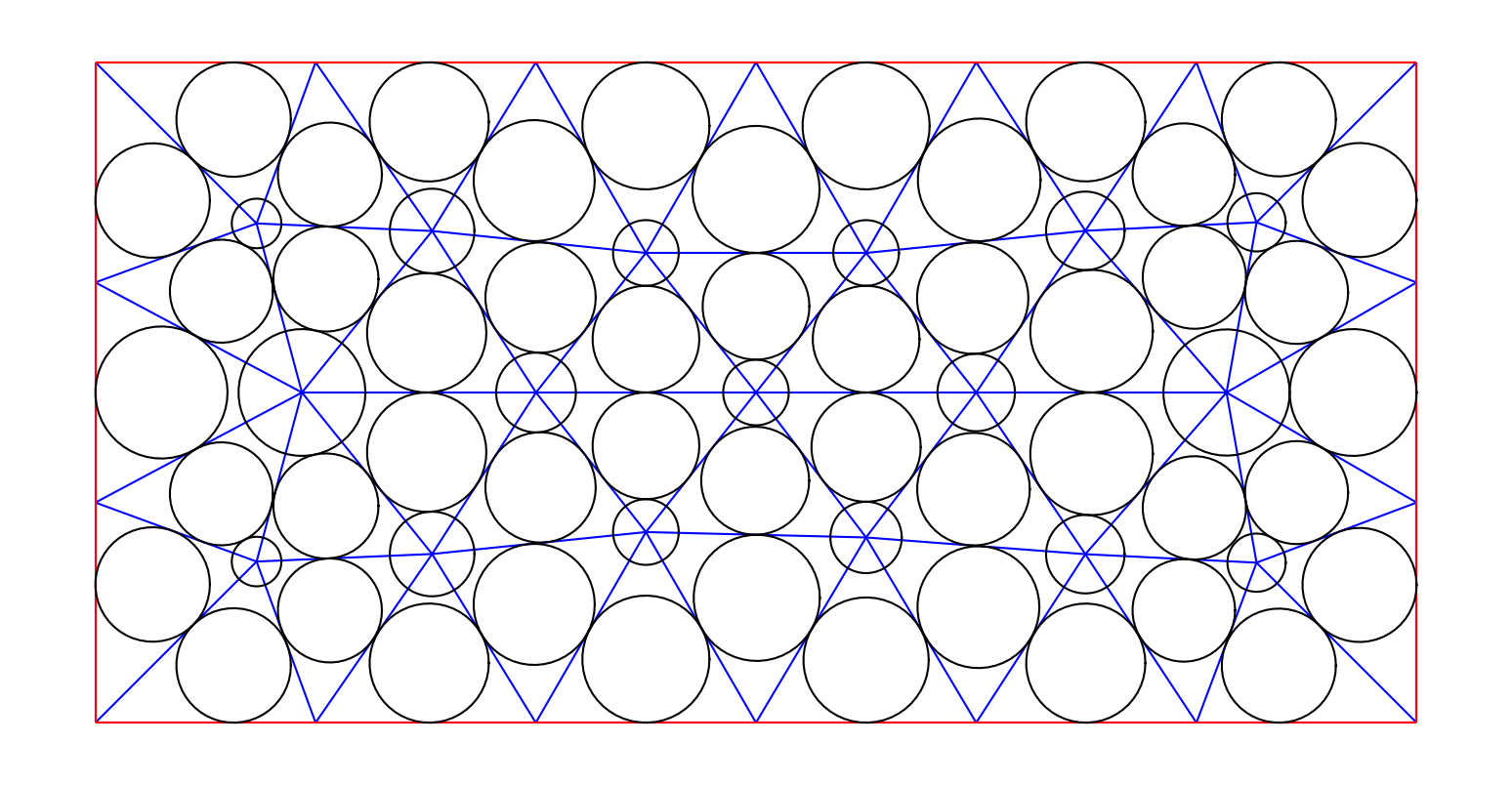}
    }
\subfloat[]{\label{fig:inc_bdry}
    \includegraphics[width=0.40\linewidth]{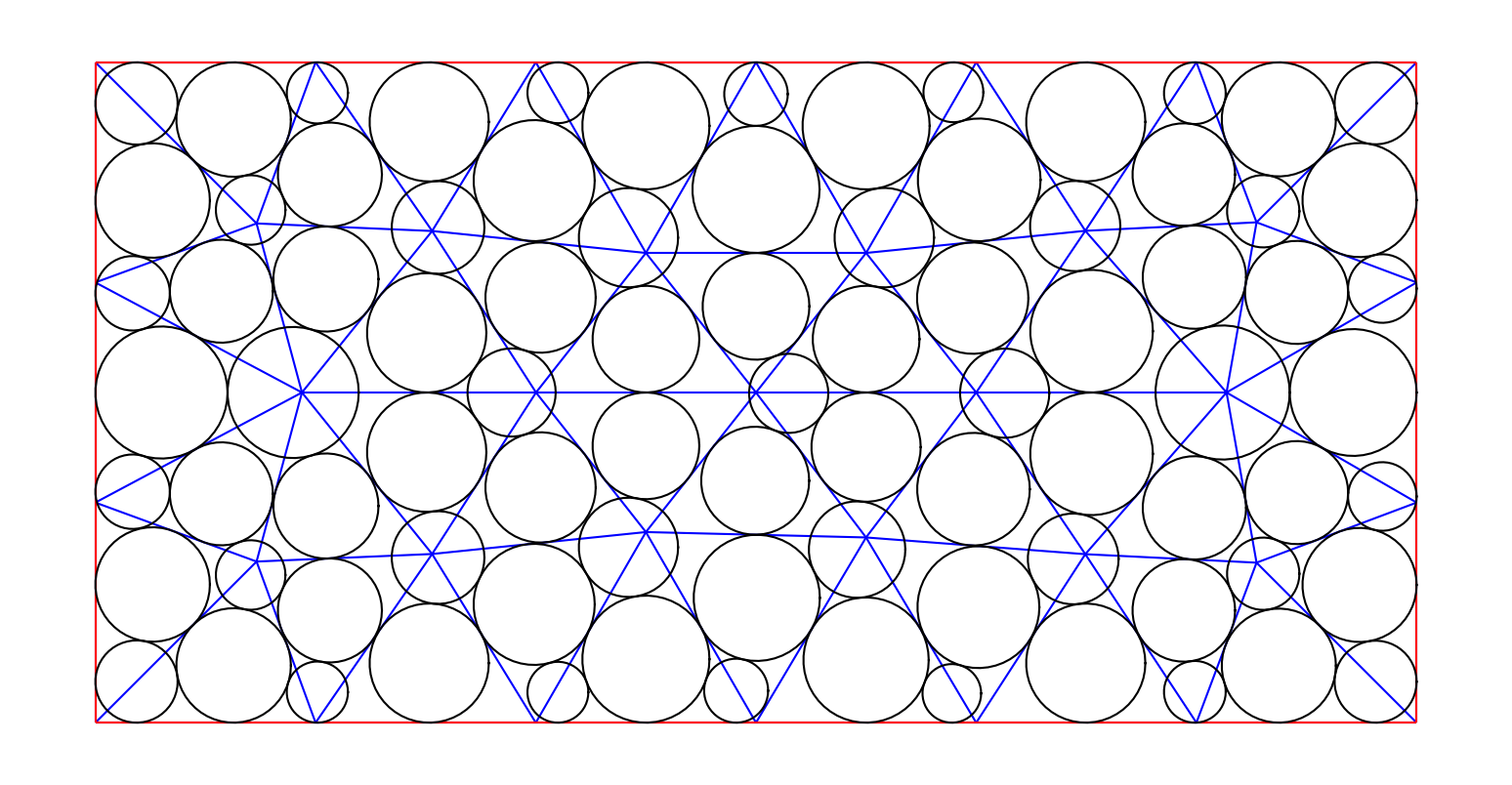}
    }\\
\caption{
    Initial particle arrangement generation.
     Triangular mesh of the wall interior and the placement of inscribed circles for the triangles and initial placement nodal circles with packing ratio 0.69 (\Cref{fig:inc_orig}).
    Final jammed configuration of security disks obtained after maximizing the radii of the nodal circles including circles associated with the boundary nodes with packing ratio 0.80 (\Cref{fig:inc_bdry}).
}
\label{fig:myfig}
\end{figure}

\label{sec:generating_particle_distribution}
We present an algorithm for generating an initial distribution of particle sizes and locations of individual particles within an arbitrary region $\Omega$ independent of particle shape. The algorithm first constructs a jammed configuration of closed security disks and then decreases all their radii by half the interaction radius $R_c$. Last the particles are placed inside the security disks such their diameters match and are rotated randomly and independently.
\begin{enumerate}
    \item Apply a prescribed discretization of $\Omega$ using finite-element mesh generating software. Here we use a triangular mesh. 
\item Construct circles that are inscribed within triangular mesh elements contained in $\Omega$. The location of the center of mass of a mesh triangle $T$ with vertices $\vv_1, \vv_2, \vv_3 \in \R^2$ is given by
$
    \cc_T = \frac{\vv_1 l_1 + \vv_2 l_2 + \vv_3 l_3}{l_1 + l_2 + l_3} 
    $
where 
$l_1 = \abs{\vv_2 - \vv_3} $, 
$l_2 = \abs{\vv_1 - \vv_3} $, and 
$l_3 = \abs{\vv_1 - \vv_2} $.
The radius is $r_T = \frac{2 \abs{T} }{l_1 + l_2 + l_3}$, where $\abs{T}$ is the area of the triangle, see \Cref{fig:inc_orig}. 

\item Create nodal circles containing each node of the mesh in $\Omega$ such that they do not intersect with any of the inscribed circles of the neighboring triangles and do not overlap with the exterior of the domain $\Omega$, see \Cref{fig:inc_bdry}. 

\item Calculate centers and radius of nodal circles delivering a jammed configuration.  Let $r(\xx,\cc)=|\xx-\cc|$ where $\cc$ is the center of any circle containing a node denoted by $\vv$ and $\xx$ is a point on the boundary of the  circle. The  constraints are 1) $r(\xx,\cc)\leq \rho_T(\cc)$ where $T$ is any triangle with vertex $\vv$ and $\rho_T(\cc)$  is the distance between $\cc$ and the inscribed circle inside $T$; 2) $r(\xx,\cc)\leq d(\cc,e)$ where $d(\cc,e)$ is the distance of $\cc$ to a boundary edge $e$. The radius $r_\vv$ and center $\cc_\vv$ of each nodal circle for a jammed configuration is given by 
\begin{align*}
\begin{split}
     r_\vv& = \max_{\cc}\{ \max_\xx
    \{ r(\xx,\cc) \wedge d(\cc, e) \}\}\hbox{ and } c_\vv=\rm{argmax}\{\max_\xx
    \{ r(\xx,\cc)  \wedge d(\cc, e) \}\}.
\end{split}
\end{align*}	
\item Reduce the radius of each security disk by $\frac{R_c}{2}$ so that contact forces are not activated.
\item Place particle of diameter equal to security disk inside and rotate randomly about center of security disk.
 \end{enumerate}   


The optimization problem 
is solved numerically. 
 In \Cref{fig:myfig}, we show the construction of a jammed packing from an arbitrary triangular mesh.

\subsection{Construction of particle shapes}%
\begin{figure}[htpb]
\centering
    \subfloat[]{\label{roundness-shape}
        \includegraphics[width=0.22\linewidth]{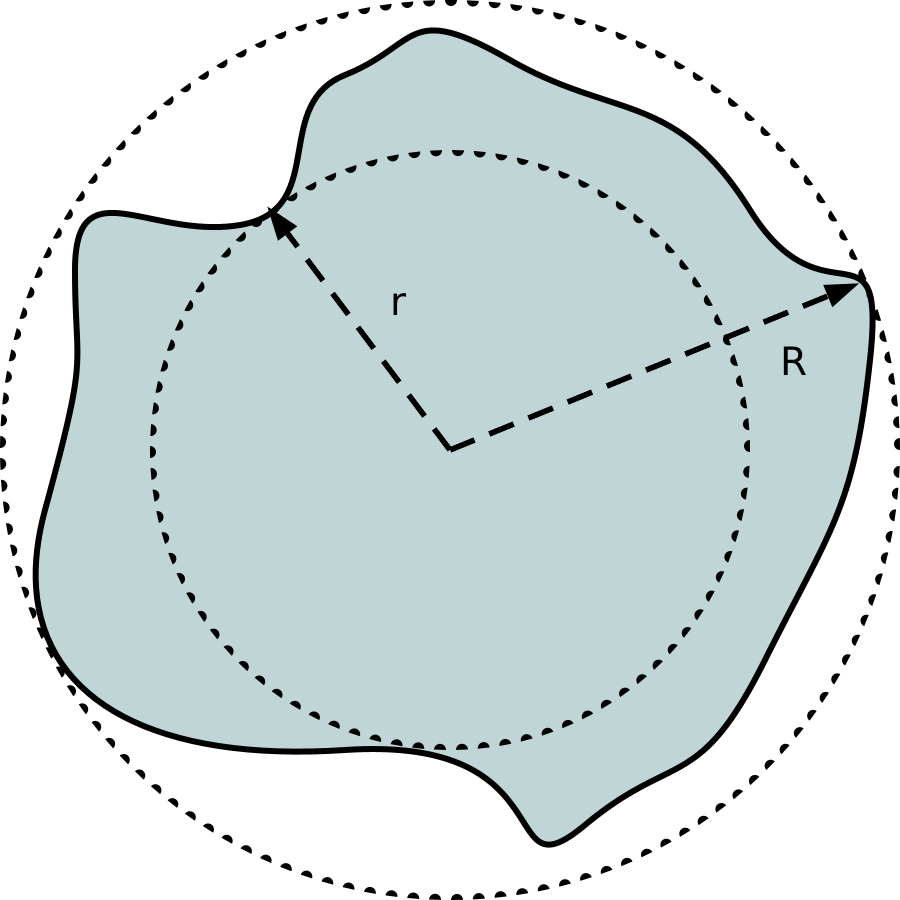}
    }
    \subfloat[]{\label{n4-shape}
     \includegraphics[width=0.22\linewidth]{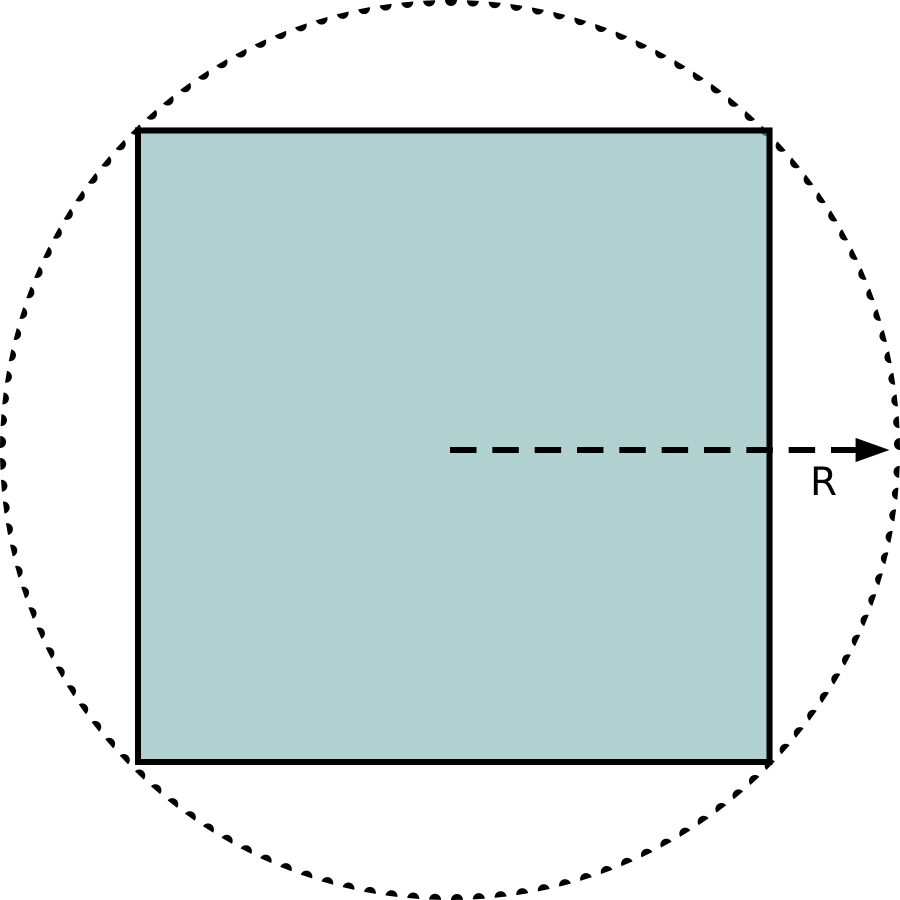}
    }
    \subfloat[]{\label{plus-shape}
        \includegraphics[width=0.22\linewidth]{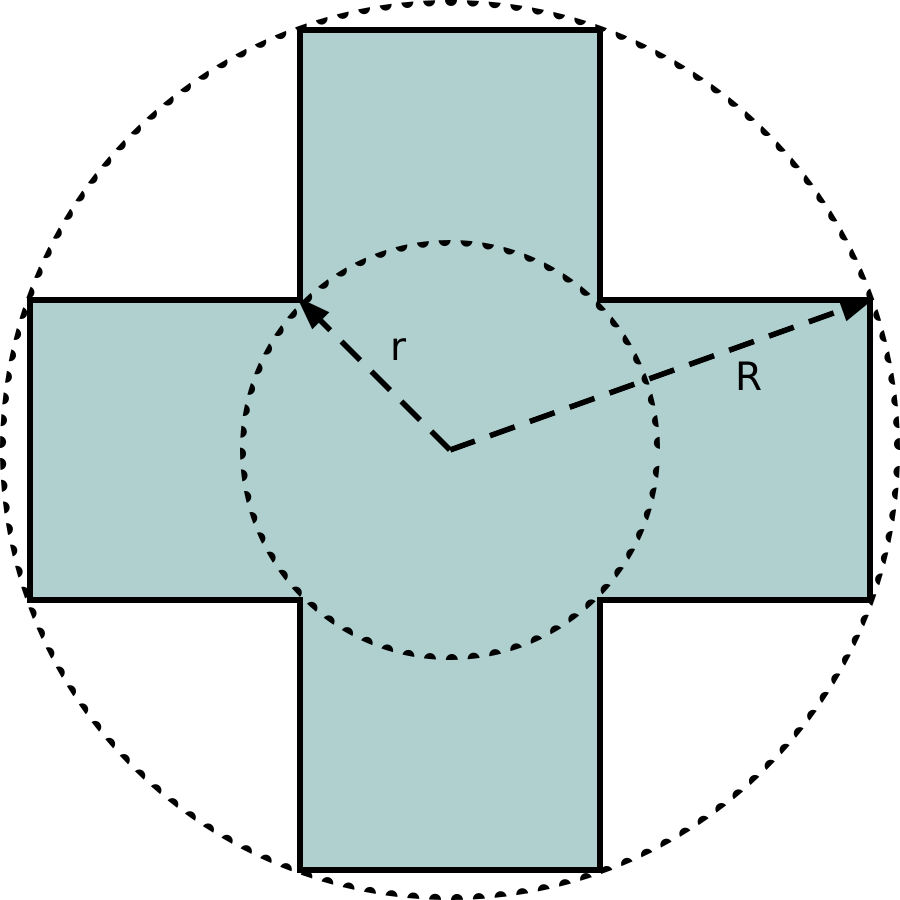}
    }
    \subfloat[]{\label{ring-shape}
        \includegraphics[width=0.22\linewidth]{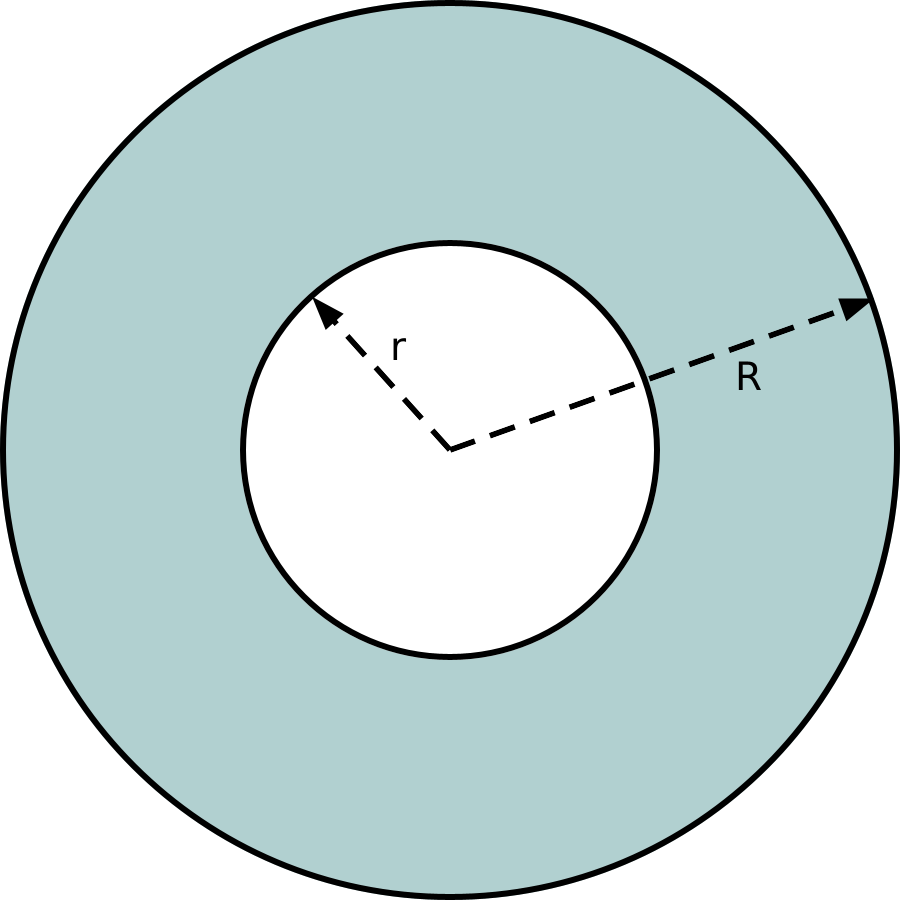}
    }
\caption{Shapes of particles considered: perturbed disk, square, plus-shaped, and annular or ring-shaped particle. All shapes are inscribed in security disks using the method of \Cref{sec:generating_particle_distribution}.} 
\label{fig:shapes}
\end{figure}

\label{sub:particle_shapes_considered}
{\rev 
We consider particle aggregates consisting of particle shapes that are perturbed disks, square-shaped, plus-shaped, and annular or ring-shaped particles (see \Cref{fig:shapes}).
The particle shapes are chosen to study the effect of convexity, symmetry, and particle topology.
The ring-shaped particles are rotationally symmetric, the square-shaped and plus-shaped particles have dihedral symmetry, and the perturbed disks are asymmetric. On the other hand, the plus-shaped particles and the perturbed disks are nonconvex, whereas the ring-shaped particles are nonconvex with a convex outer boundary.
Therefore, by considering these shapes we provide a good variability of the key geometric and topological properties that influences the bulk behavior \cite{nonsphericalPhysRevE.75.051304,rodPhysRevE.73.031306,interlockingPhysRevE.85.051307}. 
}
All shapes are inscribed in security disks using the method of \Cref{sec:generating_particle_distribution}.

The non convex particle shapes including the perturbed disks and cross shapes are constructed by perturbing the boundary of a disk of radius $R$ inward randomly.
The vertices of the piece wise polynomial approximation to the particle boundary is given by the ordered set
$
\partial D = \{  (r_i \cos \theta_i , r_i \sin \theta_i): i = 1, \dots, N\},
$ 
where $r_i = R(1 - X_i)$, where $\{X_i\}$ are independent and identically distributed uniform random variables on the interval $[0,1-\varphi]$. 
The roundness is measured by the ratio of the radii of the circles that can be inscribed within and can be circumscribed over the particle boundary. The average ``roundness,'' $\varphi$  is taken to lie in the interval $\varphi \in [0,1]$.
For particles given by perturbed disks we apply piece wise linear approximations and have taken $N = 20$ and $\varphi = 0.6$.
The square shapes are inscribed in the boundary of a security disk say of radius $R$ hence the length of the side is $\sqrt{2}R$. 
For the plus-shaped particles we choose inner radius $r = cR$ with $c = 0.2$.
For annular particles we take the inner radius to be  $\tau\times\hbox{ outer radius}$ with $\tau = 0.6$.

\subsection{Particle bed settling under gravity}%
\label{sub:settle}

\begin{figure}[htpb]
\centering
    \subfloat[]{\label{fig:data/plus_arrangement..png}
	\includegraphics[width=0.18\linewidth]{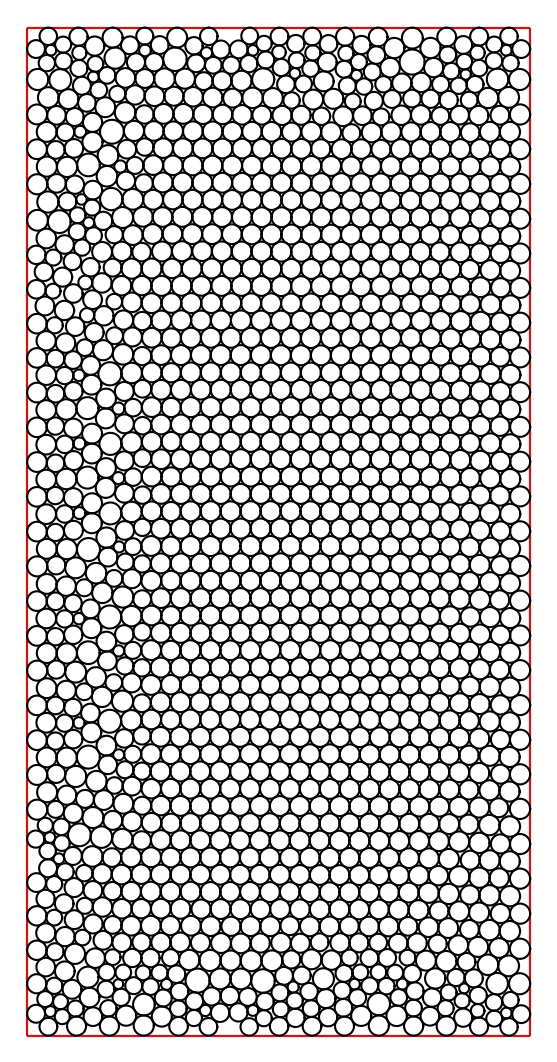}
    }
    \subfloat[]{\label{fig:settle-compact-2}
        \includegraphics[width=0.18\linewidth]{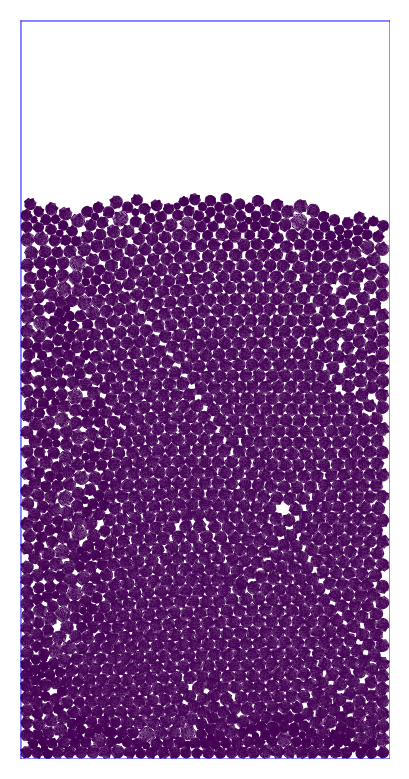}
    }
    \subfloat[]{\label{fig:settle-compact-1}
        \includegraphics[width=0.18\linewidth]{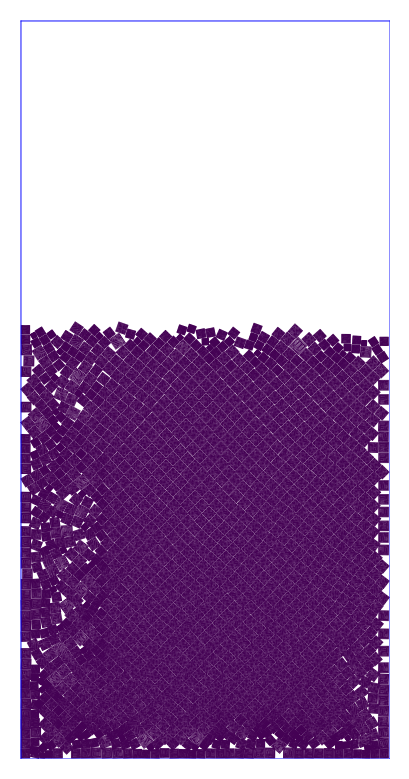}
    }
    \subfloat[]{\label{fig:settle-compact-3}
	\includegraphics[width=0.18\linewidth]{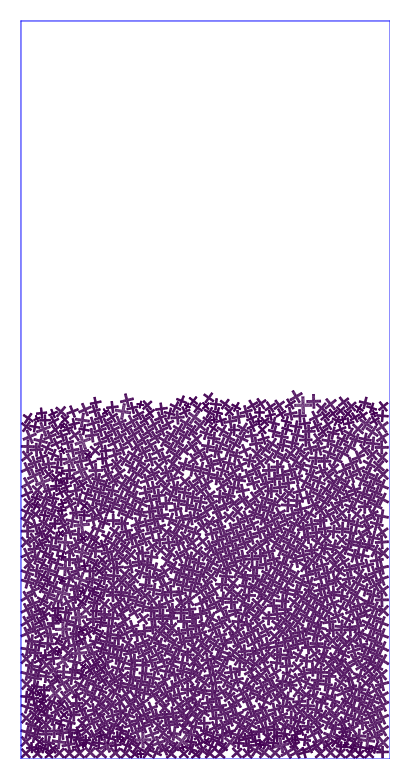}
    }
    \subfloat[]{\label{fig:settle-compact-4}
	\includegraphics[width=0.18\linewidth]{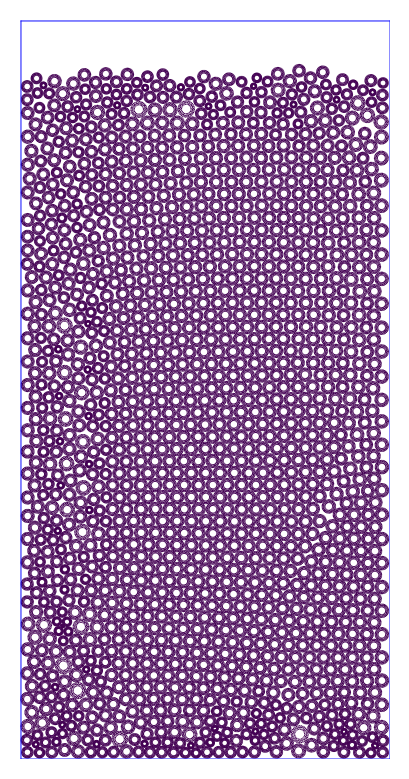}
    }
    \caption{
	The arrangement of security disks used to generate the initial particle arrangement (\Cref{fig:data/plus_arrangement..png}).
	Particles aggregates settle under gravity and in the presence of friction and damping force, and come to an equilibrium.
The particle arrangement at equilibrium for aggregates consisting of perturbed disks, squares, plus-shaped and annular particles are shown in \Cref{fig:settle-compact-2,fig:settle-compact-1,fig:settle-compact-3,fig:settle-compact-4}, respectively.
    }%
\label{fig:plus_gravity}
\end{figure}

{\rev  The dynamic settling of particle aggregates for different shapes and topology are simulated. Upon reaching equilibrium, the macroscopic properties given by height and total particle volume fraction are measured. For this case the volume fraction is the portion of the particle column occupied by particles. Here the volume of the particle column is given by the product of the height of the equilibrium configuration multiplied by the width of the container. The simulations illustrate the effect of the particle geometry and topology on the macroscopic quantity given by the particle volume fraction at equilibrium.}
Here, we consider 1490 particles of radii 0.4 mm - 1.1 mm (with mean 0.8 mm and standard deviation 0.09 mm) in a rectangular container of size 50 mm $\times$ 100 mm and study the effect of particle shapes on the packing under {\revv a gravitational acceleration of 5000 m/s$^2$}.
The initial particle arrangement is generated using the technique discussed in \Cref{sec:generating_particle_distribution} so that each particle does not experience contact force from other particles or from the wall boundaries.
After generating the positions of security disks containing particles so that they are jammed, we reduce the radii of the disks by $\frac{R_c}{2}$ so that no contact force is activated between particles. {\revvv Here, the contact radius is taken to be $R_c = \frac{\epsilon}{5}$, and the mesh size is taken to be $h = \frac{\epsilon}{8}$, where $\epsilon = 0.5$ mm.}
Next, we apply a random rotation on the particles about their centroid.
We activate the gravitational force and let the particles fall under their own weight.
The particle bed is allowed to come to equilibrium until all oscillations of the aggregate decay to zero due to damping and friction force.
Here we have taken $\mu = 0.8$ and $r_d = 0.8$.
The jammed disk packing for the initial configuration is shown in \Cref{fig:data/plus_arrangement..png} and particle aggregates after coming to equilibrium are shown in \Cref{fig:settle-compact-2,fig:settle-compact-1,fig:settle-compact-3,fig:settle-compact-4}.
{\revvv
The volume fraction $\phi$ is computed by taking the ratio of the combined particle volume to the volume enclosed by the aggregate boundary.
In \Cref{tab:settle}, we show the change in the volume fraction between the initial and the fully settled configuration for different shapes.
}
\begin{table}[ht]
    \caption{Volume of individual particles and the bulk volume fraction at equilibrium.}
    \label{tab:settle}
    \centering
\begin{tabular}{| c | c | c | c | c |}
    \hline
    Shape & Perturbed disk & Square & Plus & Annulus
    \\
    \hline
    Particle volume & $\pi R^2 \frac{(\varphi + 1)^2}{4}$ & $2 R^2$ & $c^2 R^2 ( 1 + 4 \sqrt{\frac{2}{c^2} - 1})$ &  $\pi R^2 (1  - \tau^2)$
    \\
    \hline
    Initial $\phi$ &   0.55 &  0.37 & 0.28 & 0.86
    \\
    \hline
    Equilibrium $\phi$ &   0.75 &  0.64 & 0.59 & 0.92
    \\
    \hline
\end{tabular}
\end{table}
{\rev Here the particle shapes have been chosen to illustrate the effect of convexity, non convexity and topology on packing density. We find the volume fraction occupied by packed particles is lowest for plus particles followed by squares, perturbed disks, and annuli. The simple plus particle has the least packing fraction while the perturbed sphere has a packing fraction lying above the square. The ring has the largest volume fraction due to the extra excluded volume due to the interior hole.}
{\revvv 
Among all shapes considered, the reduction of volume fraction of the loosely packed aggregate due to gravity is seen to be the maximum for the the plus-shaped particles, and the minimum for the ring-shaped particles.
}
\subsection{Bulk compaction with damage}%
\label{sub:compaction-damage}
Here, we study the effect of damage on the bulk behavior.
For each shape described in \Cref{sub:particle_shapes_considered} we consider a particle aggregate with 496 particles of radii 0.3 mm - 1.1 mm (with mean 0.7 mm and standard deviation 0.09 mm) in a rectangular container with height $h = 20$ mm and width $l= 20$ mm.
The initial position and radii of the security disks for aggregates of differently shaped particles are taken to be the same.
In each case, the top wall of the container is lowered at the speed of $v = 1$ m/s. Here, we allow the particles to experience damage, which is incorporated according to \Cref{sec:damage_model}.
Gravity is ignored here, therefore the volume fraction is determined entirely by the position of the top wall boundary.
The contact radius $R_c$, the contact parameters $K_n, \mu$, and $r_d$ are taken to be the same as in \Cref{sub:settle}.

The volume fraction is given by the ratio of the total particle volume and the volume of the container.
Therefore, the volume fraction in our simulation as a function of time is given by 
$
    \phi(t) = \frac{\sum^{N}_{i=1}  V_i }{l (h - vt)},
    $
where $V_i$ is the volume of the $i^{th}$particle and $N$ is the total number of particles.
The \textit{bulk damage} is defined as the average particle damage over all particles, whereas the \textit{particle damage} is defined as the mean damage over all points in the particle.

\subsubsection{Effect of particle shape}
\label{sub:compress-shape}
\begin{figure}[htpb]
\centering
    \subfloat[]{\label{fig:compress-column-1}
        \includegraphics[width=0.25\linewidth]{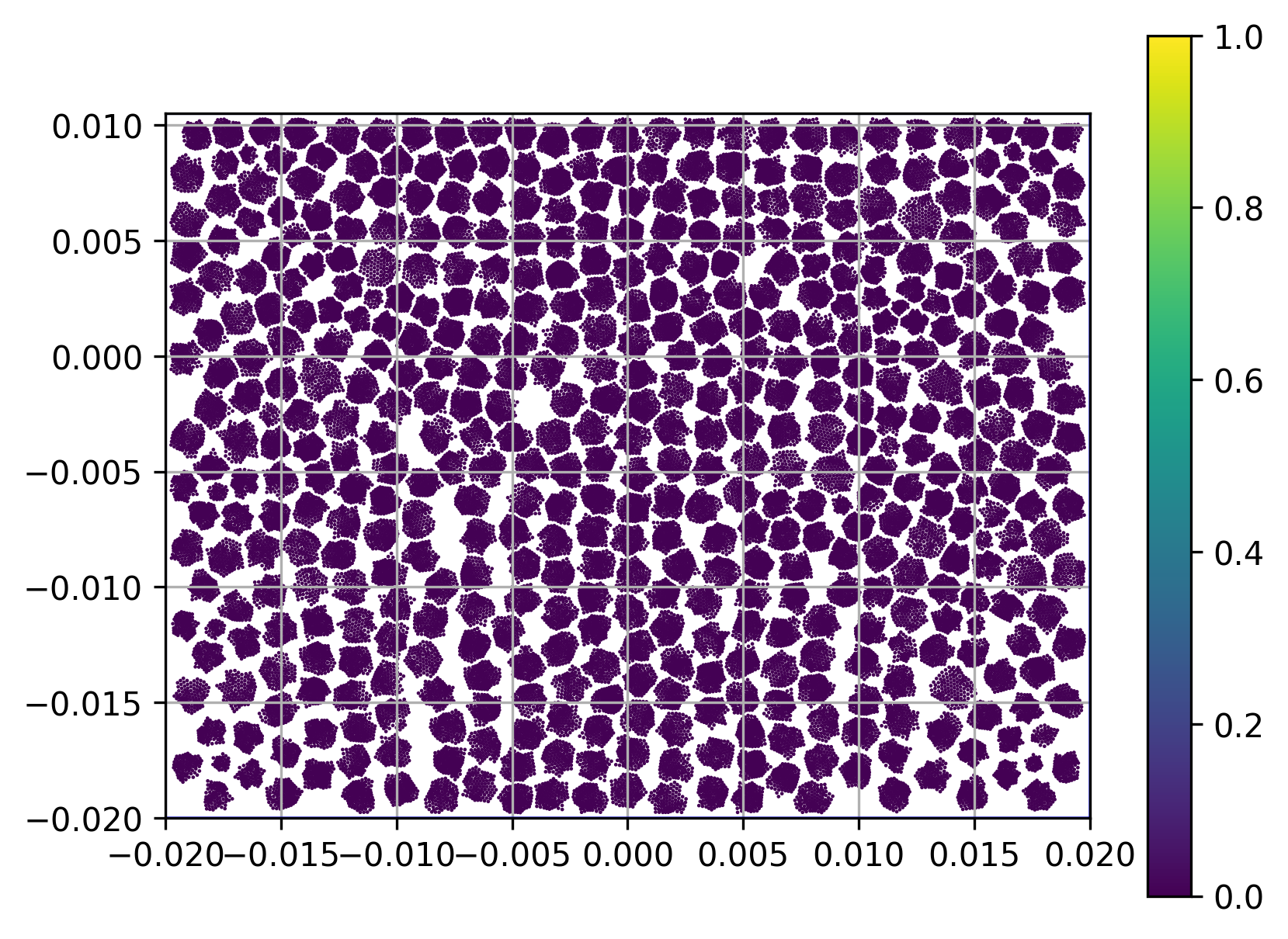}
    }
    \subfloat[]{\label{fig:compress-column-2}
        \includegraphics[width=0.25\linewidth]{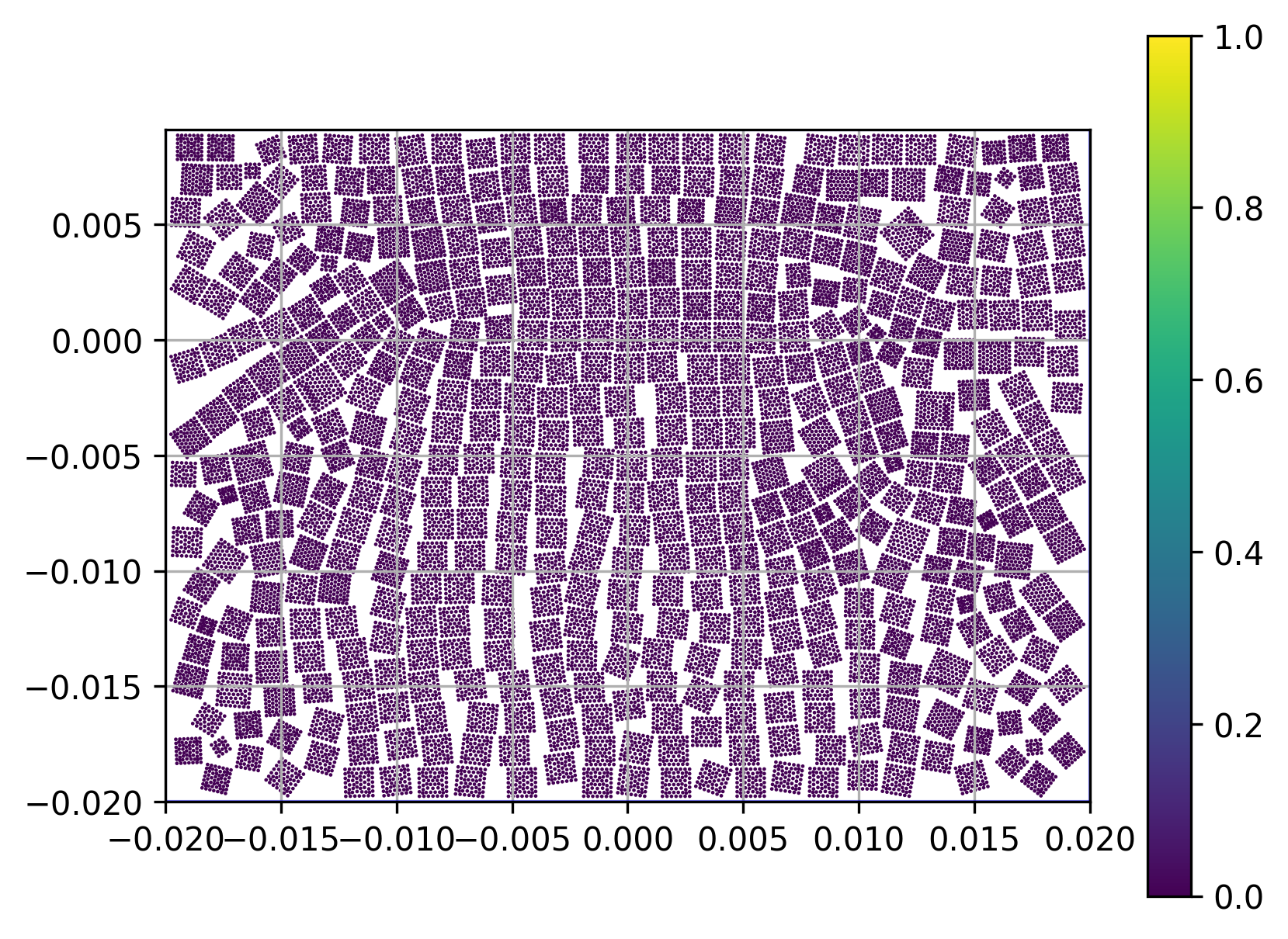}
    }
    \subfloat[]{\label{fig:compress-column-3}
        \includegraphics[width=0.25\linewidth]{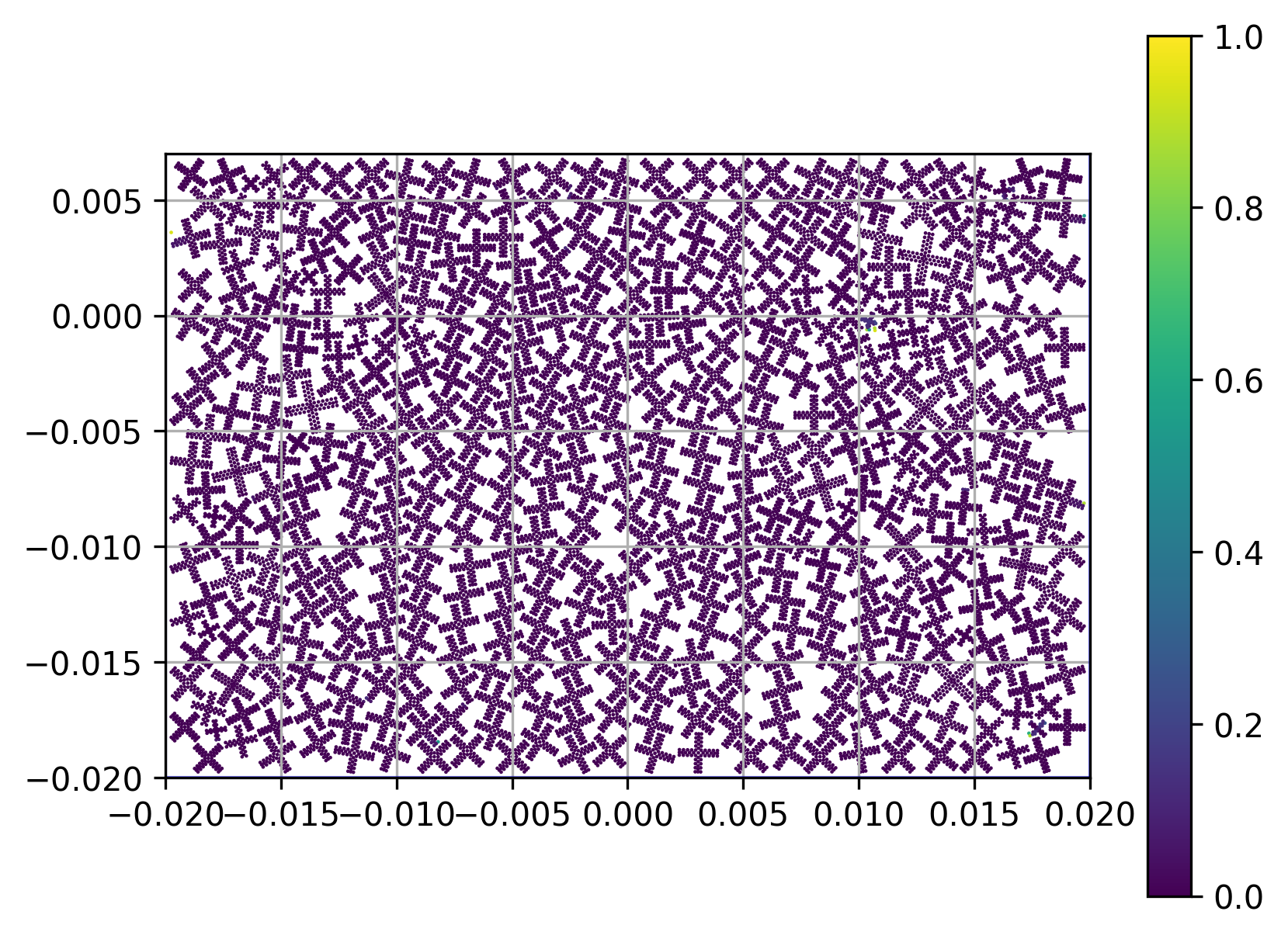}
    }
    \subfloat[]{\label{fig:compress-column-4}
        \includegraphics[width=0.25\linewidth]{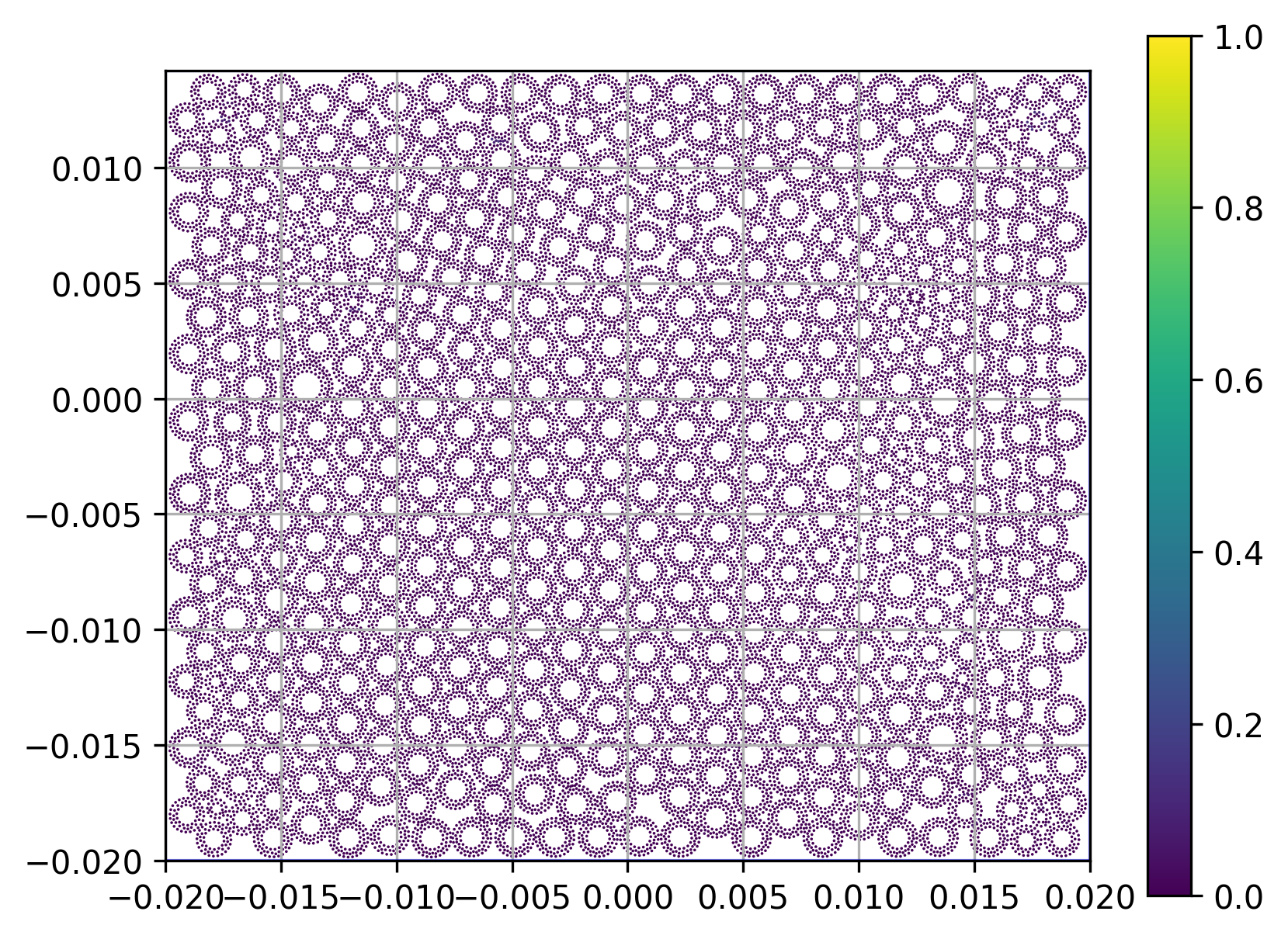}
    }
    \\
    \subfloat[]{\label{fig:compress-damage-1}
        \includegraphics[width=0.25\linewidth]{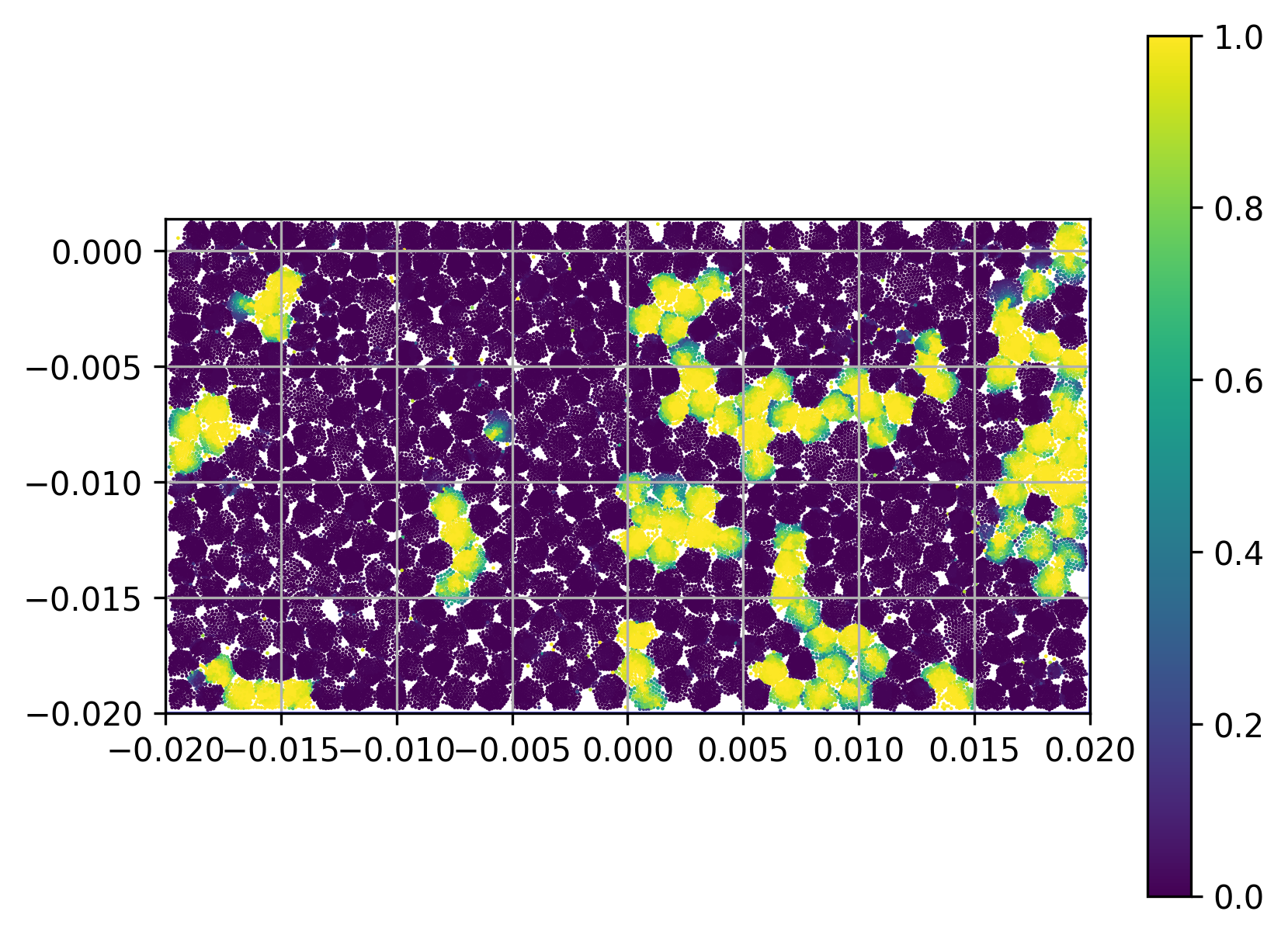}
    }
    \subfloat[]{\label{fig:compress-damage-2}
        \includegraphics[width=0.25\linewidth]{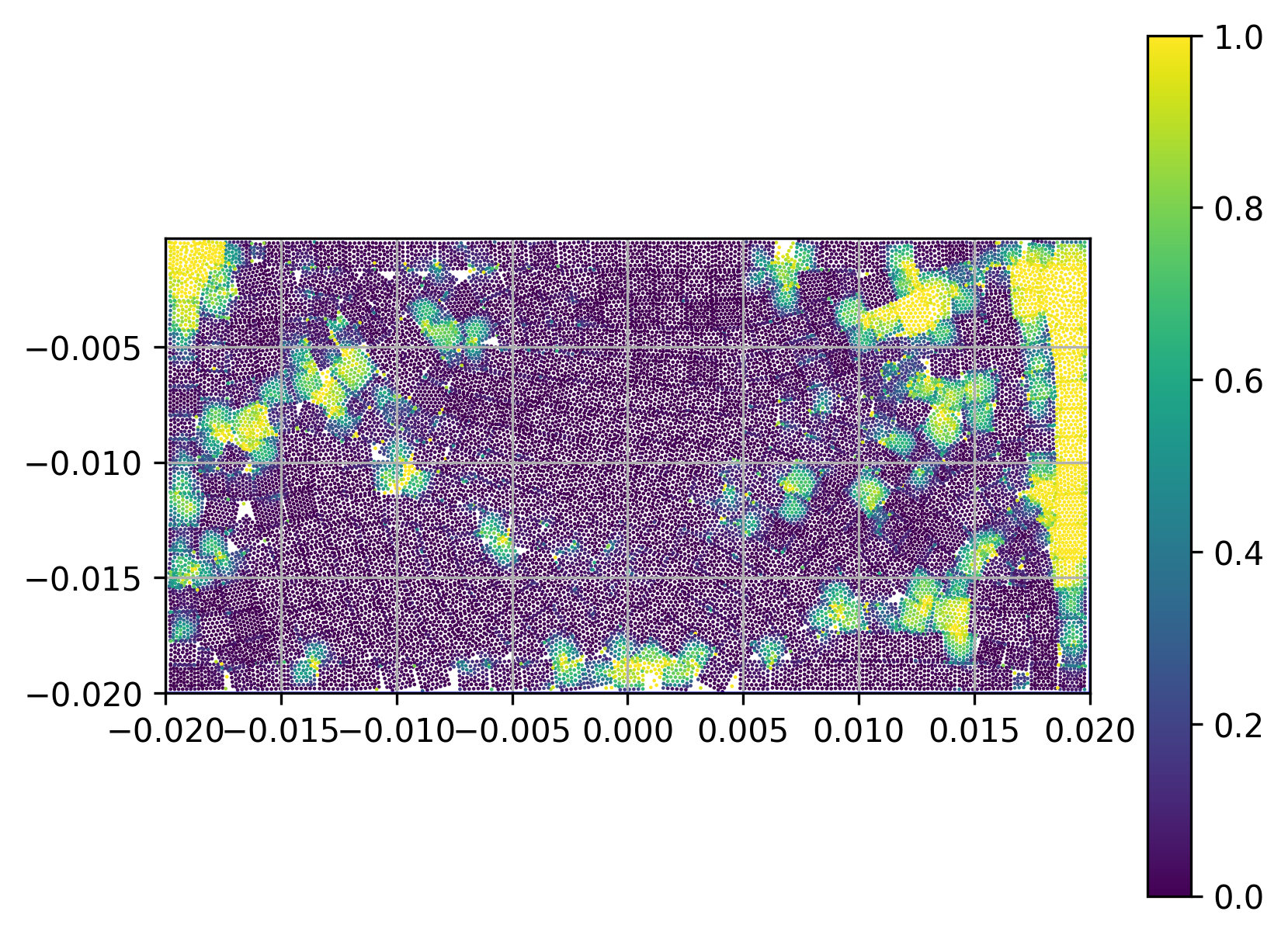}
    }
    \subfloat[]{\label{fig:compress-damage-3}
        \includegraphics[width=0.25\linewidth]{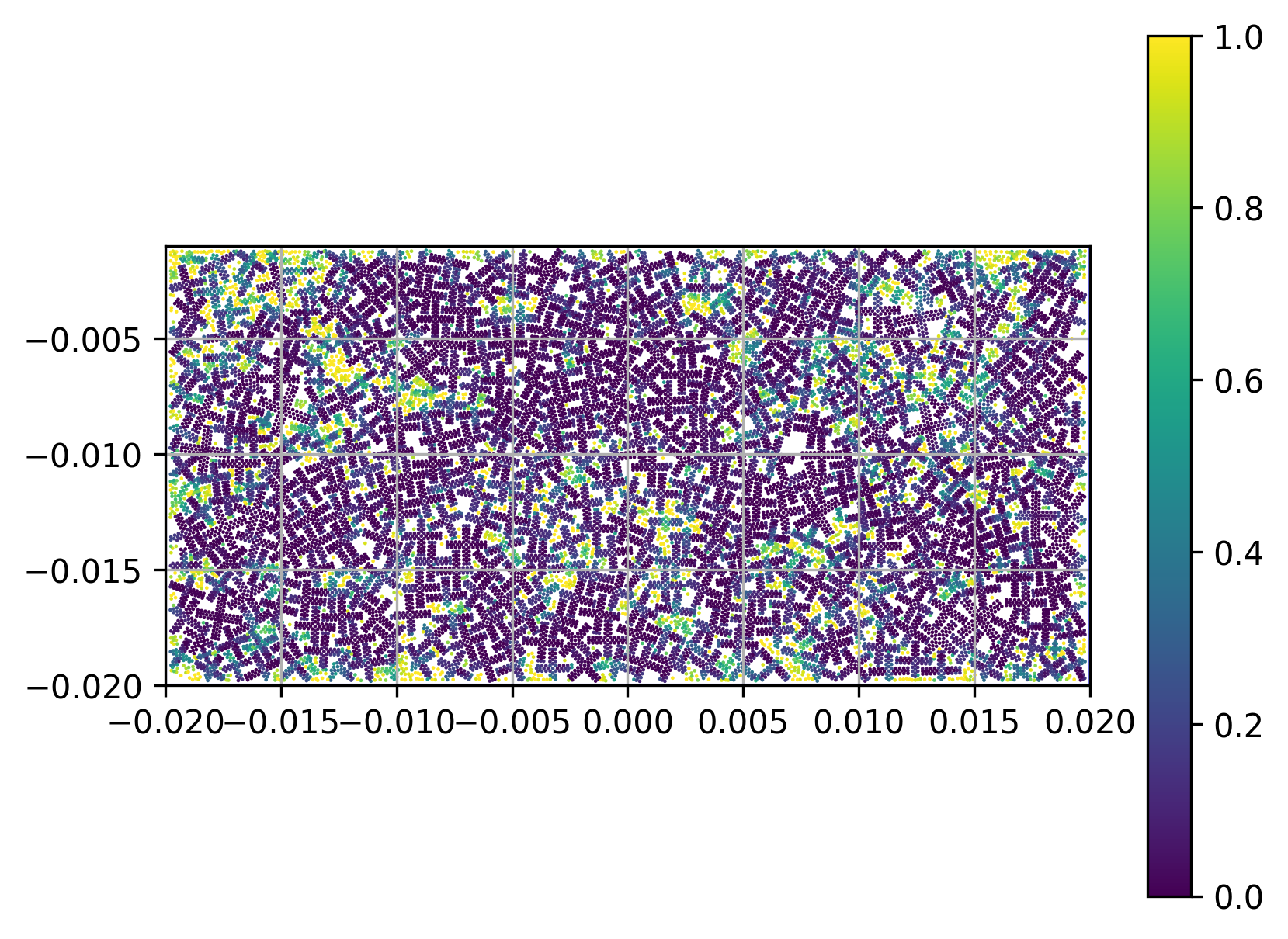}
    }
    \subfloat[]{\label{fig:compress-damage-4}
        \includegraphics[width=0.25\linewidth]{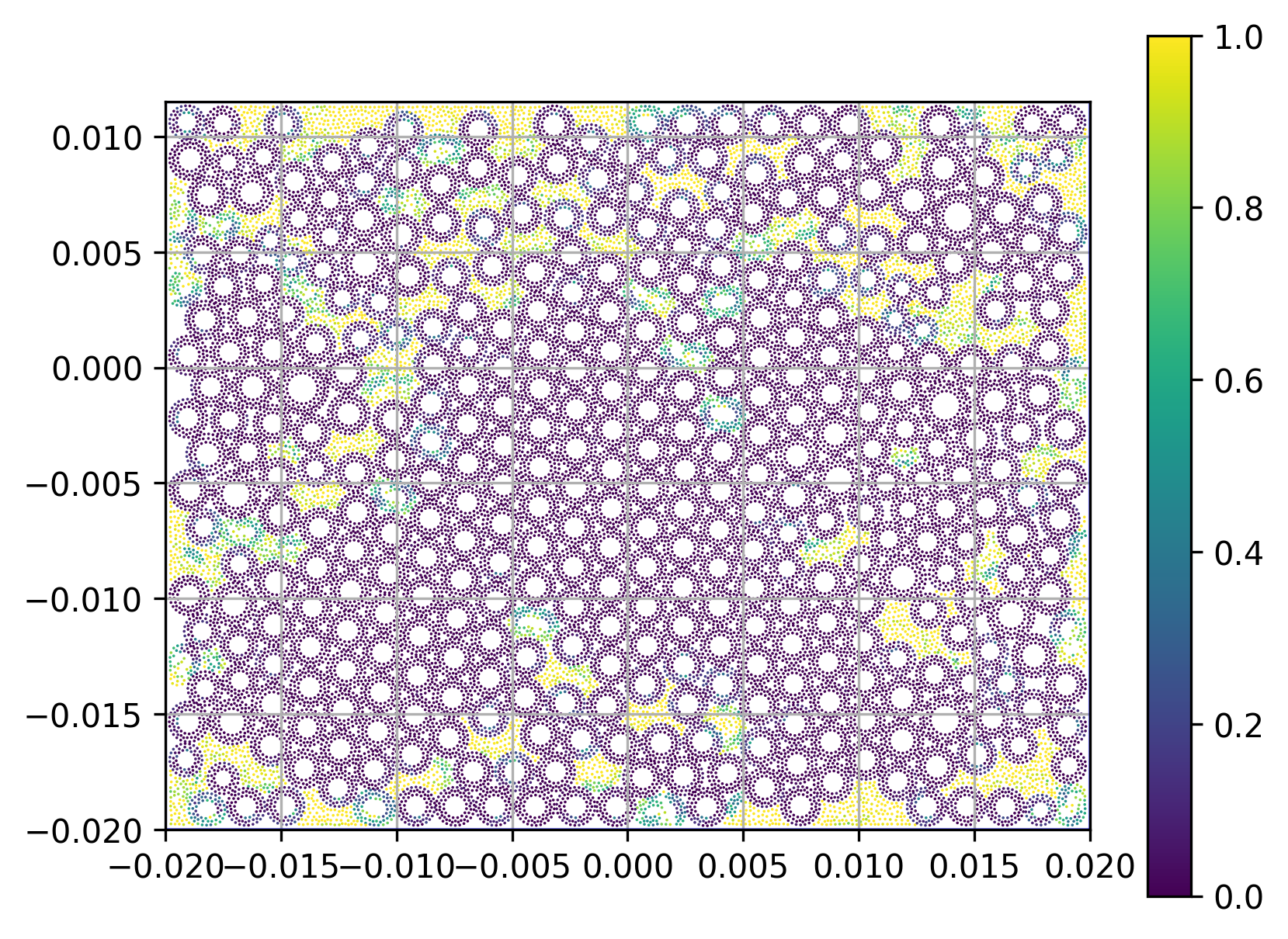}
    }
\caption{Compression test on aggregates consisting of particles of various shapes where the top wall is lowered at speed 1 m/s. 
The bottom row shows the snapshot of the aggregate when the bulk damage reaches 0.2. The top row captures the formation of particle columns before any damage has set in. 
The damage value of each node of each particle is shown in color.}%
\label{fig:compress-shapes}
\end{figure}

The initial setup and two snapshots of the simulations are shown in \Cref{fig:compress-shapes}, where the damage value of each node of each particle is plotted.
As the top wall boundary is lowered, particle aggregates become denser and eventually begin to break.
In \Cref{fig:compress-damage-1,fig:compress-damage-2,fig:compress-damage-3,fig:compress-damage-4} we show the snapshots of the aggregates when the bulk damage reaches $0.2$.
The simulation time corresponding to these snapshots are  0.018 $\mu$s, 0.020 $\mu$s, 0.021 $\mu$s, and 0.009 $\mu$s, respectively.
An intermediate configuration of each aggregate is shown in \Cref{fig:compress-column-1,fig:compress-column-2,fig:compress-column-3,fig:compress-column-4} where no damage has occurred but the particle bulk is compressed significantly. 
These snapshots are taken at 0.010 $\mu$s, 0.011 $\mu$s, 0.013 $\mu$s, and 0.006 $\mu$s after the simulation starts.  In this regime, particles form vertical columns that carry the primary mechanical load of the aggregate and eventually break down as some of the particles in these columns begin experiencing damage.
{\revv The simulations are terminated when the bulk damage reaches 1. Since aggregates of various shapes get fully damaged (i.e. when the bulk damage reaches 1) at different volume fractions, the simulations stop at different times and volume fractions.}
The series of nodes that exert contact forces on each other via the particle columns form force chains \cite{bouchaud2001force}.
We observe that damage is initiated along the force chains.

\begin{figure}[htpb]
\centering
    \subfloat[]{\label{fig:shapes-combined_forces}
     \includegraphics[width=0.4\linewidth]{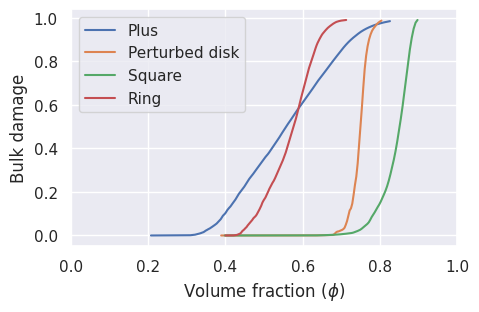}
    }
    \subfloat[]{\label{fig:shapes-combined_forces_v_damage}
     \includegraphics[width=0.4\linewidth]{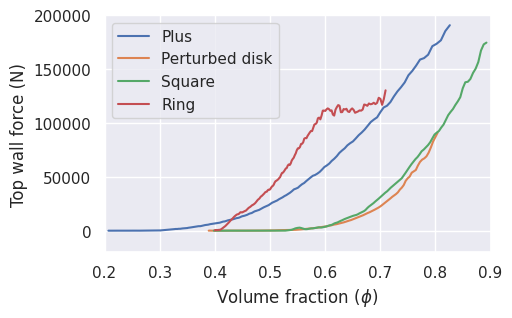}
    }
\caption{Shape effect: bulk damage and force on the top wall exerted by aggregates of difference shapes with respect to the volume fraction.}%
\label{fig:shapes-combined}
\end{figure}

The bulk damage and the force exerted by the particle aggregate on the top wall are shown in \Cref{fig:shapes-combined_forces} and \Cref{fig:shapes-combined_forces_v_damage}.
We compare the mechanical response of the bulk with respect to the volume fraction of the bulk.
 For the aggregate of plus particles the rate of damage with respect to the bulk volume fraction is the lowest among all shapes, but damage is initiated at the lowest volume fraction (at $\phi$ = 0.31) compared to other shapes.
The particle aggregate of squares experience initial damage at the highest volume fraction ($\phi$ = 0.71) and the bulk damage rate with respect to the volume fraction is the highest.
 The aggregate of ring-shaped particles provide the highest bulk strength since it exerts the highest amount of force on the top wall.
 Particles with more excluded volume (i.e., the plus and ring-shaped particles) exert more wall force compared to the ones with less excluded volume (i.e., squares and perturbed disks) at the same volume fraction both in the damaged and undamaged regime.
The aggregate with annular particles exhibits a unique non-increasing trend in the top wall force, which we investigate next.
 
\subsubsection{Particle topology and effects due to damage and excluded volume}
\label{sub:effect_of_excluded_volume}
Here, we study the effect of particle topology, in particular, the presence of holes in particles. The holes can be regarded as excluded volumes in aggregates that are unable to participate in exerting contact forces provided the particles are not crushed. 
When the excluded volume in the particle aggregate is significant, after a certain amount of damage
the wall contact force remains roughly constant even though the top wall keeps compressing the aggregate. This is observed for the aggregate of annular particles in \Cref{fig:shapes-combined_forces_v_damage} between the volume fraction $\phi=0.49$ and $0.7$. During this ``crushing''  time, broken particle fragments are able to move into the region previously enclosed by the inner circles of the annuli.
When crushing abates the wall reaction force starts increasing again.

\begin{figure}[htpb]
\centering
    \subfloat[]{\label{fig:ring-damage}
        \includegraphics[width=0.4\linewidth]{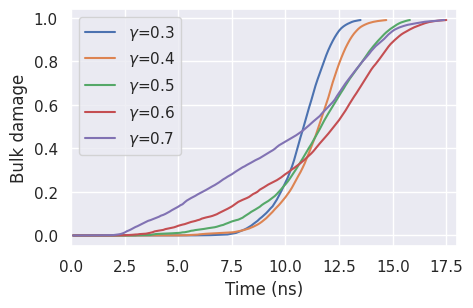}
    }
    \subfloat[]{\label{fig:ring-combined_forces}
     \includegraphics[width=0.4\linewidth]{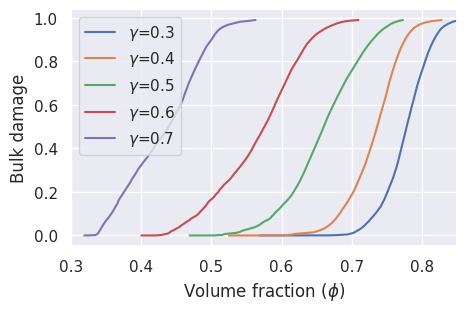}
    }
    \\
    \subfloat[]{\label{fig:ring-combined_damage}
        \includegraphics[width=0.4\linewidth]{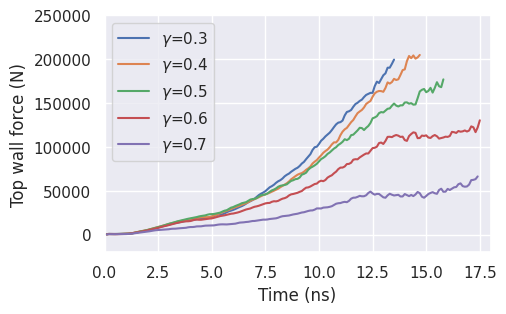}
    }
    \subfloat[]{\label{fig:ring-combined_forces_v_damage}
     \includegraphics[width=0.4\linewidth]{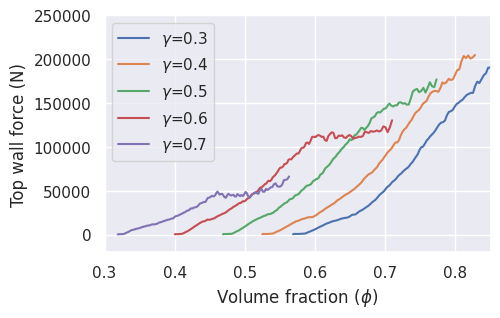}
    }
\caption{Topology effect: damage and force on the top wall by aggregates consisting of annular particles with varying thinness $\gamma$.}%
\label{fig:ring-combined}
\end{figure}

We consider annular particles with inner circle radius $r$ and outer circle radius $R$ shown in \Cref{ring-shape}.
We define the \textit{thinness} of the annular particle shown in 
\Cref{ring-shape} as
{\revv
$
    \gamma = \frac{r}{R}.
    $
}
Note that when $\gamma = 0$, the particle is a solid disk.
While keeping the outer radius $R$ fixed, we take annular particles with inner radius $r$ to be $0.3 R$, $0.4 R$, \dots, $0.7R$, which correspond to $\gamma = 0.3, \dots, 0.7$, respectively.
The wall contact force and the bulk damage for particles with various $\gamma$ values are shown in \Cref{fig:ring-combined} with respect to time in \Cref{fig:ring-combined_damage,fig:ring-damage}  and with respect to the bulk volume fraction in \Cref{fig:ring-combined_forces,fig:ring-combined_forces_v_damage}.
    The bulk damage rate with respect to both time and volume fraction is monotonically decreasing in thinness $\gamma$, i.e., particle aggregates with larger holes experience damage over a longer interval of time as well as volume fraction.  
    Moreover, aggregates with thinner particles experience initial damage earlier and vice versa, which is observed in \Cref{fig:ring-damage}.

     Since the outer radius $R$ is the same for all values of $\gamma$, the contact forces are activated at the same time across all aggregates. In general, for smaller values of $\gamma$ the wall contact force is observed to be smaller.
	However, for particles with significant thinness (e.g. $\gamma > 0.5$) the excluded volume effect is more prominent where the wall contact force remains roughly constant even though the top wall keeps compressing the bulk (observed at $t=$13 $\mu$s for $\gamma = 0.6$ and at $t = 0.012$  $\mu$s for $\gamma = 0.7$ in \Cref{fig:ring-combined_damage}). During this time, broken particle fragments are able to occupy newly available excluded volume previously enclosed by the inner circles of the annuli without exerting additional force on the top wall as the bulk is compressed.
	As the aggregates are compressed further and the excluded volumes are filled up with particle fragments, the top wall force increases again (at $t=0.017$ $\mu$s for $\gamma = 0.6 $ and $t = 0.018$ $\mu$s for $ \gamma = 0.7 $ in \Cref{fig:ring-combined_damage}).
	A similar trend is observed in \Cref{fig:ring-combined_forces_v_damage} where the top wall force is plotted with respect to the volume fraction.



\section{Conclusion}%
\label{sec:conclusion}
In this work we provide  a computational platform  with stable implementation of inter-particle damping and friction to assess aggregate motion for particles of nonconvex shape. 
 Peridynamics is coupled with DEM-like forces appropriately and the model is validated for collisions involving nonconvex domains against an experiment.
We provide a method to automate the removal of peridynamic bonds to restrict the effect of nonlocal interactions within non-convex particles of arbitrary shapes.
An algorithm to construct a jammed disk assembly is described to generate a shape-agnostic initial particle distribution for particle aggregate experiments.
Damping is introduced as a nodal interaction as opposed to an approximation using the center of mass.
Analytical expressions for wall-forces are derived to achieve higher accuracy not obtained by numerical approximation.
The effect of particle shape on settling and compaction of aggregates of deformable particles is illustrated.
Our method provides the opportunity to investigate the motion of the particle aggregate as a function of the physical properties of the individual particles including their shape, topology, elasticity, and strength.

\bibliographystyle{abbrv}
\bibliography{granular-2d}
\end{document}

%% file: ex_shared_arxiv.tex
\usepackage{lipsum}
\usepackage{amsfonts}
\usepackage{graphicx}
\usepackage{epstopdf}
\usepackage{algorithmic}
\ifpdf
  \DeclareGraphicsExtensions{.eps,.pdf,.png,.jpg}
\else
  \DeclareGraphicsExtensions{.eps}
\fi

\usepackage{xcolor}
\definecolor{azure}{rgb}{0.0, 0.5, 1.0}
\definecolor{awesome}{rgb}{1.0, 0.13, 0.32}
\definecolor{asparagus}{rgb}{0.53, 0.66, 0.42}
\definecolor{cadetgrey}{rgb}{0.57, 0.64, 0.69}

\usepackage{amsmath, amssymb, amsfonts}
\usepackage{color}
\usepackage{graphicx}
\usepackage{subfig}
\usepackage{cleveref}
\usepackage[normalem]{ulem}

\newcommand{\R}{\mathbb{R}}

\newcommand{\Z}{\mathbb{Z}}

\newcommand{\abs}[1]{\left\lvert#1\right\rvert}

\newcommand{\cc}{\mathbf{c}}
\newcommand{\ee}{\mathbf{e}}

\newcommand{\xx}{\mathbf{x}}
\newcommand{\yy}{\mathbf{y}}
\newcommand{\pp}{\mathbf{p}}
\newcommand{\qq}{\mathbf{q}}

\newcommand{\uu}{\mathbf{u}}
\newcommand{\ff}{\mathbf{f}}
\newcommand{\FF}{\mathbf{F}}
\newcommand{\Bb}{\mathbf{b}}
\newcommand{\nn}{\mathbf{n}}
\newcommand{\vv}{\mathbf{v}}

\newcommand{\xiB}{\pmb{\xi}}

\newcommand{\etaB}{\pmb{\eta}}

\newcommand{\rev}{\color{black}}
\newcommand{\revv}{\color{black}}
\newcommand{\revvv}{\color{black}}

\newtheorem{rem}{Remark}

\newsiamremark{remark}{Remark}
\newsiamremark{hypothesis}{Hypothesis}
\crefname{hypothesis}{Hypothesis}{Hypotheses}
\newsiamthm{claim}{Claim}
\headers{Grain shape effect and damage using PeriDEM}{D. Bhattacharya, R. Lipton}

\title{Simulating grain shape effects and damage in granular media using PeriDEM
\thanks{
\funding{This material is based upon work supported by the U. S. Army Research Laboratory and the U. S. Army Research Office under Contract/Grant Number W911NF-19-1-0245.}}
}

\author{Debdeep Bhattacharya\thanks{
	Department of Mathematics,
	Louisiana State University,
	Baton Rouge, Louisiana 70803
  (\email{debdeepbh@lsu.edu}).}
\and Robert P. Lipton\thanks{
Department of Mathematics,
 LSU Center of Computation \& Technology,
	Louisiana State University,
	Baton Rouge, Louisiana, 70803
  (\email{lipton@lsu.edu})}
  }

\usepackage{amsopn}
